%% file: regression_formal.tex
\documentclass[11pt,twoside]{article}

\usepackage{fullpage}

\usepackage{epsf}
\usepackage{fancyhdr}
\usepackage{graphics}
\usepackage[demo]{graphicx}
\usepackage{subfigure} 
\usepackage{psfrag}
\usepackage{microtype}
\usepackage{algorithmic}
\usepackage{color,xcolor}
\usepackage[linesnumbered,ruled]{algorithm2e}
\DontPrintSemicolon
\usepackage{color}
\newcommand{\hajek}{\mathrm{H}}

\usepackage{amsthm}
\usepackage{amsfonts}
\usepackage{amsmath}
\usepackage{amssymb,bbm}
\usepackage{comment}
\input{final_macros.tex}

\newcommand{\regu}{\lambda}
\usepackage{url}
\usepackage[colorlinks,linkcolor=magenta,citecolor=blue, pagebackref=true,backref=true]{hyperref}

\usepackage{cleveref}

\renewcommand*{\backrefalt}[4]{%
    \ifcase #1 \footnotesize{(Not cited.)}%
    \or        \footnotesize{(Cited on page~#2.)}%
    \else      \footnotesize{(Cited on pages~#2.)}%
    \fi}

\usepackage{nicefrac}
\usepackage{comment}

\usepackage{chngpage}

 \usepackage{tabularx}%

\usepackage{enumitem}
\usepackage{booktabs}
\usepackage{caption}

\usepackage{bm,bbm}
\usepackage{mathtools}

\newcommand{\simiid}{\overset{\mathrm{i.i.d.}}{\sim}}

\newcommand{\coordinate}{e}

\newcommand{\tauhat}{\widehat{\avgtreat}}

\newcommand{\bernDistr}{\mathrm{Ber}}

\newcommand{\support}{\mathrm{supp}}

\newtheorem{lemma}{Lemma}
\newtheorem{proposition}{Proposition}
\newtheorem{assumption}{Assumption}

\usepackage{mathrsfs}

\usepackage{amsmath,amsfonts,amsthm,amssymb}
\usepackage{graphicx, color, setspace}

\usepackage{comment}

\usepackage{float}
\usepackage{indentfirst}

\newtheorem{theorem}{Theorem}

\newcommand{\unifbound}{y_\infty}
\newcommand{\maxbound}{x_\infty}
\newcommand{\sparse}{k}

\newcommand{\numobs}{\ensuremath{n}}

\newcommand{\usedim}{\ensuremath{d}}

\newcommand{\cone}{\mathcal{C}}

\newcommand{\numtilde}{\widetilde{\numobs}}

\newcommand{\avgtreat}{\tau}

\newcommand{\radius}{\ensuremath{r}}

\newcommand{\fstar}{f^*}

\newcommand{\class}{\mathcal{C}}

\newcommand{\Event}{\mathscr{E}}

\newcommand{\Term}{T}

\long\def\comment#1{}

\long\def\comment#1{}

\newcommand{\funcClass}{\mathcal{F}}

\newcommand{\tauoraadj}{\tauhat_{\adj.\mathrm{oracle}}}
\newcommand{\catDistr}{\mathrm{Cat}}

\newcommand{\outcome}{y}

\newcommand{\tauora}{\tauhat_{\mathrm{oracle}}}

\newcommand{\noise}{W}

\newcommand{\resEigen}{\gamma}

\newcommand{\taustar}{\tau^*}
\newcommand{\tauhatdc}{\tauhat_{\textup{dc}}}

\newcommand{\Delhat}{\widehat{\Delta}}

\newcommand{\ftilde}{\widetilde{f}}

\newcommand{\suppset}{K}

\newcommand{\fhat}{\widehat{f}}

\newcommand{\betahat}{\widehat{\beta}}

\usepackage{scalerel}

\newcommand{\betastar}{\ensuremath{\beta^*}}

\newcommand{\Fclass}{\ensuremath{\mathcal{F}}}

\newcommand{\residual}{\ensuremath{\scaleto{\Delta}{7pt}}}

\newcommand{\convdist}{\xrightarrow{\mbox{\tiny{dist.}}}}
\newcommand{\abstractEstimator}{\mathcal{A}}

\renewcommand{\Xmat}{\ensuremath{\mathbf{X}}}

\newcommand{\treat}{T}

\newcommand{\smean}{M}

\newcommand{\fit}{R}

\newcommand{\propensity}{\pi}

\newcommand{\covariate}{x}

\newcommand{\Fun}{f}

\newcommand{\sumn}{\sum_{i=1}^n}

\newcommand{\ParDim}{d}

\newtheorem{corollary}{Corollary}

\newcommand{\emp}{\numobs}
\newcommand{\empnorm}[1]{\ensuremath{\|#1\|_{\emp}}}
\newcommand{\empinprod}[2]{\ensuremath{\inprod{#1}{#2}_{\emp}}}

\newcommand{\plainprop}{\ensuremath{\propensity_{\textup{T}}}}

\newcommand{\tauregadj}{\ensuremath{\tauhat_{\mathrm{adj}}}}

\newcommand{\Ber}{\ensuremath{\operatorname{Ber}}}

\newcommand{\fitbar}{\bar{\fit}}
\newcommand{\barfit}{\fitbar}
\newcommand{\smeanbar}{\bar{\smean}}
\newcommand{\barsmean}{\smeanbar}

\newcommand{\Treated}{\ensuremath{\mathcal{T}_\numobs}}
\newcommand{\Control}{\ensuremath{\mathcal{C}_\numobs}}

\newcommand{\myerr}{\ensuremath{\varepsilon}}

\newenvironment{carlist}
 {\begin{list}{$\bullet$}
 {\setlength{\topsep}{0in} \setlength{\partopsep}{0in}
  \setlength{\parsep}{0in} \setlength{\itemsep}{\parskip}
  \setlength{\leftmargin}{0.07in} \setlength{\rightmargin}{0.08in}
  \setlength{\listparindent}{0in} \setlength{\labelwidth}{0.08in}
  \setlength{\labelsep}{0.1in} \setlength{\itemindent}{0in}}}
 {\end{list}}

\newcommand{\bcar}{\begin{carlist}}
\newcommand{\ecar}{\end{carlist}}

\newcommand{\AsympVar}{\ensuremath{\sigma^2_\numobs}}
\newcommand{\AsympStd}{\ensuremath{\sigma_\numobs}}

\newcommand{\ErrTerm}{E}

\newcommand{\DelMSE}{\delta \mbox{MSE}}

\newcommand{\Vhat}{\ensuremath{\widehat{V}_\numobs}}
\newcommand{\Vupper}{\ensuremath{V_\numobs}}

\newcommand{\mybound}[1]{\ensuremath{\|\residual\|_{#1}}}
\newcommand{\myboundmax}[1]{\ensuremath{\|\residual\|_{\numobs}}}

\newcommand{\LevScore}{\ensuremath{\kappa^2_\numobs}}

\newcommand{\SigHat}{\ensuremath{\widehat{\mathbf{\Sigma}}}}

\makeatletter
\long\def\@makecaption#1#2{
        \vskip 0.8ex
        \setbox\@tempboxa\hbox{\small {\bf #1.} #2}
        \parindent 1.5em 
        \dimen0=\hsize
        \advance\dimen0 by -3em
        \ifdim \wd\@tempboxa >\dimen0
                \hbox to \hsize{
                        \parindent 0em
                        \hfil 
                        \parbox{\dimen0}{\def\baselinestretch{0.96}\small
                                {\bf #1.} #2
                                } 
                        \hfil}
        \else \hbox to \hsize{\hfil \box\@tempboxa \hfil}
        \fi
        }
\makeatother

\begin{document}

\begin{center}
{\bf{\LARGE{A decorrelation method for general regression adjustment
      in randomized experiments}}}

\vspace*{.2in} {\large{
 \begin{tabular}{cccc}
  Fangzhou Su$^{\dagger, \star}$ & Wenlong Mou$^{\bigtriangleup, \star}$ &
  Peng Ding$^{\dagger}$ & Martin J. Wainwright$^{\diamond, \dagger,
    \ddagger}$
 \end{tabular}
}

\vspace*{.2in}

 \begin{tabular}{c}
 Department of Electrical Engineering and Computer
 Sciences$^\diamond$\\ Department of Statistics$^\dagger$ \\ UC
 Berkeley\\
 \end{tabular}

 \medskip
 
 \begin{tabular}{c}
    Department of Electrical Engineering and Computer
    Sciences$^\ddagger$ \\ Department of Mathematics$^\ddagger$
    \\ Massachusetts Institute of Technology
 \end{tabular}
 
\medskip

 \begin{tabular}{c}
    Department of Statistical Sciences$^\bigtriangleup$
    \\ University of Toronto
 \end{tabular}
}

\begin{abstract}
We study regression adjustment with general function class
approximations for estimating the average treatment effect in the
design-based setting.  Standard regression adjustment involves bias
due to sample re-use, and this bias leads to behavior that is
sub-optimal in the sample size, and/or imposes restrictive
assumptions.  Our main contribution is to introduce a novel
decorrelation-based approach that circumvents these issues.  We prove
guarantees, both asymptotic and non-asymptotic, relative to the oracle
functions that are targeted by a given regression adjustment
procedure.  We illustrate our method by applying it to various
high-dimensional and non-parametric problems, exhibiting improved
sample complexity and weakened assumptions relative to known
approaches.

\let\thefootnote\relax\footnote{$^\star$ FS and WM contributed equally
to this work.}
\end{abstract}
\end{center}


\section{Introduction}
\label{sec:intro}

Randomized experiments are the gold standard for estimating the effect
of an intervention, as they allow for model-free inference about the
average treatment effect. Under mild
conditions, the difference of the averages over the treated and
control groups is an unbiased and consistent estimator of the average
treatment effect (ATE)~\cite{Neyman:1923,imbens2015causal}.  While
model-free inference is possible in randomized experiments, it can
be improved upon when the experimenter collects pre-treatment
covariates. In particular, if the covariates are predictive of the
potential outcomes, incorporating them in the analysis can improve the
estimation efficiency~\cite{Fisher:1935}. This line of thought leads
to procedures for regression adjustment: first, fit the outcome
functions, and then use the fitted functions to reduce the variance of
sample mean. In the asymptotic limit, the variance of the regression
adjustment estimator depends only on how well the model fits the
outcomes, which can be much lower than the variance of the outcomes
itself. The asymptotic properties of regression adjustment based on
low-dimensional linear regression are
well-understood~\cite{freedman2008regression_b,lin2013}, and provide
the foundations of causal inference with randomized controlled
experiments.

In practice, linear model classes may be restrictive, and so it is of
interest to make use of flexible non-parametric and ``black box''
methods for approximating the outcome functions.  Even when a linear
model is used, the dimension of the data can be high compared to the
sample size, so that fixed-dimension asymptotics lead to poor
approximations to the finite-sample behavior.  With these issues in
mind, a natural goal is to develop finite-sample guarantees for
regression adjustment, ones that allow for flexible choices of
function approximation.

To address these questions, a recent line of literature has been
developed to generalize the classical idea of regression adjustments
to high-dimensional and non-parametric settings. For $d$-dimensional
ordinary least square (OLS) regression, Lei and
Ding~\cite{lei2018regression} shows that classical regression
adjustment needs a sample size $\numobs \gtrsim \usedim^2$ in order
for desirable guarantees to hold\footnote{Here and elsewhere the
notation $\gtrsim$ denotes in inequality that holds up to constant
factors, and possibly logarithmic factors in dimension.}; they also
develop a de-biasing method that extends to the $\numobs \gtrsim
\usedim^{3/2}$ regime. Moving to the high-dimensional sparse models,
Bloniarz et. al.~\cite{bloniarz2015lasso} provided asymptotic
normality for Lasso-based regression adjustments in the regime
$\numobs \gtrsim s^2$, where $s$ is the sparsity level. For
non-parametric function classes, two recent
papers~\cite{cohen2023no,guo2021generalized} established theoretical
guarantees for certain non-parametric classes\footnote{Roughly
speaking, these results apply to non-parametric estimates based on
constrained least-squares within a Donsker class.} that satisfy
appropriate metric entropy conditions (see Definition 3
in~\cite{guo2021generalized}).  For general function classes, all
existing works require the outcome function (or the best approximation
thereof within some class) to be estimated at a rate faster than
$\numobs^{-1/4}$. The only exception are regression adjustments based
on ordinary least squares (OLS) estimate: in this specific setting,
Lei and Ding~\cite{lei2018regression} proposed a debiased estimator
that requires only $O(\numobs^{-1/6})$ consistency in estimating the
best linear approximations to the outcome functions. Another debiasing scheme for OLS regression adjustment was developed by~\cite{chang2021exact}, with exact unbiasedness guarantees. However, no theoretical results are derived in their paper concerning high-dimensional covariates. In a work concurrent
to this paper, Lu et al.,~\cite{lu2023debiased} further improved the debiasing
scheme under some additional assumptions, and established asymptotic normality, requiring only
consistency in estimation of the linear approximation.

In the super-population framework, the papers~\cite{wager2016high,
  list2022using} show that requirements on the rate at which the
outcome functions are estimated can be removed by imposing strong
modelling assumptions, i.e., additional probabilistic assumptions on
the covariates and outcomes, such as $\mathrm{i.i.d.}$
errors. However, in randomized controlled experiments, the only source
of randomness is the randomization of treated/control assignment, and
the modelling assumptions are usually false. Therefore, to the best of
our knowledge, the following fundamental question about regression
adjustment remains open:
\begin{quote}
Can we match oracle behavior in regression adjustment while requiring
\emph{only} $o_p(1)$ consistency from the outcome estimates?
\end{quote}
At the first glance, this goal might seem overly ambitious, especially
in light of Lei and Ding's characterization of the bias of classical
regression adjustments in the OLS
settings~\cite{lei2018regression}. Nevertheless, this paper give an
affirmative answer to this question.

We do so by developing and analyzing a new decorrelation method.  It
involves generating auxiliary random variables that split the data
into overlapping random subsets, but nonetheless have certain
independence properties.  In particular, our construction is designed
to create independence structure that affords the same benefit as
sample splitting in the super-population setting, but applies to the
design-based setting.  When regression adjustment is implemented using
our decorrelation scheme, we can show that the resulting estimates of
the treatment effect are asymptotically normal (as in a classical
analysis), and moreover, we provide finite-sample bounds relative to
an oracle estimator. Both of these results require only consistency in
estimating the best approximations of the outcome function, allowing
for arbitrarily slow rates.  Based on this point estimate, we also
develop a conservative variance estimator that can be used to develop
asymptotically valid confidence intervals.  Our theory is general, and
we instantiate it by developing its consequences for various concrete
examples, including ordinary linear regression, sparse
high-dimensional linear regression, and non-parametric regression with
smoothness-based classes and shape-constrained classes.

The remainder of the paper is organized as follows. In the remainder
of this introduce, we discuss additional related work, and then
summarize the notation used in this paper. \Cref{sec:back} introduces
the basic set-up and our estimator. \Cref{sec:main} gives point
estimation property and variance estimator. \Cref{SecExamples} gives
examples. \Cref{sec:simulation} gives simulations. \Cref{sec:proofs}
collects the proofs for \Cref{sec:main}. \Cref{sec:discussion}
concludes the paper with discussion on future work.

\paragraph{Connection to existing methodology:}
The main contribution of our paper is a simple approach to improving
the finite-sample performance of a general class of regression
adjustment procedures.  procedure. The key idea is a novel procedure
for constructing overlapping random subsets of a given dataset that
allow us to mimic a sample splitting approach, without imposing the
distributional assumptions needed for sample splitting.  Accordingly,
in this section, we compare our method to existing work on sample
splitting.

Many statistical procedures involve multiple stages, and it is
desirable to use independent randomness in each stage. When the data
are independent samples from an underlying population, a natural
approach is to use disjoint subsets at different stages. Cross-fitting
is an additional refinement, in which the role of datasets at
different stages are interchanged, thereby leading to an ensemble of
estimators that can be combined.  This classical idea has been widely
exploited in causal inference methodology with observational
studies~\cite{chernozhukov2018double,wang2020debiased,mou2023kernel}.

In the super-population setting of a randomized controlled
experiment---i.e., in which the data are assumed to be
$\mathrm{i.i.d.}$ samples from an underlying probability
distribution---the idea of sample splitting can be extended in a
natural way. Under strong probabilistic assumptions on the covariate
distribution and the outcome model, the paper~\cite{wager2016high}
studies a sample-splitting version of regression adjustment, providing
asymptotic guarantees when outcomes are fitted using high-dimensional
sparse linear regression. Their idea is further generalized to
non-parametric and machine learning-based outcome estimators by the
paper~\cite{list2022using}, which proves asymptotic normality under
super-population assumptions.

Another closely related approach is the leave-one-out method: for each
single data point, one can fit a function using the rest $(\numobs -
1)$ data points and apply to it, and the final estimator can be
obtained by averaging the outputs of this procedure on all data
points. In a series of papers, Wu and
Gagnon-Bartsch~\cite{wu2018loop,wu2021design} analyzed leave-one-out
methods for regression adjustments, providing asymptotic normality
guarantees. The resulting estimator satisfies desirable properties
such as unbiasedness.  However, in order to make leave-one-out work,
it is necessary that the correlation between fitted models at
different samples decays sufficiently fast, a condition that can be
restrictive and difficult to verify.

In contrast, our methodology uses different but overlapping subsets of
data in two stages. By careful construction of the two subsets, we can
ensure an independence structure similar to the sample splitting case,
yielding near-optimal sample complexity guarantees under mild
conditions.

\paragraph{Notation:}
Throughout this paper, we use $\{\coordinate_j\}_{j = 1}^d$ to denote
the $d$-dimensional standard bases, i.e., $\coordinate_j$ is the
vector with a one in the $j$-th coordinate, and zeros elsewhere. We
use $\vecnorm{\cdot}{p}$ to denote the $\ell^p$ norm on Euclidean
spaces, for $p \in \{1,2, \infty\}$, and we define the matrix operator
norm as $\opnorm{A} \mydefn \sup_{x \neq 0} \vecnorm{A x}{2}/
\vecnorm{x}{2}$. We also slightly abuse the notation for vector norms
with the subscript $\numobs$: for vectors $x, y \in \real^\numobs$, we
define the normalized inner product
$\inprod{x}{y}_\numobs = \numobs^{-1} \sum_{i = 1}^\numobs x_i
y_i,$
and consequently the normalized $\ell^2$-norm $\empnorm{x} \mydefn
\sqrt{\empinprod{x}{x}}$.  We use $\mathrm{Ber} (p)$ to denote the
Bernoulli distribution with parameter $p \in (0, 1)$. For a vector
$\pi$ in the $k$-dimensional probability simplex, we let $\mathrm{Cat}
(\pi)$ to be the multinomial distribution with $\Prob(X = j) = \pi_j$
for $j \in \{1,2,\cdots, k\}$. We use $\xrightarrow{\Prob}$ to denote
convergence in probability, and use $\convdist$ to denote convergence
in distribution. Given a scalar $r > 0$ and a norm
$\vecnorm{\cdot}{\nu}$, we $\ball_\nu(r) \mydefn \big\{ x ~ \mid ~
\vecnorm{x}{\nu} \leq r \big\}$ to denote the
$\vecnorm{\cdot}{\nu}$-norm ball of radius $r$.

\section{Problem set-up and the decorrelation method}
\label{sec:back}

In this section, we begin with the standard set-up of randomized
controlled trials, and the classical difference-in-means estimator.
We then describe the idea of  regression adjustment in its standard
form, before introducing a novel decorrelation scheme for performing
regression adjustment.

\subsection{Problem set-up}
\label{sec:background}

We work in the standard set-up of deterministic potential
outcomes~\cite{Neyman:1923}, along with Bernoulli assignments of
treatment over a finite population of size $\numobs$. Concretely, for
each $i \in [\numobs] \defn \{1, \ldots, \numobs \}$, let
$\outcome_i(t)$ be the deterministic potential outcome of the unit $i
\in [\numobs]$ under treatment $t \in \{0, 1\}$.  The treatments are
chosen via Bernoulli trials---that is, $\treat_i \simiid \bernDistr(
\plainprop)$.  Given the observed outcomes $\outcome_i
\defn \outcome_i(\treat_i)$ for $i = 1,2, \cdots, \numobs$, our goal
is to estimate the average treatment effect (ATE)
\begin{align}
\label{eq:ate-defn}  
\taustar \mydefn \frac{1}{\numobs} \sum_{i = 1}^\numobs \outcome_i(1)
- \frac{1}{\numobs} \sum_{i = 1}^\numobs \outcome_i(0).
\end{align}
A classical approach to doing so is via the difference-in-mean (DIM)
estimator~\cite{Neyman:1923}, given by\footnote{We divide the sums by
$\numobs_{\treatsub}$ and $\numobs_{\controlsub}$ for
simplicity. Another class of approaches, known as ``H\'{a}jek''
estimators, may be used by replacing them as the actual counts. When
regression adjustment with an intercept term is used, the two classes
of estimators are asymptotically equivalent. In
Appendix~\ref{app:hajek}, we discuss the H\'{a}jek version of our
estimators in detail.}
\begin{align}
\label{eq:diff-in-mean}  
\tauhat_{\DIM} & \defn \frac{1}{\numobs_\treatsub} \sumn \outcome_i
\cdot \treat_i - \frac{1}{\numobs_\controlsub} \sumn \outcome_i \cdot
(1-\treat_i), \qquad \mbox{where $\numobs_{\treatsub} \mydefn \numobs
  \cdot \plainprop$ and $\numobs_{\controlsub} \mydefn \numobs
  \cdot (1 - \plainprop)$.}
\end{align}
By construction, the estimate $\tauhat_{\DIM}$ is unbiased for
$\taustar$, and its $\numobs$-rescaled MSE (or variance) is given by
\begin{align}
\label{eq:var-of-dim}  
\numobs \cdot \Exs [|\tauhat_{\DIM} - \taustar|^2] = \frac{1}{\numobs}
\sum_{i = 1}^\numobs \Big\{ \frac{1 -
  \plainprop}{\plainprop} \outcome_i^2(1) +
\frac{\plainprop}{1 - \plainprop}
\outcome_i^2(0) + 2 \outcome_i(1) \outcome_i(0) \Big\}.
\end{align}
This variance can be large, so that it is natural to consider
alternative estimators that lead to reduced variances.  In many
applications, we also observe a collection of deterministic covariates
$\{\covariate_i\}_{i=1}^\numobs$ that can be used for \emph{regression
adjustment}~\cite{cassel1976some,lin2013}, a class of procedures that
we describe next.


\subsection{Standard regression adjustment}

Suppose that the observed covariates $\{x_i \}_{i=1}^\numobs$ take
values in $\real^\usedim$.  The goal of regression adjustment is to
find a pair of functions $\fstar_{\treatsub}: \real^\usedim
\rightarrow \real$ and $\fstar_{\controlsub}: \real^\usedim
\rightarrow \real$ to approximate the two potential outcomes
$\outcome_i(1)$ and $\outcome_i(0)$, respectively, such that the
induced residuals
\begin{align}
\label{eq:defn-residuals}  
  \residual_i(t) \defn \outcome_i(t) - \fstar_t(x_i), \qquad \mbox{for
    $t \in \{0, 1\}$, and $i \in [\numobs]$}
\end{align}
are small.  Given some function class $\Fclass$, a reasonable
choice---but not computable in practice---are the Euclidean
projections of the outcome functions onto $\Fclass$, given by
\begin{align}
\label{EqnOracleChoice}  
\fstar_t \in \arg \min_{f_t \in \Fclass} \big \{ \sum_{i=1}^\numobs
(y_i(t) - f_t(x_i))^2 \big \}.
\end{align}
Given such functions $\{\fstar_t \}_{t=0}^1$, we can form the
\emph{oracle adjusted estimate} of $\taustar$, given by
\begin{align}
\label{eq:oracle-classical-regadj}
\tauoraadj & \defn \frac{1}{\numobs_{\treatsub}} \sumn
\big(\outcome_i- \fstar_{\treatsub} (\covariate_i) \big) \treat_i -
\frac{1}{\numobs_{\controlsub}} \sumn \big(\outcome_i-
\fstar_{\controlsub} (\covariate_i) \big) (1-\treat_i)+
\frac{1}{\numobs} \sum_{i = 1}^\numobs \big( \fstar_{\treatsub}
(\covariate_i) - \fstar_{\controlsub} (\covariate_i) \big).
\end{align}
By construction, this estimator is unbiased, and following some
algebra, one can compute its rescaled MSE as
\begin{align}
\numobs \cdot \Exs [|\tauoraadj - \taustar|^2] = \frac{1}{\numobs}
\sum_{i = 1}^\numobs \Big\{ \frac{1 -
  \plainprop}{\plainprop} \residual_i^2(1) +
\frac{\plainprop}{1 - \plainprop}
\residual_i^2(0) + 2 \residual_i(1) \residual_i(0) \Big\}.
\end{align}
We refer to this estimator as an \emph{oracle procedure} because it is
implementable only by an oracle that knows $\{\fstar_t \}_{t=0}^1$.
In practice, since not all the outcomes are observed, the projections
$\{\fstar_t \}_{t=0}^1$ from equation~\eqref{EqnOracleChoice} cannot
be computed, but they can be targeted via a regression procedure. \\

\noindent In more detail, regression adjustment is a two-stage
procedure:
\begin{enumerate}
\item[(1)] For each $t \in \{0, 1 \}$, use the data subset $ \{ (x_i,
  y_i) \}_{i: \treat_i = t}$ to compute an estimate $\fhat_t$ of
  $\fstar_t$.
\item[(2)] Second, substitute the resulting estimates $\{ \fhat_t
  \}_{t=0}^1$ in place of $\{\fstar_t \}_{t=0}^1$ in the
  definition~\eqref{eq:oracle-classical-regadj}, thereby obtaining the
  \emph{standard regression-adjusted estimate}
\begin{align}
\label{eq:classical-regadj}
\tauregadj & \defn \frac{1}{\numobs_{\treatsub}} \sumn
\big(\outcome_i- \fhat_{\treatsub}(\covariate_i) \big) \treat_i -
\frac{1}{\numobs_{\controlsub}} \sumn \big(\outcome_i-
\fhat_{\controlsub} (\covariate_i) \big) (1-\treat_i)+
\frac{1}{\numobs} \sum_{i = 1}^\numobs \big(
\fhat_{\treatsub}(\covariate_i) - \fhat_{\controlsub}(\covariate_i)
\big).
\end{align}
\end{enumerate}

Since the function estimates $\{\fhat_t \}_{t=0}^1$ are noisy, the
estimate $\tauregadj$ has additional statistical fluctuations beyond
those present in the oracle-adjusted
procedure~\eqref{eq:oracle-classical-regadj}, and analysis is required
to understand these differences.  A line of
work~\cite{lin2013,lei2018regression,bloniarz2015lasso,guo2021generalized}
has studied the adjusted estimator $\tauregadj$ under different
conditions.  As noted by Lei and Ding~\cite{lei2018regression}, the
major challenge is correlation induced by re-using the randomness in
two stages of the procedure.  Due to this correlation, the adjustment
estimator $\tauhat_{\adj}$ can behave poorly in the finite-sample
setting.  Moreover, past finite-sample work on classical regression
adjustment has involved either stringent assumptions on the function
class, and/or sub-optimal guarantees in terms of sample complexity.
We now turn to a novel decorrelation scheme for regression adjustment
that mitigates these issues.


\subsection{Decorrelation via random subsets with overlap}
\label{subsec:estimator-construction}

In the super-population setting---in which each sample is viewed as
being drawn i.i.d. from some population distribution---one can avoid
correlations between the fitted functions $\{ \fhat_t \}_{t=0}^1$ and
the data used to estimate the treatment effect by sample splitting.
In particular, we split the data into two subsets, and using one to
fit the outcome estimates, and the other to construct the adjusted
estimator.  Cross-fitting can also be used to reduce the variance of
the resulting estimator, and has proven useful in semi-parametric
estimation (see e.g.~\cite{chernozhukov2018double}). This idea was
extended to controlled experiments assuming super-population and
randomness in the outcomes~\cite{wager2016high,list2022using}.

The validity of these methods crucially exploits probabilistic
properties of the super-population setting.  By way of contrast, our
method focuses on the finite-population framework, where the
\emph{only} randomness lies in the treatment indicators
$\{\treat_i\}_{i=1}^\numobs$.  Accordingly, we need to devise a random
subsampling scheme with the required properties.  Here we describe
such a scheme; notably, as opposed to sample splitting in the
super-population setting---in which the data is randomly split into
disjoint subsets---our procedure generates overlapping subsets.

We devise a mechanism constructing random subsets of the
data---necessarily overlapping---that induce some key independence
properties.  There are two such decompositions, one for the treated
subset $\Treated \defn \{i \in [\numobs] \mid \treat_i = 1\}$ and the
other for its complement, the control group $\Control \defn \{ i \in
[\numobs] \mid \treat_i = 0 \}$.  Let us describe at a high level some
properties of these random subsets:

\paragraph{Random subsets of treatment:}
Independently for each $i \in [\numobs]$, we generate a random pair
$(\fit_i, \smean_i) \in \{0,1 \}^2$ such that that $\max \{\fit_i,
\smean_i \} \leq \treat_i$.  This latter constraint means that we can
have $\fit_i = 1$ or $\smean_i = 1$ only if $\treat_i = 1$, so that
\begin{align*}
  \Treated(\fit) \defn \{ i \in [\numobs] \mid \fit_i = 1 \} \quad
  \mbox{and} \quad \Treated(\smean) \defn \{ i \in [\numobs] \mid
  \smean_i = 1 \}
\end{align*}
are both random subsets of $\Treated$.  We provide a construction that
ensures that $\fit_i \sim \Ber(\propensity_\fitsub)$ and $\smean_i
\sim \Ber(\propensity_\smeansub)$ are independent, where
$(\propensity_\fitsub, \propensity_\smeansub) \in (0, 1)^2$ are
probabilities that satisfy
\begin{subequations}
  \label{eqn:independence}
  \begin{align}
    \label{EqnRequire}      
    \plainprop = \propensity_{\smeansub} + \propensity_{\fitsub} -
    \propensity_{\smeansub} \propensity_{\fitsub}.
\end{align}

\paragraph{Random subsets of control:}
Similarly, for the control set $\Control$, we generate i.i.d. pairs
$(\fitbar_i, \smeanbar_i)$ such that $\max \{\fitbar_i, \smeanbar_i \}
\leq 1 - \treat_i$, and form the subsets
\begin{align*}
  \Control(\fitbar) \defn \{ i \in [\numobs] \mid \fitbar_i = 1 \} \quad
  \mbox{and} \quad \Control(\smeanbar) \defn \{ i \in [\numobs] \mid
  \smeanbar_i = 1 \},
\end{align*}  
which are random subsets of $\Control$.  As above, our construction
ensures that $\fit_i \sim \Ber(\propensity_\barfitsub)$ and $\smean_i
\sim \Ber(\propensity_{\barsmeansub})$ are independent, where the
probabilities $(\propensity_\barfitsub, \propensity_\barsmeansub)$
satisfy
\begin{align}
\label{EqnRequireBar}  
1 - \plainprop = \propensity_{\barsmeansub} +
\propensity_{\barfitsub} - \propensity_{\barsmeansub}
\propensity_{\barfitsub}.
\end{align}
\end{subequations}

\paragraph{Decorrelated regression adjustment:}

We make use of these random subsets to implement a decorrelated form
of regression adjustment.  It is a generic approach, one that applies
to any type of regression procedure $\abstractEstimator$ for
approximating the outcome functions.

\begin{itemize}
\item \textbf{Step I:} Use the data indexed by $\Treated(\fit)$ and
  $\Control(\fit)$ to estimate the outcome functions:
  \begin{subequations}
\label{eqs:dec-regadj-procedure}    
    \begin{align}
        \fhat_{\fitsub} \mydefn \abstractEstimator \Big( \big\{
        (\covariate_i, \outcome_i) \mid i \in \Treated(\fit)
        \big\}\Big), \quad \mbox{and} \quad \fhat_{\barfitsub} \mydefn
        \abstractEstimator \Big( \big\{ (\covariate_i, \outcome_i) \,
        \mid \, i \in \Control(\fitbar) \big\}\Big).
    \end{align}
    \item \textbf{Step II:} Compute the decorrelated estimator
    \begin{align}
   \tauhatdc \mydefn \frac{1}{\numobs_{\smeansub}} \sumn
   \big(\outcome_i- \fhat_{\fitsub} (\covariate_i) \big) \smean_i -
   \frac{1}{\numobs_{\barsmeansub}} \sumn \big(\outcome_i-
   \fhat_{\barfitsub} (\covariate_i) \big) \barsmean_i +
   \frac{1}{\numobs} \sum_{i = 1}^\numobs \big( \fhat_{\fitsub}
   (\covariate_i) - \fhat_{\barfitsub} (\covariate_i) \big).
    \end{align}
\end{subequations}
where $\numobs_{\smeansub} \defn \propensity_\smeansub \numobs$ and
$\numobs_{\barsmeansub} \defn \propensity_{\barsmeansub} \numobs$.
\end{itemize}
The ``decorrelation'' property of this estimator arises from the fact
that $\fit_i$ and $\smean_i$ are independent (and so are $\fitbar_i$
and $\smeanbar_i$).  Consequently, we have
\begin{align*}
\Exs \Big[\big(\outcome_i- \fhat_{\fitsub} (\covariate_i) \big)
  \smean_i \Big] & = \big (\outcome_i -
\Exs[\fhat_{\fitsub}(\covariate_i)] \big) \Exs[\fit_i] \; = \; \big
(\outcome_i - \Exs[\fhat_{\fitsub}(\covariate_i)] \big) \propensity_i,
\end{align*}
along with an analogous property for the terms involving $(\barfit_i,
\barsmean_i)$.  Thus, the estimate $\tauhatdc$ is unbiased for
$\taustar$, and moreover, we can decouple the regression error
associated with $(\fhat_\fitsub, \fhat_{\barfitsub})$ from the
statistical fluctuations associated with estimating $\taustar$.

\subsection{Properties of decorrelating sequences}

\noindent For future reference, we summarize here the properties of
the random variables used in our analysis of the DC estimator to
follow, and then describe a particular procedure that generates random
variables with these properties.
\begin{lemma}
  \label{LemConstruction}
For a Bernoulli RV $\treat \sim \Ber(\propensity_\treat)$ and any
pairs of probabilities $(\propensity_\fitsub, \propensity_\smeansub)$
and $(\propensity_\barfitsub, \propensity_\barsmeansub)$ satisfying
equations~\eqref{EqnRequire} and~\eqref{EqnRequireBar}, we can
generate a quadruple \mbox{$(\fit, \smean, \fitbar, \smeanbar) \in
  \{0,1\}^4$} of random variables such that:
  \begin{enumerate}
  \item[(a)] The random variables $\fit$ and $\smean$ are independent
    Bernoulli with parameters $\propensity_\fitsub$ and
    $\propensity_\smeansub$, respectively.
  \item[(b)] The random variables $\fitbar$ and $\smeanbar$ are
    independent Bernoulli with parameters $\propensity_\barfitsub$ and
    $\propensity_\barsmeansub$, respectively.
  \item[(c)] We have $\max \{\fit, \smean \} \leq \treat$
    and $\max \{ \fitbar, \smeanbar \} \leq 1 - \treat$.
  \item[(d)] We have $\cov \{ \smean, \smeanbar \} = -
    \propensity_\smeansub \propensity_\barsmeansub$.
  \end{enumerate}
\end{lemma}

Let us describe a particular procedure (and the one used in our
implementation of our decorrelation procedure) that generates random
variables with the asserted properties.  For scalars $a$ and $b$ in
$(0,1)$, we use $\catDistr(a, b, 1-a -b)$ denote the multinomial
distribution over the choices $\{1, 2, 3\}$ with probabilities $(a, b,
1 - a - b)$, respectively.  With this notation, our procedure consists
of the following steps:
\begin{itemize}
\item If $\treat = 1$, we sample
    \begin{align*}
    Z &\sim \catDistr \Big( \frac{\propensity_{\smeansub}
      \propensity_{\fitsub}}{\plainprop},
    \frac{\propensity_{\smeansub} - \propensity_{\smeansub}
      \propensity_{\fitsub}}{\plainprop}, \frac{ \propensity_{\fitsub}
      - \propensity_{\smeansub} \propensity_{\fitsub}}{\plainprop}
    \Big),
    \end{align*}
 and then set
\begin{align*}
 \smean & \mydefn \begin{cases} 1 & Z \in \{1, 2\},\\ 0 & Z =
   3,
        \end{cases} \quad  \fit \mydefn \begin{cases} 1 & Z \in \{1, 3\},\\
        0 & Z = 2,
        \end{cases}
        \quad\bar{\smean} = \bar{\fit} = 0.
    \end{align*}
\item If $\treat = 0$, we sample
        \begin{align*}
            Z &\sim \catDistr \Big( \frac{\propensity_{\barsmeansub}
              \propensity_{\barfitsub}}{1 - \plainprop},
            \frac{\propensity_{\barsmeansub} -
              \propensity_{\barsmeansub} \propensity_{\barfitsub}}{1 -
              \plainprop}, \frac{\propensity_{\barfitsub} -
              \propensity_{\barsmeansub} \propensity_{\barfitsub} }{1
              - \plainprop}\Big),
        \end{align*}
        and then set
        \begin{align*}
          \barsmean &\mydefn \begin{cases} 1 & Z \in \{1, 2\},\\ 0 &
            Z = 3,
        \end{cases} \quad \barfit \mydefn \begin{cases} 1 & Z \in \{1, 3\},\\
        0 & Z = 2,
        \end{cases}
        \quad \smean = \fit = 0.
    \end{align*}
\end{itemize}

In~\Cref{AppConstruction}, we prove that the quadruples $(\fit,
\smean, \fitbar, \smeanbar)$ generated from this procedure satisfy
each of the four properties (a)--(d) stated in~\Cref{LemConstruction}.


\section{Main guarantees}
\label{sec:main}

In this section, we present a series of guarantees on the DC
estimator.  We begin in~\Cref{SecOracleProp} by considering the
properties of an oracle version of the DC estimator, in which the
target functions $\{ \fstar_t \}_{t=0}^1$ are known.  We establish a
bound on the difference between the DC oracle and the regression
adjustment oracle.  In~\Cref{SecNonAsymp}, we state a non-asymptotic
bound on the difference between the DC estimate and its oracle
version.  Finally, ~\Cref{SecAsymp} is devoted to asymptotic analysis
of the DC estimator, including a guarantee of asymptotic normality and
a procedure for valid confidence intervals.


\subsection{Properties of the oracle DC-estimator}
\label{SecOracleProp}

Recall the oracle estimator~\eqref{eq:oracle-classical-regadj}
associated with a standard regression adjustment procedure.  In an
analogous fashion, we define an oracle associated with the
DC-estimator~\eqref{eqs:dec-regadj-procedure} as 
\begin{align}
\label{eqn:oracle}
\tauoradc \mydefn \frac{1}{\numobs_{\smeansub}} \sumn \big(\outcome_i-
\fstar_{\treatsub} (\covariate_i) \big) \smean_i -
\frac{1}{\numobs_{\barsmeansub}} \sumn \big(\outcome_i-
\fstar_{\controlsub} (\covariate_i) \big) \barsmean_i +
\frac{1}{\numobs} \sum_{i = 1}^\numobs \big( \fstar_{\treatsub}
(\covariate_i) - \fstar_{\controlsub} (\covariate_i) \big).
\end{align}
As with the classical oracle~\eqref{eq:oracle-classical-regadj}, the
DC oracle presumes knowledge of the target functions $\{\fstar_t
\}_{t=0}^1$; it differs from the classical oracle in using the
Bernoulli variables $(\smean_i, \smeanbar_i)$, chosen to according
to~\Cref{LemConstruction}, as opposed to the original treatment
variables $(\treat_i, 1- \treat_i)$.

For $t \in \{0,1 \}$, define the residual $\residual_i(t) \defn
\outcome_i(t) - \fstar_t(x_i)$ associated with the $i^{th}$ unit.  A
straightforward calculation yields
\begin{align}
  \numobs \cdot \Exs \big[ |\tauoradc - \taustar|^2 \big] &=
  \frac{1 - \propensity_\smeansub}{\propensity_\smeansub}
  \vecnorm{\residual (1)}{\numobs}^2 + \frac{1 -
    \propensity_{\barsmeansub}}{\propensity_{\barsmeansub}}
  \vecnorm{\residual (0)}{\numobs}^2 + 2 \inprod{\residual
    (1)}{\residual
    (0)}_\numobs. \label{eq:tauora-variance-exact}
\end{align}
Condition~\eqref{eqn:independence} implies that $\propensity_\smeansub
< \plainprop$ and $\propensity_{\barsmeansub} \leq 1 -
\plainprop$. As a result, the oracle estimator $\tauoradc$ under the
decorrelated scheme has slightly larger variance than the oracle
$\tauoraadj$ discussed in the previous section. In particular, define
the $\numobs$-rescaled difference in mean-squared errors
\begin{subequations}
\begin{align}
\DelMSE(\tauoradc, \tauoraadj) \defn \numobs \Big \{ \Exs \big[
  |\tauoradc - \taustar|^2 \big] - \Exs \big[ |\tauoraadj -
  \taustar|^2 \big] \Big \}.
\end{align}
Some calculation then shows that
\begin{align}
\label{eq:efficiency-loss-of-dc-oracle}  
\DelMSE(\tauoradc, \tauoraadj) & = \frac{1 -
  \propensity_\smeansub}{\propensity_\smeansub}
\propensity_\fitsub \vecnorm{\residual (1)}{\numobs}^2 + \frac{1 -
  \propensity_{\barsmeansub}}{\propensity_{\barsmeansub} ( 1 -
  \plainprop)} \propensity_{\barfitsub}
\vecnorm{\residual(0)}{\numobs}^2.
\end{align}
\end{subequations}
Consequently, as long as $\plainprop$ remains bounded away from $0$
and $1$, by choosing the probabilities $\propensity_\fitsub$ and
$\propensity_\barfitsub$ sufficiently close to zero, we can ensure
that the efficiency loss of the DC oracle compared to the standard
oracle is arbitrarily small.  However, small choices of these
probabilities mean that the computable DC estimator uses smaller
(random) subsets of the data to approximate the oracle functions.
This tradeoff effectively disappears in the asymptotic limit as
$\numobs \rightarrow +\infty$, but is important in the finite-sample
regime.  Our non-asymptotic theory to follow makes this trade-off
precise.


\subsection{Non-asymptotic bounds on $\tauhatdc$}
\label{SecNonAsymp}

In this section, we provide a non-asymptotic bound on the difference
$\abss{\tauhatdc - \tauoradc}$ between the DC estimator and its oracle
version.  Recall that the first step
(cf. equation~\eqref{eqs:dec-regadj-procedure}) in the DC estimator is
to compute estimates $\fhat_{\fitsub}$ and $\fhat_{\barfitsub}$ of the
functions $\fstar_1$ and $\fstar_0$.  We assume that the estimate
$\fhat_\fitsub$ is accurate in the following sense.  Given an error
probability $\delta \in (0, 1)$, if the estimator is applied using the
observed outcomes indexed by $\Treated(\fit)$ with $\fit_i \sim
\Ber(\propensity_\fitsub)$, then we have
\begin{align}
\label{EqnErrAssumption}  
  \vecnorm{\fhat_{\fitsub} - \fstar_1}{\numobs} & \leq
  \myerr_\treatsub(\propensity_{\fitsub}, \delta) \quad \mbox{with
    probability $1 - \delta$},
\end{align}
where $(\propensity_\fitsub, \delta) \mapsto \myerr_\treatsub$ is an
error function.  We also impose the analogous assumption for the
estimate $\fhat_{\barfitsub}$ in terms of an error function
$\myerr_\controlsub(\propensity_{\barfitsub}, \delta)$.

\begin{theorem}
\label{ThmMain}
Suppose that we implement the DC-estimator using:
\bcar
\item a sequence of i.i.d. quadruples $(\smean_i, \barsmean_i,
  \fit_i, \barfit_i)_{i=1}^\numobs$ satisfying the properties
  of~\Cref{LemConstruction}
\item a function estimate $\fhat_\fitsub$ satisfying the
  bound~\eqref{EqnErrAssumption}, with the analogous condition for the
  estimate $\fhat_\barfitsub$.
\ecar
Then for any $\delta \in (0, 1)$, we have
\begin{align}
\label{eq:main-theorem}  
\sqrt{\numobs} \abss{\tauhatdc - \tauoradc} \leq \underbrace{\Biggr \{
  \frac{\myerr_\treatsub (\propensity_{\fitsub}, \delta/ 4
    \big)}{\propensity_\smeansub} +
  \frac{\myerr_\controlsub\big(\propensity_{\barfitsub}, \delta / 4
    \big)}{ \propensity_{\barsmeansub}} \Biggr \}}_{\mbox{Function
    estimation error}} \cdot \sqrt{\log (4 / \delta)}
\end{align}
with probability at least $1 - \delta$.
\end{theorem}
\noindent See~\Cref{SecProofThmMain} for the proof. \\

\Cref{ThmMain} establishes an non-asymptotic error bound for
$\sqrt{\numobs} (\tauhatdc - \tauoradc)$, which (for fixed $\delta$)
is proportional to the error in estimating the functions
$\{\fstar_t\}_{t=0}^1$.  Consequently, as long as this function
estimation error converges to zero as $\numobs \rightarrow \infty$,
then we have guaranteed to have $\sqrt{\numobs} |\tauhatdc -
\tauoradc| = o_{\Prob} (1)$, a quantity that is asymptotically smaller
than the rescaled oracle error $\sqrt{\numobs} |\tauoradc - \taustar|$
itself. Second, note that
equation~\eqref{eq:efficiency-loss-of-dc-oracle} implies that, for
some universal constant $c > 0$, we have
\begin{align*}
\sqrt{\numobs} |\tauoradc - \taustar| \leq \sqrt{\numobs} |\tauoraadj
- \taustar| + c \Big \{ \vecnorm{\residual(1)}{\numobs}
\sqrt{\propensity_\fitsub} + \vecnorm{\residual(0)}{\numobs}
\sqrt{\propensity_{\barfitsub}} \Big \},
\end{align*}
with high probability. Combining with equation~\eqref{eq:main-theorem}
and the triangle inequality, we find that
\begin{align}
\label{eq:final-err-decomp}  
\sqrt{\numobs} \Big \{|\tauhatdc - \taustar| - |\tauoraadj - \taustar|
\Big \} & \leq c' \Big\{ \vecnorm{\residual(1)}{\numobs}
\sqrt{\propensity_{\fitsub}} +\myerr_\treatsub (\propensity_{\fitsub},
\delta ) \Big\} + c' \Big\{ \vecnorm{\residual (0)}{\numobs}
\sqrt{\propensity_{\barfitsub}} + \myerr_\controlsub
(\propensity_{\barfitsub}, \delta ) \Big\}.
\end{align}
The estimation error $\myerr(\propensity, \delta)$ is a decreasing
function of $\propensity$, since smaller values of $\propensity$ lead
to smaller subsets of data used for function estimation.  The final
bound~\eqref{eq:final-err-decomp} also involves a term of form
$\sqrt{\propensity}$, so that we see that there is a natural trade-off
in an optimal finite-sample choice of the pair $(\propensity_\fitsub,
\propensity_\barfitsub)$.  In~\Cref{SecExamples} to follow, we discuss
optimal choices of this pair for concrete models.

Although choosing this pair can yield improved finite-sample
guarantees, we note that the asymptotic behavior is quite robust to
these choices.  As shown in our asymptotic analysis to follow, our
scheme can match the desirable properties of the oracle estimator
$\tauoraadj$ as long as both $\propensity_\fitsub$ and
$\propensity_{\barfitsub}$ converge to $0$, and the errors in
estimating the pair $\{\fstar_t\}_{t=0}^1$ also go to zero.


\subsection{Asymptotic normality and inference}
\label{SecAsymp}

We now turn to some asymptotic guarantees for the DC estimator.  In
order to do so, we consider a sequence of models indexed by $\numobs$.
Various quantities in this model sequence---including the outcome
functions $\{\fstar_t \}_{t=0}^1$ and the treatment probability
$\plainprop$---may depend on $\numobs$, but we omit this dependence so
as to keep the notation stream-lined.


\subsubsection{Guarantee of asymptotic normality}

Recall the residuals $\residual_i(t) \defn \outcome_i(t) -
\fstar_t(x_i)$ as previously defined~\eqref{eq:defn-residuals}.  Our
asymptotic guarantee involves the variance
\begin{align}
\label{EqnDefnAsympVar}  
\AsympVar & \defn \frac{1}{\numobs} \sumn \left\{
\frac{1-\propensity_{\smeansub}}{\propensity_{\smeansub}}
\residual^2_i(1) + \frac{1 -
  \propensity_{\barsmeansub}}{\propensity_{\barsmeansub}}
\residual^2_i(0) + 2 \residual_i(1) \residual_i(0) \right\}.
\end{align}
We assume that
\begin{subequations}
\begin{align}
\label{EqnAsympVar}    
\lim \inf_{\numobs \rightarrow \infty} \AsympVar > 0, \quad \mbox{and} \quad
\frac{1}{\numobs} \sum_{i=1}^\numobs \residual_i^4(t) \leq C \qquad
\mbox{for $t \in \{0, 1\}$,}
\end{align}
both of which are natural conditions in proving a central limit
theorem.  Finally, we assume that the treatment probability
$\propensity_\treat$ remains uniformly bounded away from $0$ and
$1$---that is, there exists some $\alpha \in (0, 1)$ independent of
$\numobs$ such that
\begin{align}
\label{EqnOverlap}
\plainprop \in \big [\alpha, 1 - \alpha \big].
\end{align}
Finally, we assume our function estimates $\{\fhat_\fitsub,
\fhat_\barfitsub \}$ are consistent in probability---viz.
\begin{align}
\label{EqnConsistent}
\vecnorm{\fhat_{\fitsub} - \fstar_\treatsub}{\numobs} =
o_{\mathbb{P}}(1), \quad \mbox{and} \quad \vecnorm{ \fhat_{\barfitsub}
  - \fstar_\controlsub}{\numobs} = o_{\mathbb{P}}(1).
\end{align}
\end{subequations}
\begin{proposition}
\label{PropCLT}
Under the conditions~\eqref{EqnAsympVar}, the overlap
condition~\eqref{EqnOverlap} and the consistency
condition~\eqref{EqnConsistent}, we have
\begin{align}
\frac{\sqrt{\numobs}(\tauhatdc - \taustar) }{\AsympStd} & \convdist
\mathcal{N} (0, 1).
\end{align}
\end{proposition}
\noindent See~\Cref{SecProofPropCLT} for the proof. \\

At a high level, \Cref{PropCLT} guarantees that the DC estimator
$\tauhatdc$ matches the desirable asymptotic behavior of the oracle
$\tauoradc$.  A key fact is that it requires only
$o_{\mathbb{P}}(1)$-consistency ~\eqref{EqnConsistent} of the function
estimates, as opposed to a specific rate as a function of sample size.
This mild requirement affords more general applicability for the DC
procedure compared to analogous guarantees for different methods from
past work~\cite{guo2021generalized,cohen2023no,bloniarz2015lasso}.


\subsubsection{Confidence intervals}

In order to use~\Cref{PropCLT} to construct confidence intervals for
$\taustar$, it is necessary to construct an estimator of the variance
$\AsympVar$ from equation~\eqref{EqnDefnAsympVar}.  Unfortunately,
this is not possible, since the quantity $\AsympVar$ is not
identifiable based on the observations.  However, from classical
theory, there is an identifiable upper bound (cf.
Neyman~\cite{Neyman:1923}), namely
\begin{align}
\label{EqnVupper}  
\AsympVar \leq \Vupper \defn \frac{1}{\numobs} \sumn
\left\{\frac{\residual^2_i(1)}{\propensity_{\smeansub}}
+\frac{\residual^2_i(0)}{\propensity_{\barsmeansub}} \right\}.
\end{align}
We can estimate this upper bound, leading to a conservative confidence
interval. In particular, we define the estimator
\begin{align*}
   \Vhat^2 = \frac{1}{\numobs \propensity_{\smeansub}^2 }\sumn
   (\outcome_i - \fhat_{\fitsub} (\covariate_i) )^2 \smean_i +
   \frac{1}{\numobs \propensity_{\barsmeansub}^2} \sumn (\outcome_i -
   \fhat_{\barfitsub} (\covariate_i) )^2 \bar{\smean}_i.
\end{align*}
The following theorem gives a confidence interval based on $\Vhat^2$
with an asymptotic guarantee on its coverage.
\begin{theorem}
\label{ThmInference}
For any $\alpha \in (0, 1/2)$, let $z_\alpha$ denote the $1 -
(\alpha/2)$-quantile of a standard normal variate.  Under the
conditions of~\Cref{PropCLT}, the interval
\begin{align}
\label{EqnConservativeConf}  
\big[\tauhatdc - \Vhat z_\alpha / \sqrt{\numobs}, \tauhatdc + \Vhat z_\alpha / \sqrt{\numobs} \big] 
\end{align}
has asymptotic coverage of at least $1 - \alpha$.
\end{theorem}
\noindent See~\Cref{SecProofThmInference} for the proof of this claim.

\subsubsection{Comparison to the difference-in-means estimator}

It is also useful to compare the asymptotic variance $\AsympVar$ of
the decorrelated regression adjustment estimator $\tauhatdc$ with that
of the difference-in-means (DIM) estimator. Lin~\cite{lin2013} showed
that for low-dimensional ordinary least squares, regression adjustment
will never harm asymptotic efficiency.  This ``no-harm'' property need
not hold in general for standard regression adjustment, because it is
based on fitting the outcome models separately, without taking into
account their correlations in the finite population.
%
%
Concretely, the variance of the DIM estimator and regression adjusted
estimators depend on:
\begin{itemize}
\item DIM estimator: the quadratic terms $\{ \outcome_i^2(t)
  \}_{t=0}^1$ and cross-terms $\outcome_i(0) \outcome_i(1)$
\item Regression adjustment: the quadratic terms $\{ \residual_i^2(t)
  \}_{t=0}^1$ and cross-terms $\residual_i(0) \residual_1(1)$.
\end{itemize}
A method that estimates the functions $\fstar_t$ via least-squares
will reduce the quadratic terms.  However, the cross-terms are
incomparable, except for the special case of linear function classes
(c.f.~\cite{li2017general}, Example 9, where a decomposition result is
established for the asymptotic variance in the OLS case). Note that
the efficiency improvement can be achieved by a modified regression
adjustment approach~\cite{cohen2023no}. However, this method does not
directly extend to the decorrelated framework, so that it is an
interesting open question how to combine our decorrelation methods
with such modified regression procedures so as to improve efficiency.

That being said, as discussed preceding~\Cref{ThmInference}, it is not
actually possible to construct confidence intervals using the exact
asymptotic variance of either the DIM or regression-adjusted
procedures.  Instead, due to the lack of identifiability, we must make
use of the upper bound $\Vupper$ from equation~\eqref{EqnVupper} for a
regression-adjusted procedure, and its analog (defined in terms of the
outcomes $\{y_i(t)\}_{t=0}^1$ for the DIM estimator.  Consequently,
whenever the functions $\{\fstar_t \}_{t=0}^1$ are defined via
least-squares regression~\eqref{EqnOracleChoice}, then the
conservative confidence interval~\eqref{EqnConservativeConf}
constructed from~\Cref{ThmInference} is (asymptotically) shorter than
Neyman's conservative confidence interval based on the
difference-in-mean estimator.  This attractive property has been noted
for OLS adjustments~\cite{li2019rerandomization}, but our results
apply to much wider class of models, requiring only consistency in the
function estimation.


\section{Some consequences for specific procedures}
\label{SecExamples}

In this section, we illustrate some consequences of our general theory
for specific forms of regression adjustment. We begin
in~\Cref{sec::OLS} with the classical setting of ordinary
least-squares (OLS) regression.  We then discuss consequences for
sparse linear regression (\Cref{subsec:lasso}) and non-parametric
regression (\Cref{subsec:nonpar}).


\subsection{Ordinary least-squares linear regression}
\label{sec::OLS}

Let us begin regression adjustments based on OLS.  Given the covariates $(\covariate_i)_{i = 1}^\numobs \subseteq
\real^\usedim$, we define the covariate matrix $\Xmat
\defn \begin{bmatrix} x_1 & x_2 & \cdots & x_\numobs
\end{bmatrix}^\top \in \real^{\numobs \times \usedim}$.
We assume that the covariates $(\covariate_i)_{i = 1}^\numobs$ are
orthonormal, satisfying $\numobs^{-1} \Xmat^\top \Xmat = I_\usedim$.
We also assume without loss of generality that the covariates include
an intercept term.

For $t \in \{0, 1 \}$, we define the population least-square
coefficients
\begin{align}
\label{EqnPopulationLS}
\betastar_t \mydefn \arg \min_{\beta \in \real^\usedim}
\vecnorm{\outcome(t) - \Xmat \beta}{\numobs}^2,
\end{align}
so that the outcome functions from our general theory take the form
$\fstar_t(x) = \inprod{\betastar_t}{x}$.  Given the sequences
$(\fit_i)_{i = 1}^\numobs$ and $(\barfit_i)_{i = 1}^\numobs$, we
implement the DC estimate $\tauhatdc$ based on the least-squares
estimates
\begin{align*}
  \betahat_{\fitsub} \mydefn \arg\min_{\beta \in \real^\usedim} \Big\{
  \frac{1}{\numobs \propensity_{\fitsub}} \sum_{i = 1}^\numobs
  (\outcome_i - \inprod{\beta}{\covariate_i})^2 \fit_i \Big\}, \quad
  \mbox{and} \quad \betahat_{\barfitsub} \mydefn \arg\min_{\beta \in
    \real^\usedim} \Big\{ \frac{1}{\numobs \propensity_{\barfitsub}}
  \sum_{i = 1}^\numobs (\outcome_i - \inprod{\beta}{\covariate_i})^2
  \barfit_i \Big\}.
\end{align*}
Introducing the shorthand $\myboundmax{2} \defn \max_{t \in \{0, 1\}}
\|\residual(t)\|_\numobs$ and $\mybound{\infty} \defn \max_{t \in
  \{0,1 \}} \|\residual(t)\|_\infty$, our lower bound on the
sub-sampling probabilities involves the parameter
\begin{align}
  \mu_\numobs \defn
\tfrac{\mybound{\infty}^2}{\numobs \myboundmax{2}^2} \in \big [
  \tfrac{1}{\numobs}, 1 \big]
\end{align}
that measures the uniformity of the residuals.  When $\mu_\numobs =
1/\numobs$, then the residuals are all equal in magnitude, whereas we
have $\mu_\numobs = 1$ in the worst case.  Finally, as in the analysis
of Lei and Ding~\cite{lei2018regression}, the \emph{maximum leverage
score} $\LevScore \mydefn \max \limits_{i=1, \ldots, \numobs}
\vecnorm{\covariate_i}{2}^2/\numobs$ plays a central role in our analysis.

With this notation, we have the following result:
\begin{corollary}
  \label{cor:ols}
Given $\plainprop \in [\alpha, 1 - \alpha]$ and $\delta \in (0,1)$,
there are universal constants $c, c'$ such that choosing $\propensity_{\fitsub}
= \propensity_{\barfitsub} \in \big( c' ( \mu_\numobs + \log \usedim)
\log(1/\delta), \; \plainprop - \alpha \big)$ ensures that
\begin{align}
\label{eqn:tau-OLS-error}
\sqrt{\numobs} \abss{\tauhatdc - \tauoradctextup} \leq c \: \sqrt{
  \frac{\kappa_\numobs^2 \; \myboundmax{2}^2 \: \log
    \usedim}{\propensity_\fitsub}} \; \log(1 / \delta)
\end{align}
with probability at least $1 - \delta$.
\end{corollary}
\noindent See~\Cref{SecProofCorOLS} for the proof. \\

\paragraph{Asymptotic optimality:}
To understand the implications of the bound~\eqref{eqn:tau-OLS-error},
consider covariate vectors $x_i \in \real^\usedim$ that have uniformly
bounded entries, so that $\LevScore \precsim \usedim / \numobs$, and a
sequence of residual vectors such that $\|\residual\|_\numobs^2
\precsim 1$.  In this case, the bound~\eqref{eqn:tau-OLS-error}
ensures that
\begin{align*}
\sqrt{\numobs} \abss{\tauhatdc - \tauora^{\dc}} \leq c
\sqrt{\frac{\usedim \log \usedim}{\numobs \, \propensity_\fitsub}} \,
\log(1/\delta)
\end{align*}
for some universal constant $c$.

The permissible values of $\propensity_\fitsub$ depends on the
residual heterogeneity $\mu_\numobs$.  One reasonable scaling
\footnote{For example, these bounds are satisfied with high
probability if $(\covariate_i, \outcome_i)_{i = 1}^\numobs$ are
$\mathrm{i.i.d.}$ random, satisfying appropriate moment assumptions.}
is $\mu_{\numobs} \lesssim \tfrac{\usedim \log \usedim}{\numobs}$.  In
this case, we can allow $\propensity_{\fitsub}$ to go to zero at the
rate $\tfrac{\usedim \log \usedim}{\numobs}$, and ensure that
$\tauhatdc$ has the same asymptotic distribution as the oracle
$\tauoraadj$.

\paragraph{Finite-sample guidance:}
If we want a more refined finite-sample result, then we need to study
the trade-off involved in the
decomposition~\eqref{eq:final-err-decomp}.  Inspecting it for this OLS
case, we find that taking $\propensity_{\fitsub},
\propensity_{\barfitsub} \asymp \sqrt{\tfrac{\usedim \log
    \usedim}{\numobs}}$ leads to the error bound
\begin{align*}
\sqrt{\numobs} \Big \{ |\tauhatdc - \taustar| - |\tauoraadj -
\taustar| \Big \} & \leq c \: \Big( \frac{\usedim \log
  \usedim}{\numobs} \Big)^{1/4},
\end{align*}
with high probability, where $c$ is a universal constant. \\

Our result provides a useful bound even in the regime $\numobs \gtrsim
\usedim \log \usedim$.  This improves upon a previous guarantee, due
to Lei and Ding~\cite{lei2018regression}, that required $\numobs
\gtrsim \usedim^{3/2}$ in order for asymptotic normality to hold.  The
scaling required by our method matches that given in concurrent work
by Lu et al.~\cite{lu2023debiased}, but their analysis imposes some
non-standard tail conditions on the empirical tail behavior of the
potential outcomes (see Assumption 5 in their paper), whereas the
results given here only use standard assumptions (e.g., those in the
paper~\cite{lei2018regression}).  Moreover, our approach is not
limited to OLS: it is conceptually simpler and more generalizable than
the approach~\cite{lu2023debiased}, as shown by examples in the
sequel.

\subsection{Sparse high-dimensional linear regression}
\label{subsec:lasso}

When the problem dimension $\usedim$ is much larger than the sample
size $\numobs$, it becomes necessary to impose additional structure on
the problem in order to obtain consistent estimates.  For linear
prediction, sparsity of the regression vector is a widely studied
condition.  Accordingly, this section is devoted to analysis of
regression adjustment using sparse linear functions, and in
particular, outcome estimation via a variant of the Lasso estimator.

For simplicity in exposition, we assume in this section that both the
covariates and outcome are uniformly bounded
\begin{align}
\max_{i \in [\numobs]} \vecnorm{\covariate_i}{\infty} &\leq \maxbound,
\quad \mbox{and} \quad \max_{i \in [\numobs]} \max_{t \in \{0,1 \}}
|\outcome_i(t)| \leq \unifbound.
\end{align}
Note that we establish non-asymptotic guarantees with explicit
dependence on the pair $(\maxbound, \unifbound)$.  Thus, when
converted to asymptotic results, we are able to consider scalings in
which this pair diverges as the sample size $\numobs$ increases.

Recalling the population least-squares
coefficients~\eqref{EqnPopulationLS} $\{ \betastar_t \}_{t=0}^1$, we
assume that each vector is $k$-sparse, for some sparsity $k \ll d$.
Moreover, we assume that
\begin{align}
\label{EqnInnerProdBound}
\max_{i=1, \ldots, \numobs} |\inprod{x_i}{\betastar_t}| & \leq
\unifbound \qquad \mbox{for $t \in \{0,1 \}$,}
\end{align}
which is reasonable since the outcomes to be approximated also satisfy
this bound.  As we see momentarily, our result has a mild dependence
on $\unifbound$, so it can be chosen rather coarsely.

Given this set-up, we estimate $\betastar_1$ via the Lasso-type
program
\begin{align}
  \label{eq:lasso-program-modified}
\betahat_\fitsub & \in \arg \min_{\beta} \left \{ \frac{1}{2
  \numobs_{\fitsub}} \sum_{i = 1}^\numobs \fit_i \big(
\inprod{\covariate_i}{\beta} - \outcome_i \big)^2 + \regu_\numobs
\vecnorm{\beta}{1} \right \} \qquad \mbox{subject to $\max
  \limits_{i=1, \ldots, \numobs} |\inprod{\covariate_i}{\beta} | \leq
  \unifbound$.}
\end{align}
The estimate $\betahat_{\barfitsub}$ of $\betastar_0$ is defined in an
analogous way.  Observe that the optimization
problem~\eqref{eq:lasso-program-modified} is a minor variation of the
standard Lasso program, which does not involve the additional
constraints on the inner products $\inprod{x_i}{\beta}$.  These
side-constraints are needed to stabilize the estimator so that the sub-sampled matrix satisfies a
restricted eigenvalue (RE) condition (see equation~\eqref{EqnRE}
in~\Cref{SecProofCorLasso} for more detail).

Our guarantee applies to the estimator implemented with any
regularization parameter lower bounded as
\begin{align}
\label{eq:regularization-req-lasso}  
  \regu_\numobs \geq \frac{2}{\numobs} \vecnorm{\Xmat^T (\outcome -
    \Xmat \betastar_t) }{\infty} + 16 \covariate_\infty
  \outcome_\infty \sqrt{\frac{\log (\usedim
      /\delta)}{\propensity_\fitsub\numobs}} \qquad \mbox{for $t \in
    \{0,1 \}$.}
\end{align}
This type of choice is standard in the Lasso literature.
\begin{corollary}
  \label{CorLasso}
Suppose that $\plainprop \in [0.1, 0.5]$, the the vectors
$\{\betastar_t\}_{t=0}^1$ are $k$-sparse and satisfy the
bound~\eqref{EqnInnerProdBound}, and the covariate matrix $\Xmat$
satisfies the RE condition~\eqref{EqnRE} with parameter $\gamma > 0$.
Then for any $\delta \in (0, 1)$, the sparse DC estimate with
$\propensity_{\fitsub} = \propensity_{\barfitsub} \in \big({\log
  (\usedim /\delta)}/{\numobs}, \plainprop - 0.1 \big)$ and a
regularization parameter~\eqref{eq:regularization-req-lasso} satisfies
the bound
\begin{align}
\label{eq:lasso-corollary}  
\sqrt{\numobs} |\tauhatdc - \tauoradctextup| \leq c \left\{
\frac{\covariate_\infty \outcome_\infty}{\propensity_{\fitsub}}
\sqrt{\frac{\sparse \log (\usedim / \delta)}{\resEigen \numobs}} +
\regu_\numobs \sqrt{\frac{\sparse}{\resEigen}} \right\} \cdot
\sqrt{\log (1 / \delta)}
\end{align}
with probability at least $1 - \delta$.
\end{corollary}
\noindent See~\Cref{SecProofCorLasso} for the proof. \\

Let us discuss some implications of this guarantee, in both
the asymptotic and non-asymptotic settings.

\paragraph{Asymptotic optimality:}
Suppose that we regard $\resEigen$, $\maxbound$ and $\unifbound$ as
constants.  Then with $\regu_\numobs$ from
equation~\eqref{eq:regularization-req-lasso}, we obtain
\begin{align*}
\sqrt{\numobs} |\tauhatdc - \tauoradc| \lesssim \max_{t \in \{0,1\}}
\left\{ \frac{1}{\propensity_{\fitsub}} \sqrt{\frac{\sparse \log
    (\usedim / \delta)}{\numobs}} + \sqrt{\frac{\sparse}{\numobs}} \;
\Big \| \frac{\Xmat^T (\outcome - \Xmat \betastar_t)}{\sqrt{\numobs}}
\Big \|_\infty \right\} \cdot \sqrt{\log (1 / \delta)}.
\end{align*}
For asymptotic normality at the oracle rate, we need to ensure that
$\sqrt{\numobs} |\tauhatdc - \tauoradc| = o_{\Prob}(1)$.  It is
sufficient to require that
\begin{align}
\label{eq:lasso-regadj-sufficient-cond}  
\frac{\sparse \log \usedim}{\numobs} \overset{(a)}{\longrightarrow} 0,
\quad \mbox{and} \quad \sqrt{\frac{\sparse}{\numobs}} \Big\|
\frac{\Xmat^T(\outcome - \Xmat \betastar_t)}{\sqrt{\numobs}} \Big
  \|_\infty \overset{(b)}{\longrightarrow} 0, \quad \mbox{for $t \in
    \{0,1\}$.}
\end{align}
Under this condition, a choice such that $\propensity_\fitsub
\rightarrow 0$ and $\propensity_\fitsub / \sqrt{{\sparse \log
    d}/{\numobs}} \rightarrow + \infty$ ensures that the estimator
$\tauhatdc$ has the ideal asymptotic behavior of $\tauoraadj$.

Condition~\eqref{eq:lasso-regadj-sufficient-cond}(a) is a
standard requirement for consistency in high-dimensional sparse models
e.g., see Chap. 7 in the book~\cite{wainwright2019high}).  As stated,
equation~\eqref{eq:lasso-regadj-sufficient-cond}(b) is an asymptotic
scaling condition for a deterministic sequence of problems, but we can
verify its validity under various probabilistic modeling assumptions
on the covariates and outcomes.

It is useful to compare our guarantee in~\Cref{CorLasso} with previous
results on Lasso regression adjustment from the
paper~\cite{bloniarz2015lasso}. Rewritten in our notation, the scaling
condition 4 in the paper~\cite{bloniarz2015lasso} becomes
\begin{align}
\frac{\sparse^2 \log^2 \usedim}{\numobs}
\overset{(a')}{\longrightarrow} 0, \quad \mbox{and} \quad \sparse
\sqrt{ \frac{\log \usedim}{\numobs}} \; \Big\| \frac{\Xmat^T (\outcome
  - \Xmat \betastar_t)}{\sqrt{\numobs}} \Big\|_\infty
\overset{(b')}{\longrightarrow} 0,~ \mbox{for $t \in \{0, 1 \}$.}\label{eq:lasso-regadj-stringent-cond}
\end{align}
Comparing equation~\eqref{eq:lasso-regadj-stringent-cond} to the
conditions in equation~\eqref{eq:lasso-regadj-sufficient-cond}, both
items in the condition induce more stringent requirement on the
sparsity level $\sparse$. This difference highlights an advantage of
our decorrelation strategy: similar to the OLS case, we avoid any
quadratic scalings in the sample complexity while preserving the
oracle asymptotic properties.

\paragraph{Non-asymptotic guidance:}
Our result also suggests non-asymptotic choices of the tuning
parameters $(\propensity_{\fitsub}, \propensity_{\barfitsub})$, in
particular those that minimize
decomposition~\eqref{eq:final-err-decomp}.  The optimal choices of
these parameters should scale as $\big(\frac{\sparse \log
  \usedim}{\numobs} \big)^{1/3}$.  With this choice, if we assume a
standard scaling\footnote{For instance, this scaling holds for various
random ensembles of $(x,y)$ pairs.}  $\big \| \frac{\Xmat^T (y - \Xmat
  \betastar)}{\sqrt{\numobs}} \big \|_\infty \lesssim \sqrt{\log
  \usedim}$, we conclude that
\begin{align*}
\sqrt{\numobs} \big \{ |\tauhatdc - \taustar| - |\tauoraadj -
\taustar| \big \} & \leq c \Big( \frac{\sparse \log \usedim}{\numobs}
\Big)^{1/6},
\end{align*}
with high probability.


\subsection{Non-parametric function classes}
\label{subsec:nonpar}

Non-parametric function classes provide great flexibility in
approximating the relation between covariates and outcomes. In this
section, we show that the oracle properties in Theorem~\ref{ThmMain}
can be achieved in a fairly general setting.

Given a convex set\footnote{Here under the finite-population set-up,
we directly define the function class $\funcClass$ as a subset of
$\real^\numobs$.  When we have a class $\widetilde{\funcClass}$ of
functions mapping from the space of covariates to reals, we define
$\funcClass \mydefn \big\{ (f (x_i))_{i = 1}^\numobs ~:~ f \in
\widetilde\funcClass \big\}$.} $\funcClass \subseteq \real^\numobs$,
we consider regression adjustment based on the constrained least
squares estimate
\begin{align}
\label{eq:nonpar-regression}
\fhat_{\fitsub} \mydefn \arg \min_{\Fun \in \funcClass} \left \{
\frac{1}{\numobs_{\fitsub}} \sum_{i = 1}^\numobs \fit_i (\Fun_i -
\outcome_i)^2 \right \},
\end{align}
with $\fhat_\barfitsub$ defined in an analogous manner.  The relevant
oracle functions $\{\fstar_t\}_{t=0}^1$ are based on Euclidean
projection of the outcome vector onto $\funcClass$; in particular, see
equation~\eqref{EqnOracleChoice}.

In this case, as in standard in non-parametric analysis, our guarantee
involves the solution to a certain fixed point equation For any
compact set $\class$, we use $N(\varepsilon; \class)$ to denote its
$\varepsilon$-covering number under the norm
$\vecnorm{\cdot}{\numobs}$.  Let $\radius_{\numobs, \delta} > 0$ be
the largest positive solution to the fixed-point equation
\begin{align}
\label{eq:critical-radius}  
\radius^2 = \frac{64}{\propensity_\fitsub} \inf_{\gamma \geq 0} \left
\{ \gamma + \frac{1}{\sqrt{\numobs}} \int_{\gamma / 4}^{2 \radius}
\sqrt{\log N (t, \funcClass \cap \ball_\numobs (\radius) )} dt +
\radius \sqrt{\frac{\log (1 / \delta)}{\numobs}} \right \}.
\end{align}

\begin{corollary}
  \label{CorNonPara}
  Suppose that $\plainprop \in (0.1, 0.5)$, and we implement
  non-parametric regression adjustment with $\propensity_{\fitsub} =
  \propensity_{\barfitsub}$ belonging to the interval $(0, \plainprop
  - 0.1)$.  Then for any $\delta \in (0, 1)$, we have
\begin{align}
\label{EqnNonPara}  
\sqrt{\numobs} |\tauhatdc - \tauoradctextup| \leq \radius_{\numobs,
  \delta} \sqrt{\log (1 / \delta)}
\end{align}
with probability $1 - \delta$.
\end{corollary}
\noindent See~\Cref{SecProofCorNonPara} for the proof. \\

Let us consider some concrete instantiations of this
corollary. Consider a function class $\funcClass$ whose metric entropy
is bounded as
\begin{align}
\label{eq:entropy-condition}  
\log N(t, \funcClass ) & \leq c \, \Big(\frac{1}{t} \Big)^\alpha,
\quad \mbox{for some $\alpha > 0$.}
\end{align}
These types of entropies arise for Sobolev spaces, and other types of
smoothness classes.  Under this scaling, it can be verified that a
solution to the critical inequality~\eqref{eq:critical-radius} takes
the form
\begin{align*}
\radius_{\numobs, \delta} \leq c' \; \begin{cases}
  \big(\frac{1}{\propensity_\fitsub^2 \numobs} \big)^{\frac{1}{2 +
      \alpha}} + \sqrt{\tfrac{\log (1 / \delta)}{\propensity_\fitsub^2
      \numobs}}, & \mbox{if $\alpha < 2$,} \\
  \big(\tfrac{\log^2 \numobs}{\propensity_\fitsub^2 \numobs}
  \big)^{\frac{1}{4}} + \sqrt{\tfrac{\log (1 /
      \delta)}{\propensity_\fitsub^2 \numobs}}, & \mbox{if $\alpha =
    2$,} \\
  \big(\frac{1}{\propensity_\fitsub^2 \numobs}
    \big)^{\frac{1}{2\alpha}} + \sqrt{\tfrac{\log (1 /
        \delta)}{\propensity_\fitsub^2 \numobs}} & \mbox{if $\alpha >
      2$.}
    \end{cases}
\end{align*}
Let us consider some implications of this fact.

\paragraph{Asymptotic optimality:}
Beginning with the asymptotic view, any choice of
$\propensity_\fitsub$ (and $\propensity_\barfitsub$) such that
$\propensity_\fitsub \rightarrow 0$ while $\propensity_\fitsub \numobs
\rightarrow \infty$ ensures that the DC estimator $\tauhatdc$ is
asymptotically equivalent to $\tauoradc$, with finite-sample control
on the error guaranteed by~\eqref{EqnNonPara}.

\paragraph{Non-asymptotic guidance:}

Furthermore, combining the bound with the error
decomposition~\eqref{eq:final-err-decomp}, we can choose the optimal
values
\begin{align*}
\propensity_{\fitsub}, \propensity_{\barfitsub} \asymp \begin{cases}
  \numobs^{- \frac{2}{\alpha + 6}} & \alpha < 2,\\ \numobs^{-1/4}
  \sqrt{\log \numobs} & \alpha = 2,\\ \numobs^{- \frac{1}{\alpha + 2}}
  & \alpha > 2.
  \end{cases}
\end{align*}
Resulting in the high-order convergence rate
\begin{align}
\label{eq:nonpar-final-error-rate}  
  |\tauhatdc - \taustar| - |\tauoraadj - \taustar| \lesssim \numobs^{-
    1 / 2} \cdot \begin{cases} \numobs^{- \frac{1}{\alpha + 6}} &
    \alpha < 2,\\ \numobs^{-1/8} (\log \numobs)^{1/4} & \alpha =
    2,\\ \numobs^{- \frac{1}{2\alpha + 4}} & \alpha > 2.
\end{cases}
\end{align}
Condition~\eqref{eq:entropy-condition} only requires the metric
entropy to be controlled by a polynomial of $1/t$, but does \emph{not}
require the entropy integral to converge (which corresponds to the
$\alpha < 2$ case). This is in sharp contrast with the previous
work~\cite{guo2021generalized,cohen2023no}, and accommodates important
functions classes, including (among others) (a) $\usedim$-dimensional
$s$-order H\"{o}lder functions, where $\alpha = \usedim / s$; and (b)
$\usedim$-dimensional convex functions, where $\alpha = \usedim /
2$. (see~\cite{guntuboyina2012covering}).

This relaxed condition is achieved by our decorrelation
strategy. Indeed, it is
observed~\cite{lei2018regression,guo2021generalized} that in order to
make the bias of lower order, the standard regression adjustment
method would require the functions $\fstar_\treatsub,
\fstar_\controlsub$ to be estimated at a rate faster than
$\numobs^{-1/4}$, which rules out non-Donsker classes. By way of
contrast, our decorrelation method automatically removes the bias.


\section{Simulations}
\label{sec:simulation}

In order to illustrate and complement our theory, we present a suite
of numerical experiments designed to expose the finite-sample
performance of our estimator, and compare it with existing methods in
literature.  In~\Cref{subsec:simu-ols}, we present regression
adjustment based on ordinary least squares with $(\numobs, \usedim)$
both growing, whereas~\Cref{subsec:non-parametric-simulation} is
devoted to adjustments based on non-parametric estimates.


\subsection{Ordinary linear regression}
\label{subsec:simu-ols}

We begin with regression adjustment based on ordinary least-squares
(OLS), as discussed in~\Cref{sec::OLS}.  In order to expose
dimensional aspects of the problem, we study sequences of problems
with increasing sample size $\numobs$ and dimension $\usedim$,
according the scaling $\usedim = \lceil \numobs^\gamma \rceil$ for an
exponent $\gamma \in (0, 1)$.  We compare the following four
estimators: the decorrelated estimator $\tauhat_{\dc}$ from
equation~\eqref{eqs:dec-regadj-procedure}, the classical regression
adjustment estimator $\tauhat_{\adj}$; the difference-in-means
estimator $\tauhat_{\DIM}$ from equation~\eqref{eq:diff-in-mean}; and
the debiased estimator $\tauhat_{\mathrm{debias}}$ studied by Lei and
Ding~\cite{lei2018regression}.  We consider the balanced case with
$\plainprop = 1/2$ throughout our study.  For the decorrelated
estimator $\tauhat_{\dc}$, we follow the theoretical guidance
from~\Cref{sec::OLS}, and set $\propensity_\fitsub =
\propensity_{\barfitsub} = \min (\sqrt{d / \numobs}, 1/4)$.  When
multiple minimizers exist in the OLS problem, we pick the solution
with minimal Euclidean norm, which can be computed by replacing the
matrix inverse with the pseudo-inverse.

Additionally, we also provide numerical comparisons of the width and
coverage of confidence intervals (CIs) constructed from the point
estimators. We focus on CIs of the form
\begin{align*}
[\tauhat_\diamond - z_\alpha \widehat{V}_\diamond / \sqrt{\numobs},
  \tauhat_\diamond + z_\alpha \widehat{V}_\diamond / \sqrt{\numobs}],
\end{align*}
where $z_\alpha$ is the upper $\alpha / 2$ quantile of standard normal
distribution, and for a given method $\diamond \in \{\dc, \adj, \DIM,
\mathrm{debias}\}$, the quantity $\widehat{V}_\diamond$ is an
associated variance estimate.

Throughout our simulation studies, we take the
confidence level $\alpha = 5\%$ so that $z_\alpha \approx 1.96$. The
variance estimator $\widehat{V}_\dc^2$ is given by
Theorem~\ref{ThmInference}. For the classical regression adjustment
method, we take $\widehat{V}_{\adj}$ to be the Huber--White-2 standard
error~\cite{mackinnon1985some}, and for $\widehat{V}_{\dc}$, we use
the Huber--White-type variance estimator constructed in the
paper~\cite{lei2018regression}. For the difference-in-means method, we
use a simple plug-in variance estimator
\begin{align*}
  \widehat{V}_{\DIM}^2 = \frac{1}{\numobs \plainprop^2 }\sumn
   \outcome_i^2 \treat_i +
   \frac{1}{\numobs (1 - \plainprop)^2} \sumn \outcome_i^2 (1 - \treat_i).
\end{align*}

\paragraph{Construction of the simulation instances:}
For each $(\numobs, \usedim)$-pair, we first draw $\mathrm{i.i.d.}$ samples $\big(\widetilde{\covariate}_{ij} \big)_{ i \in
  [\numobs], j \in [\usedim]} \overset{\mathrm{i.i.d.}}{\sim}
t (2)$ from the Student's $t$-distribution with degree-of-freedom 2, re-center them, and append an intercept term, i.e., the covariates are given by
\begin{align*}
  \covariate_i = \begin{bmatrix}1 \\ \widetilde{\covariate}_i - \numobs^{-1} \sum_{\ell =
  1}^\numobs \widetilde{\covariate}_\ell \end{bmatrix} \in \real^{\usedim + 1}, \quad \mbox{for } i=1,2,\cdots, \numobs.
\end{align*}
Based on the covariates, we
generate the potential outcomes by $\outcome_i (1) = 4 \inprod{\betastar}{x_i} + \varepsilon_i$ and
$\outcome_i (0) = 0$, where $\betastar = \begin{bmatrix}0 &
  1& 1& \cdots &1
\end{bmatrix} / \sqrt{\usedim} \in \real^{d + 1}$ the vector $\varepsilon$ is computed
following the bias-maximizing strategy in Section 4.4 of the
paper~\cite{lei2018regression}; see details
in~\Cref{app:additional-simulation}.

\begin{figure}[ht!]
\begin{tabular}{ccc}
  \widgraph{0.45\textwidth}{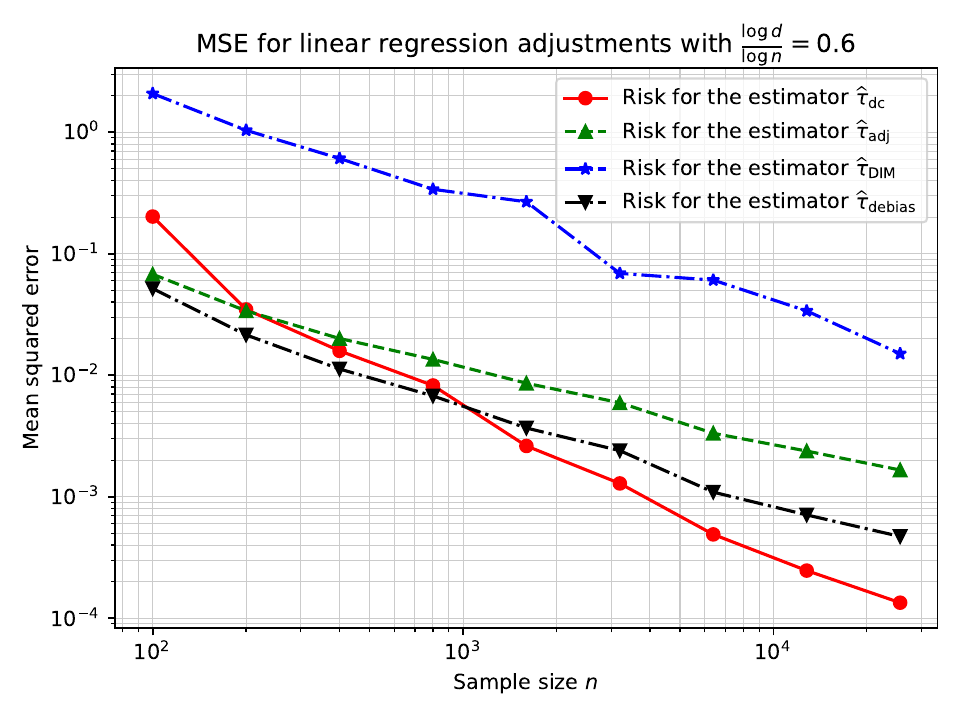} &&
  \widgraph{0.45\textwidth}{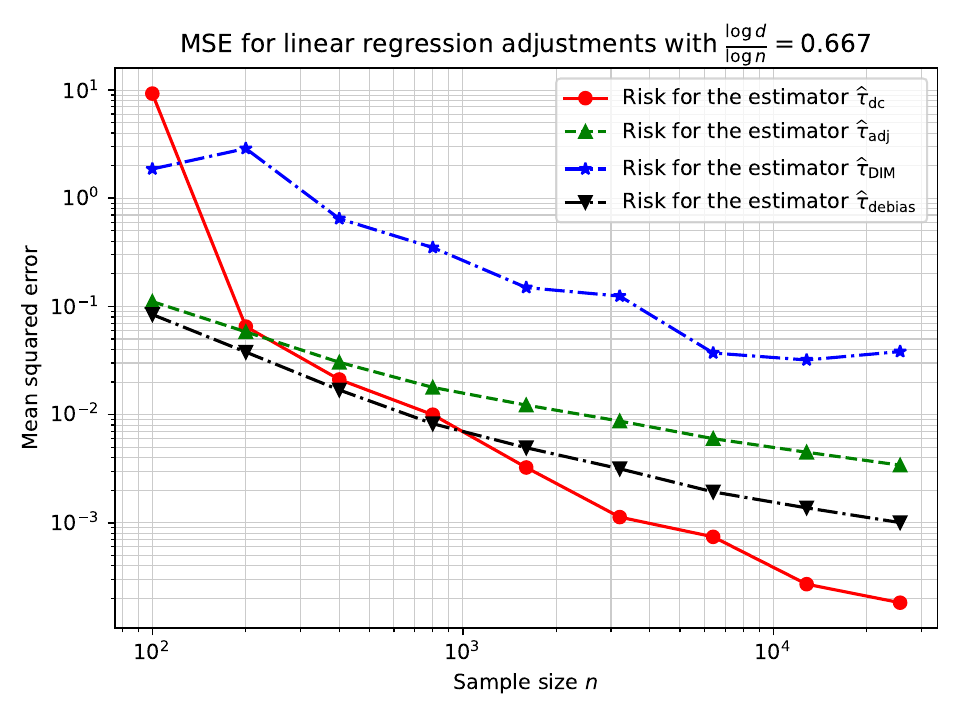} \\ (a) $\usedim
  \sim \numobs^{0.6}$ && (b) $\usedim \sim \numobs^{0.667}$ \\
  \widgraph{0.45\textwidth}{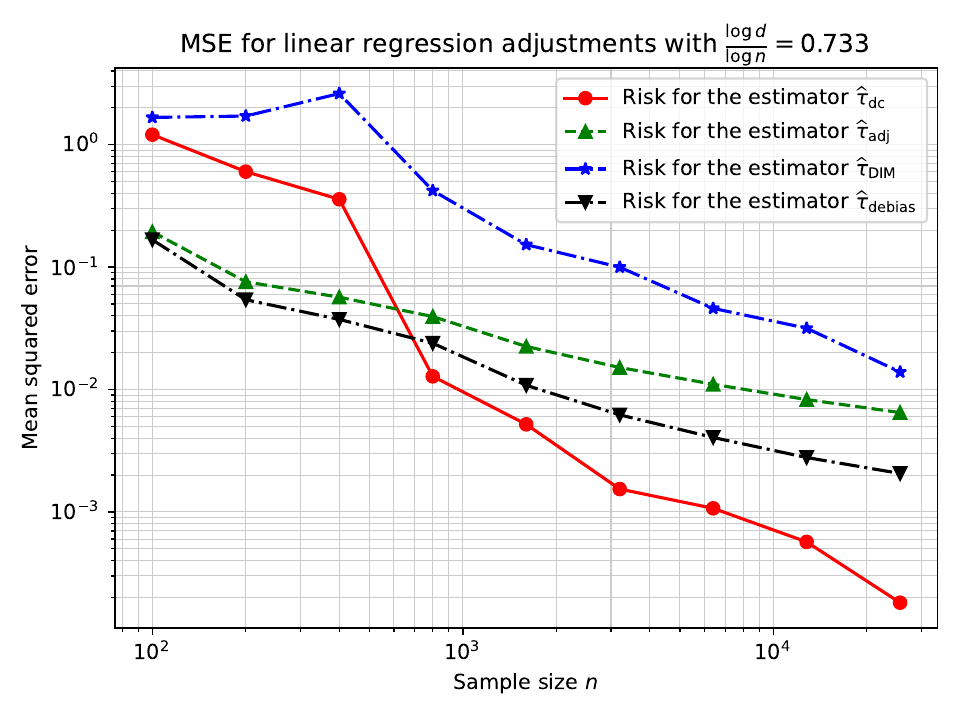} &&
  \widgraph{0.45\textwidth}{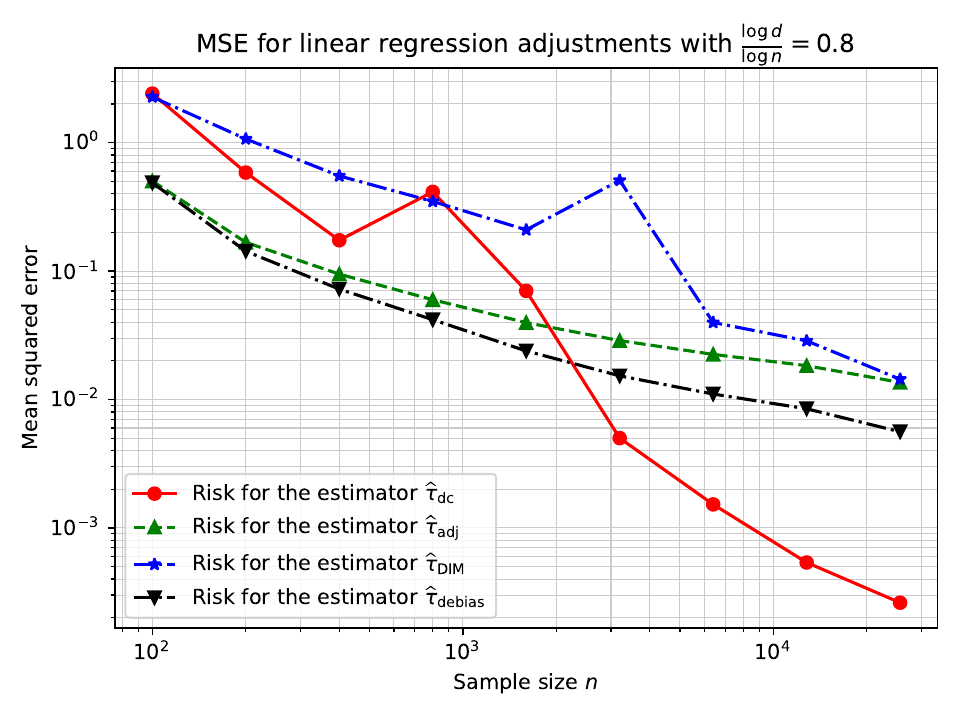} \\
(c) $\usedim \sim \numobs^{0.733}$ && (d) $\usedim \sim \numobs^{0.8}$
  \\
\end{tabular}
\caption{Plots of the mean-squared error $\Exs \big[
    \abss{\tauhat_{\diamond} - \taustar}^2 \big]$ versus sample size
  $\numobs$.  Each curve corresponds to a different algorithm
  $\diamond \in \big\{ \mathrm{dc}, \mathrm{adj}, \mathrm{DIM},
  \mathrm{debias} \big\}$. Each marker corresponds to a Monte Carlo
  estimate based on the empirical average of $100$ independent
  runs. As indicated by the sub-figure titles, each panel corresponds
  to a scaling regime in terms of the pair $(\numobs, \usedim)$. Both
  axes in the plots are given by logarithmic scales. Some of the
  curves may overlap with each other.}
\label{fig:simulation-linear}
\end{figure}

\begin{figure}[ht!]
\begin{tabular}{ccc}
  \widgraph{0.45\textwidth}{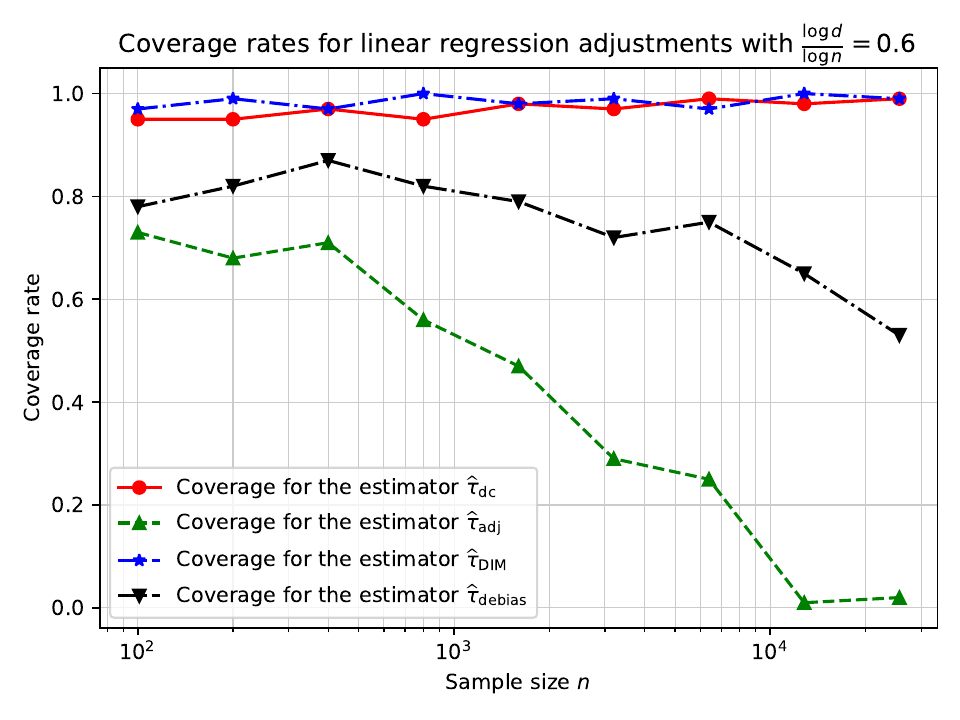} &&
  \widgraph{0.45\textwidth}{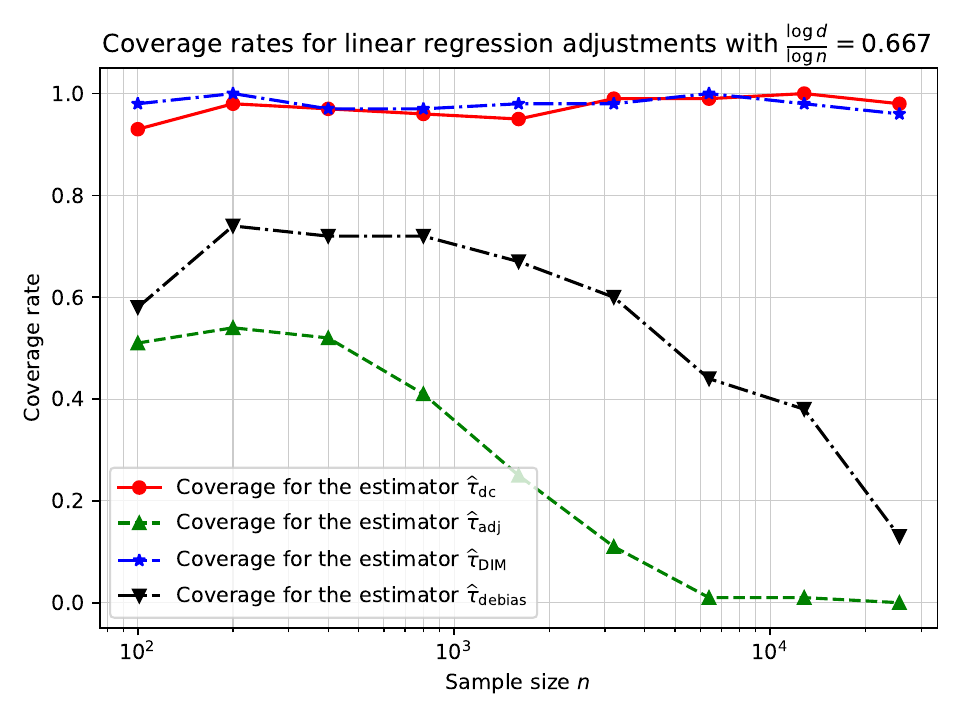} \\ (a) $\usedim
  \sim \numobs^{0.6}$ && (b) $\usedim \sim \numobs^{0.667}$ \\
  \widgraph{0.45\textwidth}{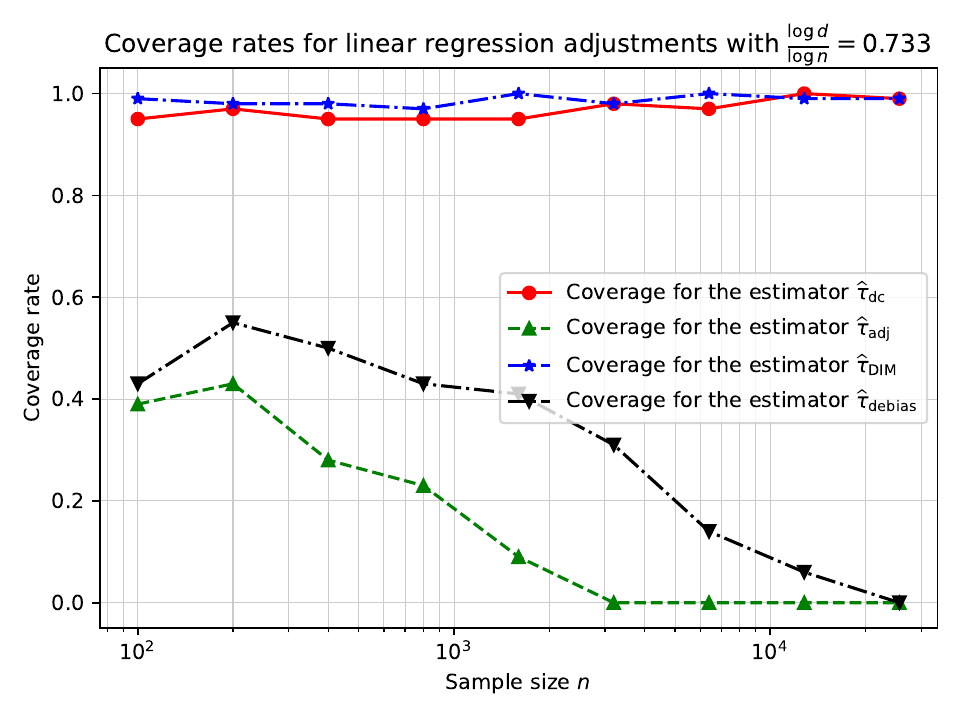} &&
  \widgraph{0.45\textwidth}{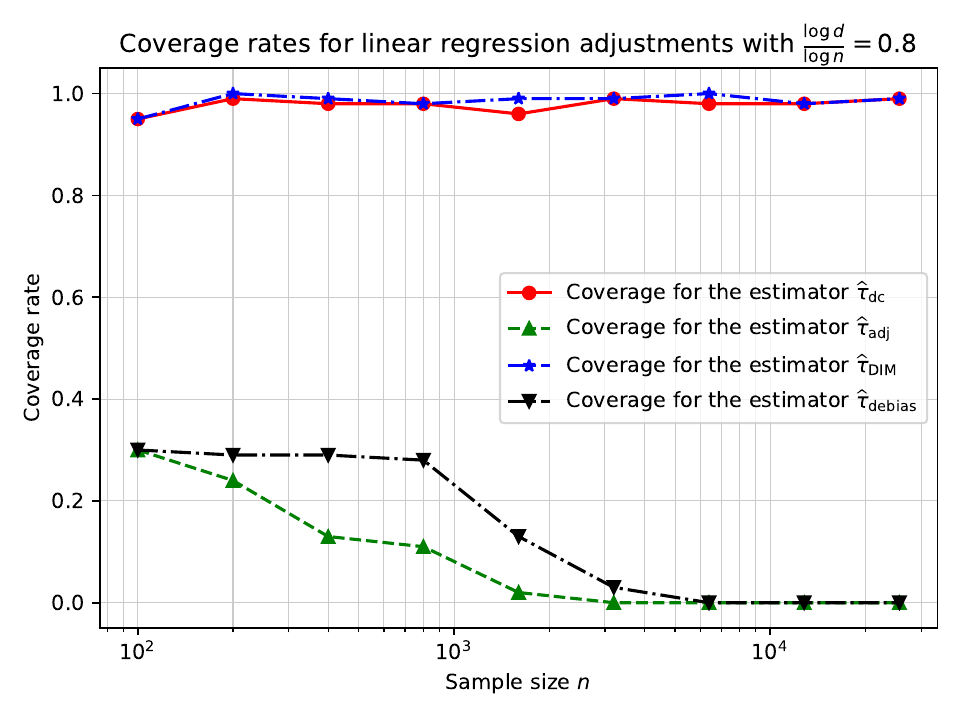} \\
(c) $\usedim \sim \numobs^{0.733}$ && (d) $\usedim \sim \numobs^{0.8}$
  \\
\end{tabular}
\caption{Plots of the coverage rate $\Prob \big(
   \abss{ \taustar - \tauhat_{\diamond} } \leq z_\alpha \widehat{V}_{\diamond} / \sqrt{\numobs} \big)$ versus sample size
  $\numobs$.  Each curve corresponds to a different algorithm
  $\diamond \in \big\{ \mathrm{dc}, \mathrm{adj}, \mathrm{DIM},
  \mathrm{debias} \big\}$. Each marker corresponds to a Monte Carlo
  estimate based on the empirical average of $100$ independent
  runs. As indicated by the sub-figure titles, each panel corresponds
  to a scaling regime in terms of the pair $(\numobs, \usedim)$. The sample size is given by logarithmic scale and the coverage rate is given by linear scale. Some of the
  curves may overlap with each other.}
\label{fig:simulation-linear-coverage}
\end{figure}

\begin{figure}[ht!]
\begin{tabular}{ccc}
  \widgraph{0.45\textwidth}{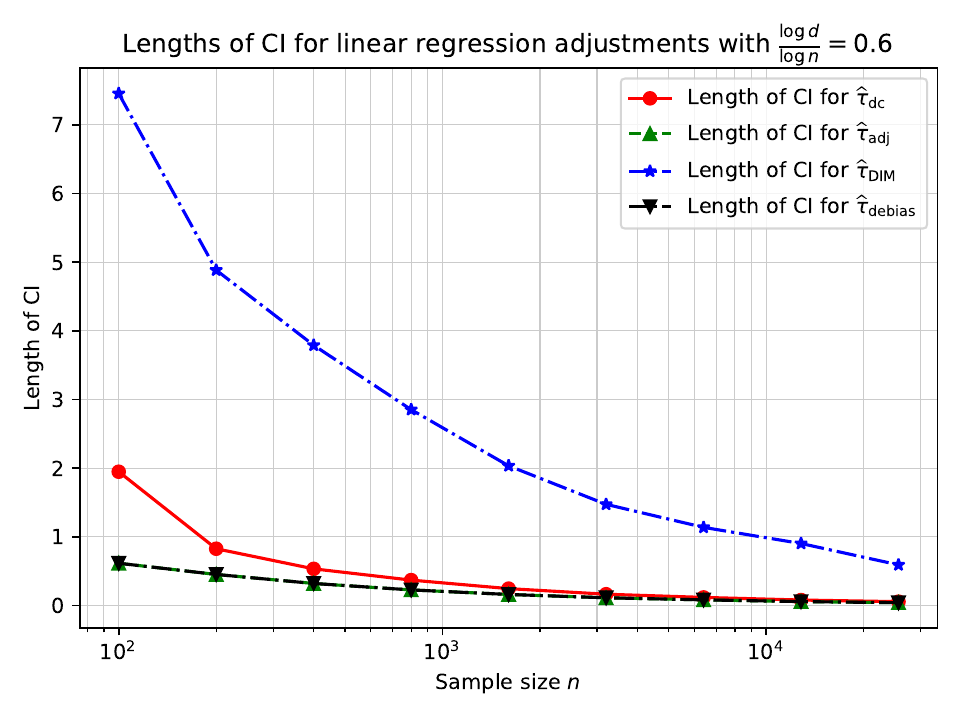} &&
  \widgraph{0.45\textwidth}{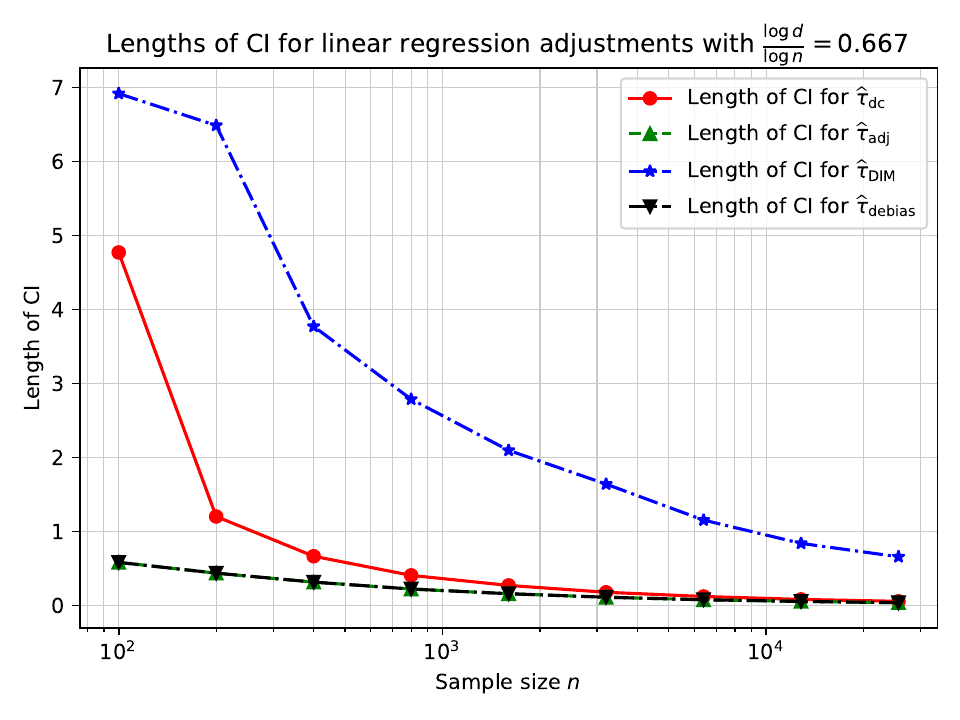} \\ (a) $\usedim
  \sim \numobs^{0.6}$ && (b) $\usedim \sim \numobs^{0.667}$ \\
  \widgraph{0.45\textwidth}{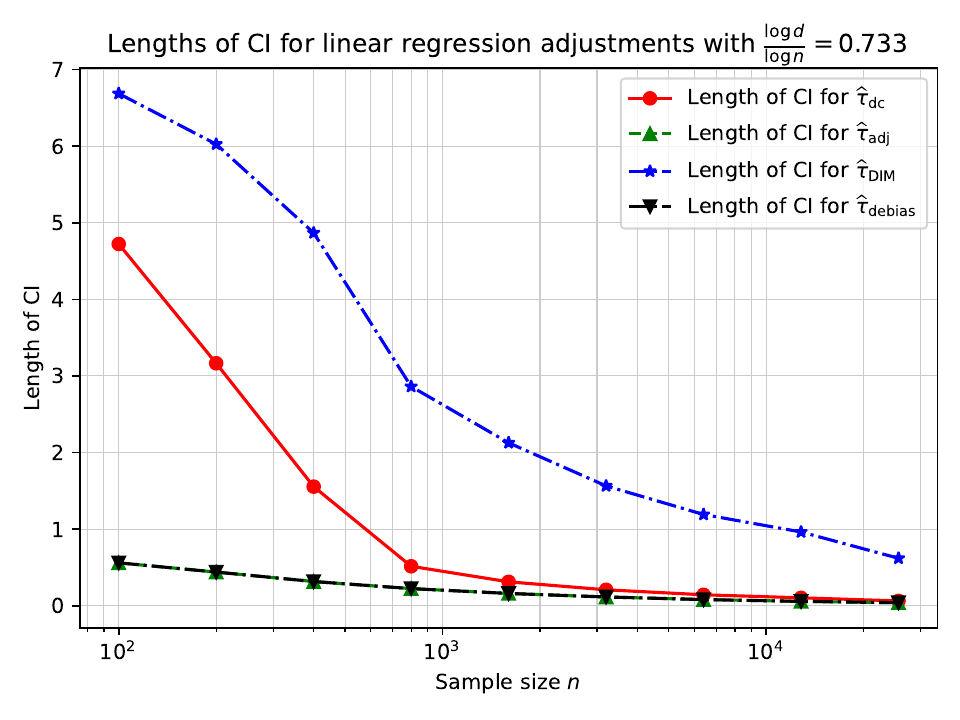} &&
  \widgraph{0.45\textwidth}{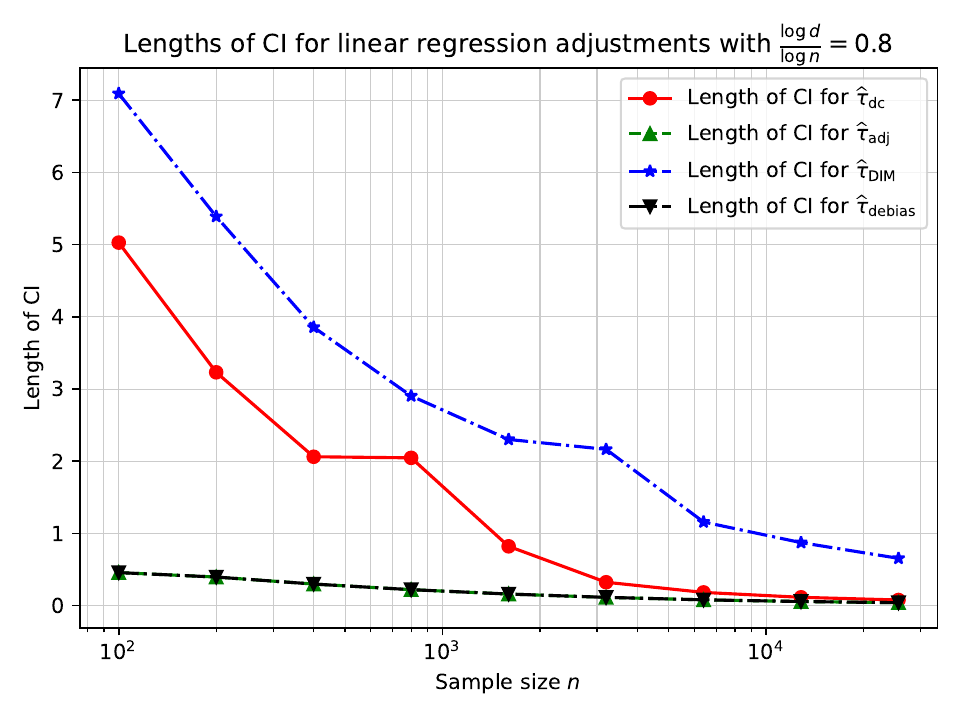} \\
(c) $\usedim \sim \numobs^{0.733}$ && (d) $\usedim \sim \numobs^{0.8}$
  \\
\end{tabular}
\caption{Plots of the length of the confidence interval $2 z_\alpha \Exs \big[ \widehat{V}_{\diamond} / \sqrt{\numobs} \big]$ versus sample size
  $\numobs$.  Each curve corresponds to a different algorithm
  $\diamond \in \big\{ \mathrm{dc}, \mathrm{adj}, \mathrm{DIM},
  \mathrm{debias} \big\}$. Each marker corresponds to a Monte Carlo
  estimate based on the empirical average of $100$ independent
  runs. As indicated by the sub-figure titles, each panel corresponds
  to a scaling regime in terms of the pair $(\numobs, \usedim)$. The sample size is given by logarithmic scale and the length is given by linear scale. Some of the
  curves may overlap with each other.}
\label{fig:simulation-linear-length}
\end{figure}

\paragraph{Simulation results:}

In
Figures~\ref{fig:simulation-linear},~\ref{fig:simulation-linear-coverage},
and~\ref{fig:simulation-linear-length}, we present the simulation
results for the four estimators on the synthetic problem instances
described above. We consider four scaling regimes, with $\usedim =
\lceil \numobs^{\gamma} \rceil$ for $\gamma \in \{0.6, 0.667, 0.733,
0.8\}$. We estimate the mean-squared errors (MSE) of point estimators,
coverage rates, and lengths of confidence intervals by averaging $100$
independent Monte Carlo trials.

From Figure~\ref{fig:simulation-linear}, we can observe that when the
sample size is large (e.g. $\numobs \gg 1000$), the decorrelated
estimator $\tauhatdc$ consistently achieves the best empirical
performance under all four regimes. With very sample sizes, it can
suffer from high variance due to insufficient data used in OLS
regression. In particular, the spike in the panel (d) at $(\numobs =
800, \usedim = 211)$ is due to the ``double-descent''
phenomenon\footnote{Concretely, for a $\usedim$-dimensional OLS based
on $m$ samples, the risk is maximized when $m \approx \usedim$, and
decreases from this maximum when the sample size $m$ becomes either
larger or smaller. In our experimental set-up, we choose
$\propensity_{\fitsub} = \propensity_{\barfitsub} = 1/4$ in this case,
leading to a sample size around $200$ being used in the first stage of
de-correlated estimator, which is close to the dimension of
covariates. The other two regression-adjustment-based estimators use
the full treatment/control groups of size around $400$, and so do not
suffer from this spike.}  in high-dimensional linear
regression~\cite{belkin2019reconciling}.

For a sufficiently large sample size, the MSE curves for $\tauhatdc$
and $\tauhat_{\DIM}$ both have slope approximately equal to $-1$,
while the difference-in-mean estimator $\tauhat_{\DIM}$ exhibits a
larger pre-factor, as shown by the large intercept in these plots with
a logarithmic scale. The cause of this large pre-factor is the strong
signal term $4 \inprod{\betastar}{x_i}$ in the data-generating
process; it can captured by a regression adjust procedure like
$\tauhatdc$, but not by the difference-in-means. The debiased
estimator $\tauhat_{\mathrm{debias}}$ consistently outperforms the
standard forms of regression adjustment, while the slopes for both
estimators become worse as the exponent $\gamma$ grows. In theory,
both $\tauhatdc$ and $\tauhat_{\DIM}$ converge at a $(\numobs^{-1/2})$
rate for any value of $\gamma$, while the debiased estimator
$\tauhat_{\mathrm{debias}}$ requires $\gamma \leq 2/3$ for such a rate
of convergence. For $\gamma \in (0.5, 1)$, the standard regression
adjustment estimator $\tauhat_\adj$ converges at a slower
$\numobs^{\gamma - 1}$ rate due to its large bias, which is consistent
with our experimental observations. Note that in our simulation, even
with a slower rate of convergence, the methods $\tauhat_\adj$ and
$\tauhat_{\mathrm{debias}}$ still outperform $\tauhat_{\DIM}$. This is
because gap in the constant factor is large while the sample size is
not large enough to counter-effect the gap. In panel (d) of
Figure~\ref{fig:simulation-linear}, we can see that the MSE curves for
$\tauhat_\adj$ and $\tauhat_\DIM$ intersects at sample size $\numobs =
25600$, and we anticipate $\tauhat_\DIM$ to outperform $\tauhat_\adj$
(and $\tauhat_{\mathrm{debias}}$ as well) with a larger sample size.

Turning to the confidence intervals, in
Figures~\ref{fig:simulation-linear-coverage}, we observe that the
coverage rates for $\tauhat_{\dc}$ and $\tauhat_{\DIM}$ are
consistently at the level of 95\%, in all regimes and with all sample
sizes. On the other hand, the estimates $\tauhat_{\adj}$ and
$\tauhat_{\mathrm{debias}}$ have degraded coverage for large sample
sizes. This is consistent with the theoretical predictions: when the
biases in $\tauhat_{\adj}$ and $\tauhat_{\mathrm{debias}}$ are
dominating, the standard deviation estimator serves as a poor proxy
for the actual error. With a finite sample size, the debiased
estimator $\tauhat_{\mathrm{debias}}$ can remove bias to some extent,
so that its confidence interval enjoys better coverage guarantees than
the standard adjusted estimator $\tauhat_{\adj}$. In the hard regime
of $\gamma \in \{ 0.733, 0.8\}$, the coverage rates for both
$\tauhat_{\mathrm{debias}}$ and $\tauhat_\adj$ go to zero as the
sample size $\numobs$ grows. In the easy regime of $\gamma = 0.6$,
however, the finite-sample coverage of $\tauhat_{\mathrm{debias}}$ is
still significantly worse than $\tauhat_{\dc}$ and
$\tauhat_{\DIM}$. This is because the latter two estimators are
unbiased in an exact sense, while $\tauhat_{\mathrm{debias}}$ only
eliminates the leading-order bias.

Finally, in Figure~\ref{fig:simulation-linear-length}, we observe that
$\tauhat_{\dc}$ the length of CIs constructed from $\tauhatdc$
exhibits a phase transition: with a small sample size (so that the
typical number of samples used to fit OLS is insufficient), the CIs
are long; when the sample size becomes large, the lengths of its CIs
become comparable to that of $\tauhat_{\adj}$ and
$\tauhat_{\mathrm{debias}}$, though the latter two exhibit poor
coverage. In contrast, the valid CIs constructed from $\tauhat_\DIM$
are significantly longer. To conclude, $\tauhatdc$ enjoys a
best-of-both-worlds performance in statistical inference --- it is
almost always valid, while being short whenever some meaningful
regression can be done.


\subsection{Non-parametric regression with interpolation}
\label{subsec:non-parametric-simulation}

We now turn to simulation studies for regression adjustment based on
non-parametric regression. In particular, we focus on a class of
non-parametric estimators that interpolate---that is, fit the training
data perfectly.  Interpolating estimators are common when using neural
networks~\cite{zhang2021understanding}, and there is associated
theory~\cite{belkin2019does,bartlett2020benign} on their
performance. In this section, we show through simulation studies that
the standard regression adjustment estimator $\tauhat_\adj$ exhibits
significantly worse error due to the large bias, while the
decorrelated estimator $\tauhat_\dc$ is $\sqrt{\numobs}$-consistent.

A regression method is said to \emph{interpolate} the training data
$(\covariate_i, \outcome_i)_{i \in \mathcal{I}}$ if it produces a
function estimate $\fhat$ such that $\fhat (\covariate_i) =
\outcome_i$ for any $i \in \mathcal{I}$. When an interpolating
estimator is used for classical regression adjustment, the resulting
ATE estimator takes the form
\begin{align*}
\tauhat_\adj = \frac{1}{\numobs} \sum_{i = 1}^\numobs \big(
\fhat_{\treatsub} (\covariate_i) - \fhat_\controlsub (\covariate_i)
\big),
\end{align*}
so that it corresponds to an outcome regression estimator. As a
result, any biases in the estimated functions $(\fhat_\treatsub,
\fhat_\controlsub)$ will lead to a biased estimator $\tauhat_\adj$,
and the rate of convergence can be slower than $\numobs^{-1/2}$. Our
decorrelated method, on the other hand, always achieves
$\numobs^{-1/2}$-rate as well as the efficiency of the oracle
estimator.

To illustrate this fact, we consider a one-dimensional non-parametric
estimation problem, with the covariates $ \covariate_i \in [0, 1]$ for
$i = 1, \ldots, \numobs$. Given a H\"{o}lder smoothness exponent $\alpha \in (0,
1]$, the non-parametric estimator is given by a shrinkage version of
the regressogram estimator, along with a normalization step. In
particular, we split the interval evenly $[0, 1]$ into $B_\numobs =
\lceil \numobs^{\frac{1}{1 + 2 \alpha}} \rceil$ segments $(I_b)_{b \in
  [B]}$ with equal lengths. Given the data $(\covariate_i,
\outcome_i)_{i\in \mathcal{I}}$, we define
\begin{align*}
\ftilde(x) \mydefn \big(1 - \numobs^{- \frac{\alpha}{2\alpha + 1}}
\big) \sum_{b = 1}^B \bm{1}_{x \in I_b} \cdot \frac{\sum_{i \in
    \mathcal{I}} y_i \bm{1}_{x_i \in I_b}}{\sum_{i \in \mathcal{I}}
  \bm{1}_{x_i \in I_b}}.
\end{align*}
Note that only one term in the summation $\sum_{b = 1}^B$ is non-zero,
which corresponds to the line segment where the point $x$ lies.

Given the estimated pair $(\ftilde_\fitsub, \ftilde_{\barfitsub})$,
we further define the interpolating functions $\fhat_\fitsub,
\fhat_{\barfitsub}$ as
\begin{align*}
\fhat_{\star} (x_i) \mydefn \begin{cases} \outcome_i & \star_i = 1 \\
\ftilde_{\star} (x_i) & \mbox{otherwise}
  \end{cases}, \quad
  \mbox{for $\star \in \{\fitsub, \barfitsub\}$ and $i \in
    [\numobs]$.}
\end{align*}
In the settings of random design and outcomes, we can verify that the
estimators $(\fhat_{\fitsub}, \fhat_{\barfitsub})$ achieve the
$\numobs^{- \frac{\alpha}{1 + 2\alpha}}$ minimax rate of convergence
for $\alpha$-H\"{o}lder
functions~\cite{tsybakov2008introduction}. Following arguments similar
to those used to prove~\Cref{CorNonPara}, we can also provide
guarantees for the finite-design setting considered here.
\footnote{To be clear, however, we have chosen this estimator simply
for illustrative purposes, not because of any strong preference for it
over other alternatives.}

Similar to Section~\ref{subsec:simu-ols}, we also study the confidence
sets constructed from the point estimators being considered. We use
the variance estimator in Theorem~\ref{ThmInference} for $\tauhatdc$,
and the plug-in methods to estimate the variances of $\tauhat_\adj$
and $\tauhat_\DIM$. In particular, for standard regression adjustment,
we use the variance estimator
\begin{align}
     \widehat{V}_\adj^2 = \frac{1}{\numobs \plainprop^2 }\sumn
   (\outcome_i - \fhat_{\treatsub} (\covariate_i) )^2 \treat_i +
   \frac{1}{\numobs (1 - \plainprop)^2} \sumn (\outcome_i -
   \fhat_{\controlsub} (\covariate_i) )^2 (1 - {\treat}_i).\label{eq:regadj-classical-var-est}
\end{align}

\paragraph{Construction of simulation instances:}
For a given sample size $\numobs$, we use equi-spaced design points
$x_i = (i - 1) / \numobs$ for $i = 1,2, \ldots, \numobs$, and consider
the outcomes
\begin{align*}
y_i(1) = |2 x_i - 1| + \varepsilon_i / 3, \quad \mbox{and} \quad
y_i(0) = 0, \quad \mbox{where $\varepsilon_i \simiid \mathcal{N} (0,
  1)$}.
\end{align*}
On this problem instance, we compare three candidate estimators: the
decorrelated estimator $\tauhatdc$, the classical regression
adjustment method $\tauhat_\adj$, and difference-in-means
$\tauhat_{\DIM}$.
\begin{figure}[ht!]
\begin{tabular}{ccc}
  \widgraph{0.45\textwidth}{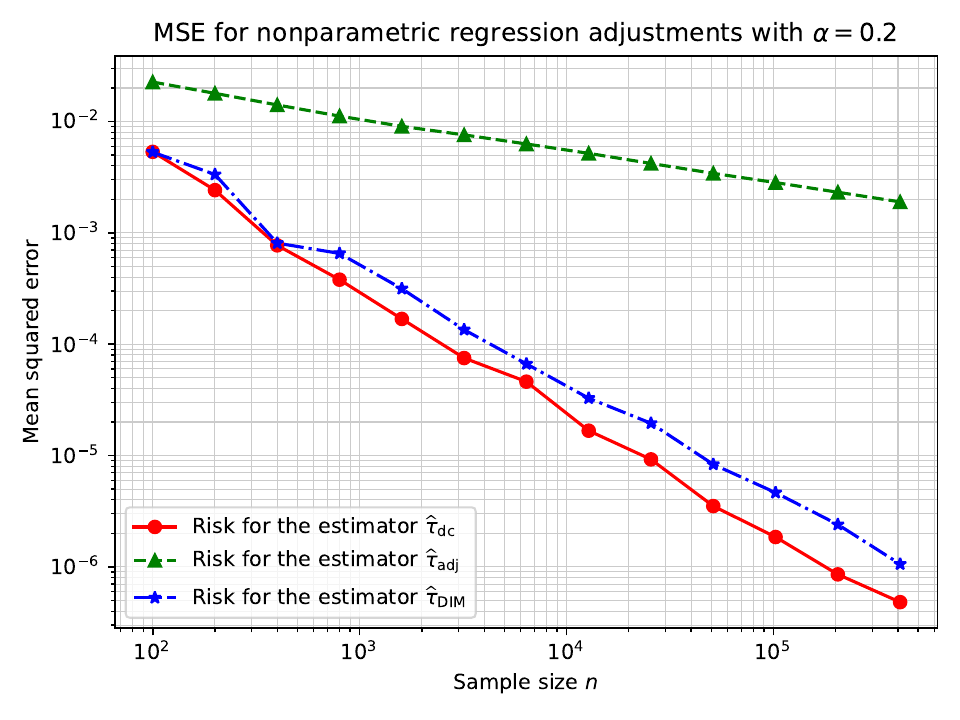} &&
  \widgraph{0.45\textwidth}{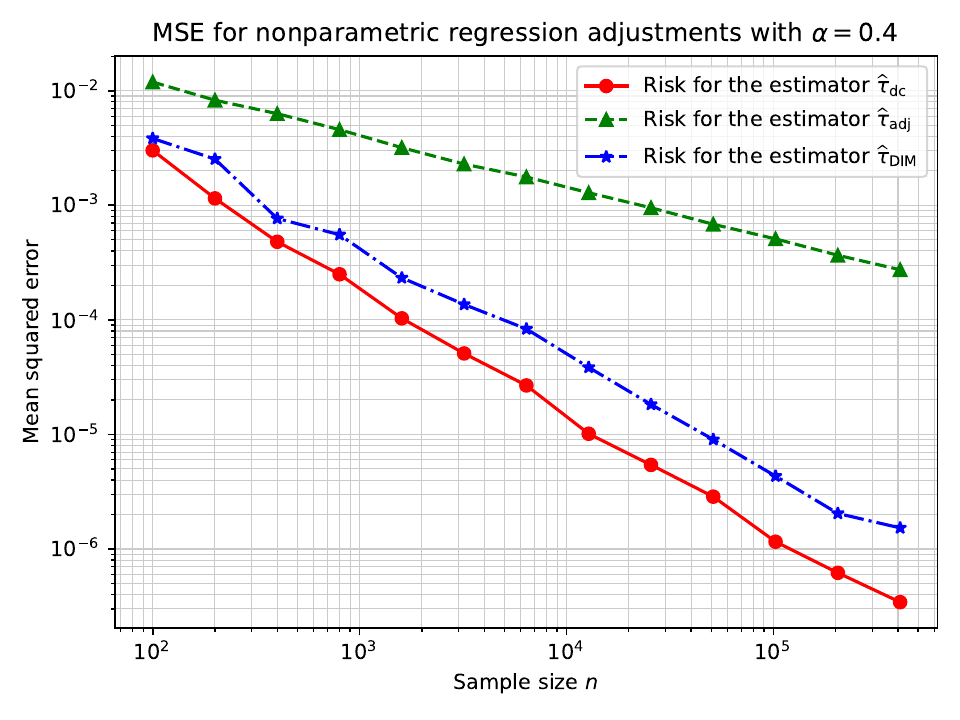} \\ (a) $\alpha =
  0.2$ && (b) $\alpha = 0.4$ \\
  \widgraph{0.45\textwidth}{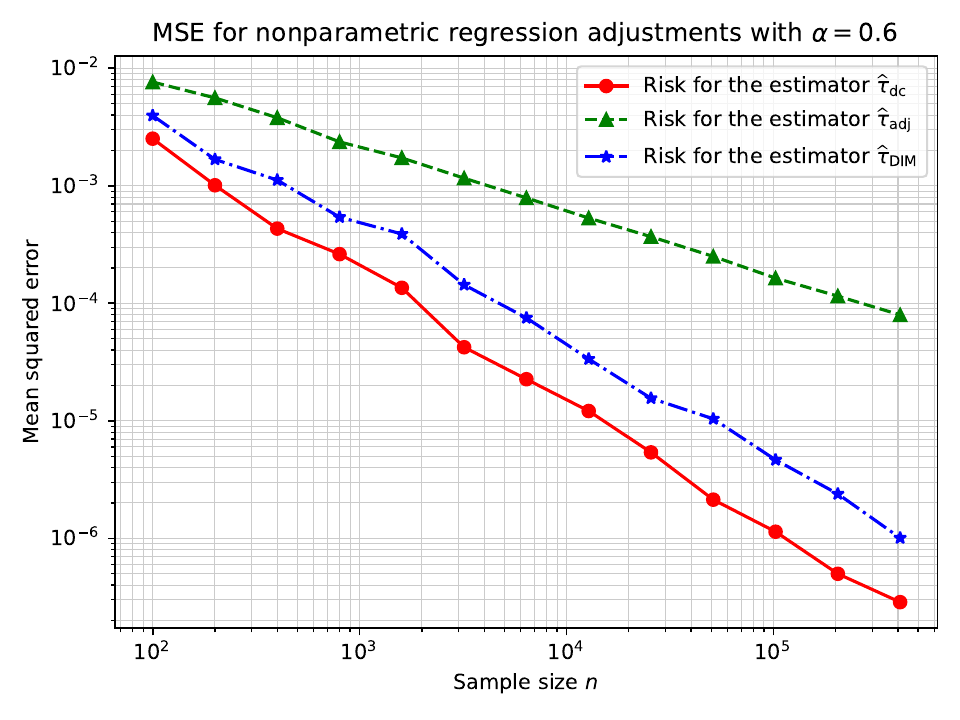}
  &&
  \widgraph{0.45\textwidth}{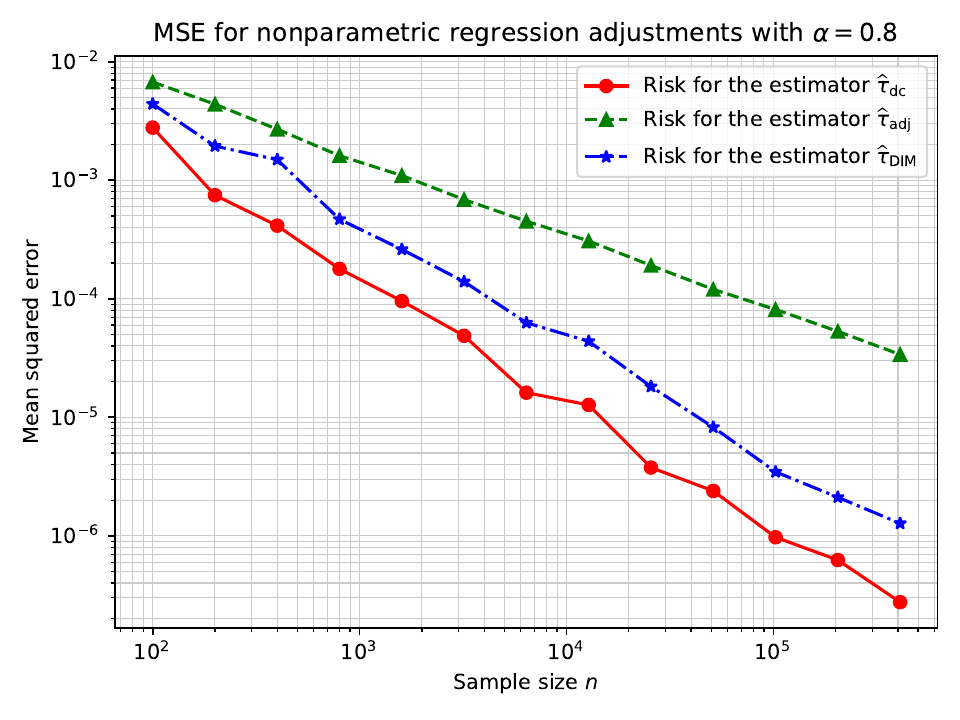}
  \\
(c) $\alpha = 0.6$ && (d) $\alpha = 0.8$ \\
\end{tabular}
\caption{Log-log plots of the mean-squared error $\Exs \big[
    \abss{\tauhat_{\diamond} - \taustar}^2 \big]$ versus sample size
  $\numobs$.  Each curve corresponds to a different algorithm
  $\diamond \in \big\{ \mathrm{dc}, \mathrm{adj},
  \mathrm{DIM}\big\}$. Each marker corresponds to a Monte Carlo
  estimate based on the empirical average of $100$ independent
  runs. As indicated by the sub-figure titles, each panel corresponds
  to a different smoothness exponent $\alpha$.  Some of the curves may
  overlap with each other.}
\label{fig:simulation-holder}
\end{figure}

\begin{figure}[ht!]
\begin{tabular}{ccc}
  \widgraph{0.45\textwidth}{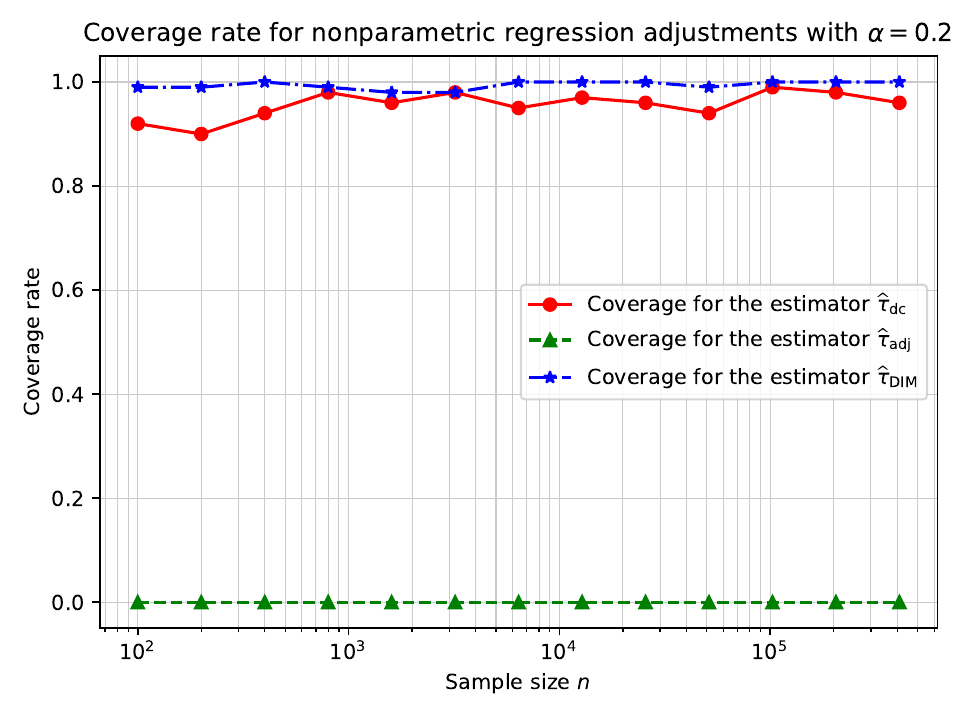} &&
  \widgraph{0.45\textwidth}{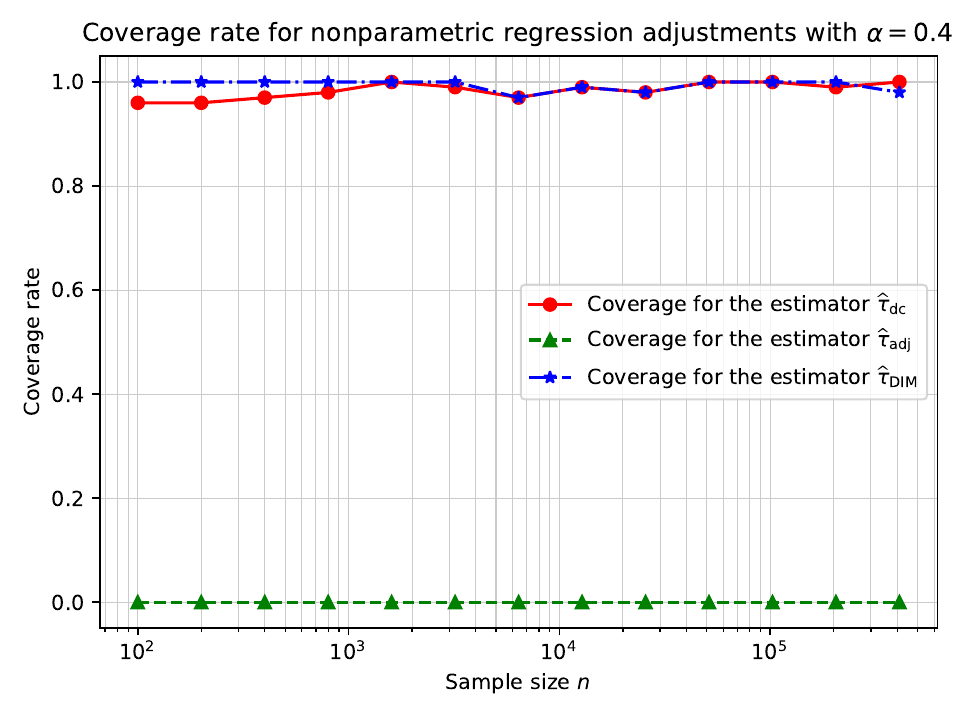} \\ (a) $\alpha =
  0.2$ && (b) $\alpha = 0.4$ \\
  \widgraph{0.45\textwidth}{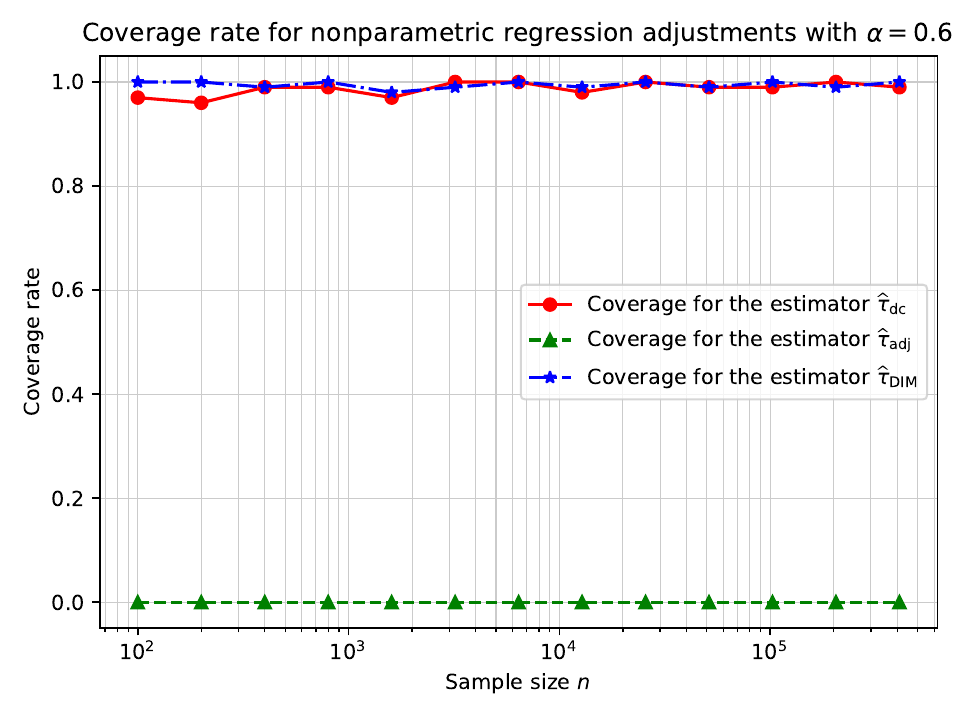}
  &&
  \widgraph{0.45\textwidth}{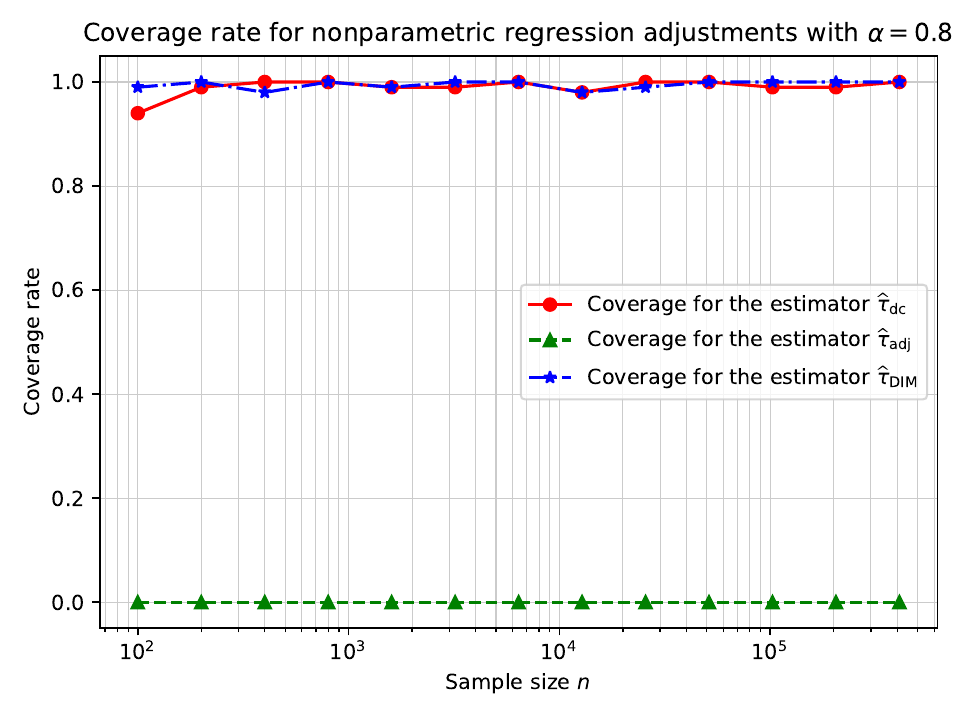}
  \\
(c) $\alpha = 0.6$ && (d) $\alpha = 0.8$ \\
\end{tabular}
\caption{Plots of the coverage rate $\Prob \big(
   \abss{ \taustar - \tauhat_{\diamond} } \leq z_\alpha \widehat{V}_{\diamond} / \sqrt{\numobs} \big)$ versus sample size
  $\numobs$.  Each curve corresponds to a different algorithm
  $\diamond \in \big\{ \mathrm{dc}, \mathrm{adj},
  \mathrm{DIM}\big\}$. Each marker corresponds to a Monte Carlo
  estimate based on the empirical average of $100$ independent
  runs. As indicated by the sub-figure titles, each panel corresponds
  to a different H\"{o}lder exponent $\alpha$. The sample size is given by logarithmic scale and the coverage rate is given by linear scale. Some of the curves may overlap with
  each other.}  
\label{fig:simulation-holder-coverage}
\end{figure}

\begin{figure}[ht!]
\begin{tabular}{ccc}
  \widgraph{0.45\textwidth}{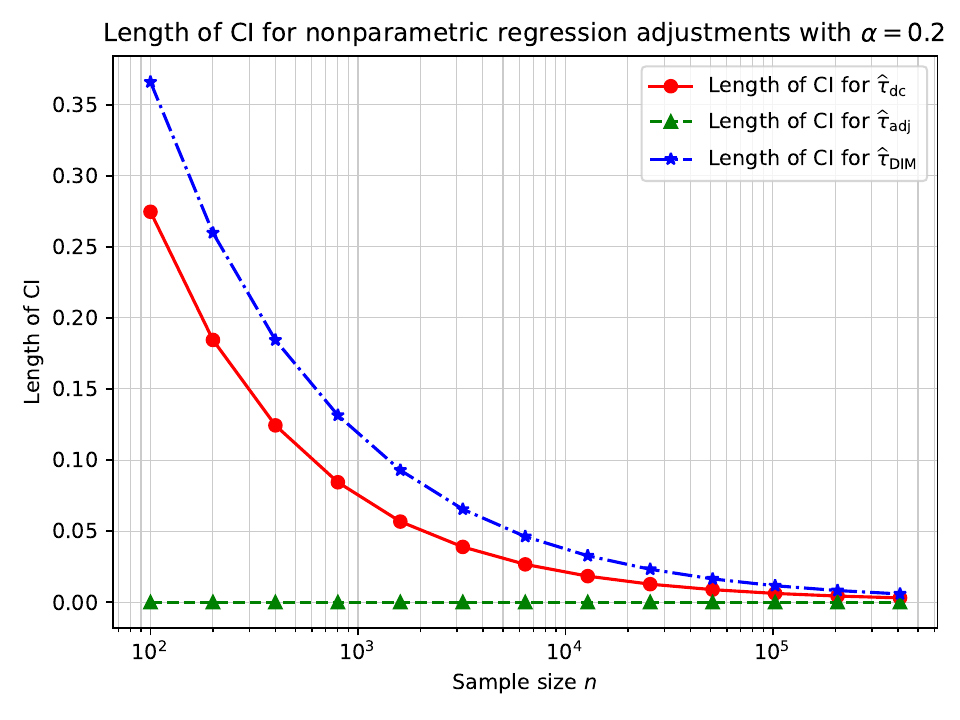} &&
  \widgraph{0.45\textwidth}{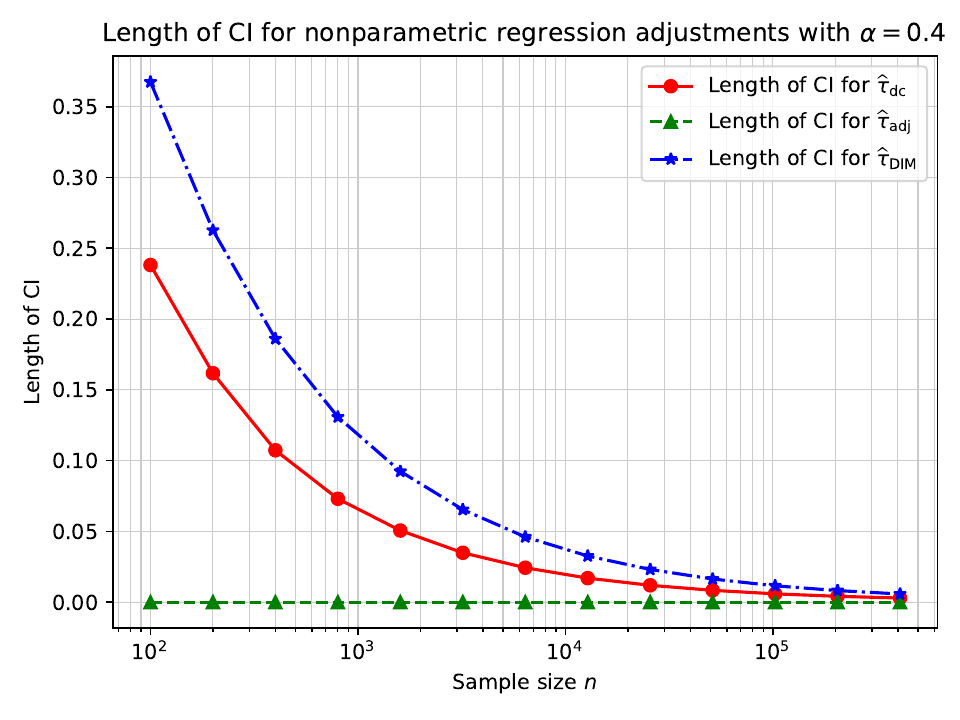} \\ (a) $\alpha =
  0.2$ && (b) $\alpha = 0.4$ \\
  \widgraph{0.45\textwidth}{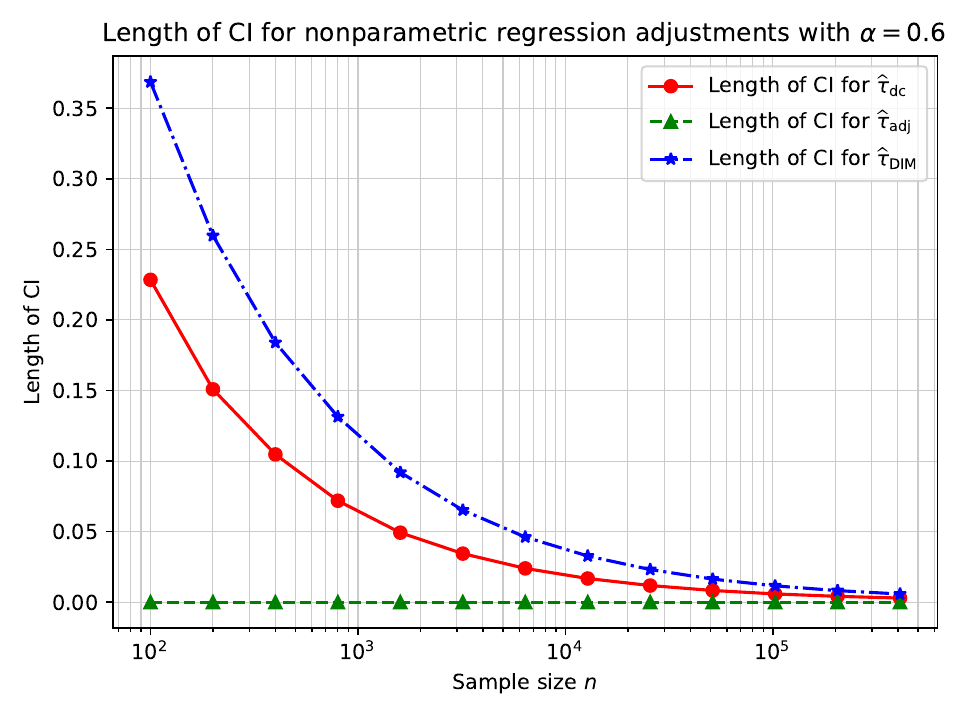}
  &&
  \widgraph{0.45\textwidth}{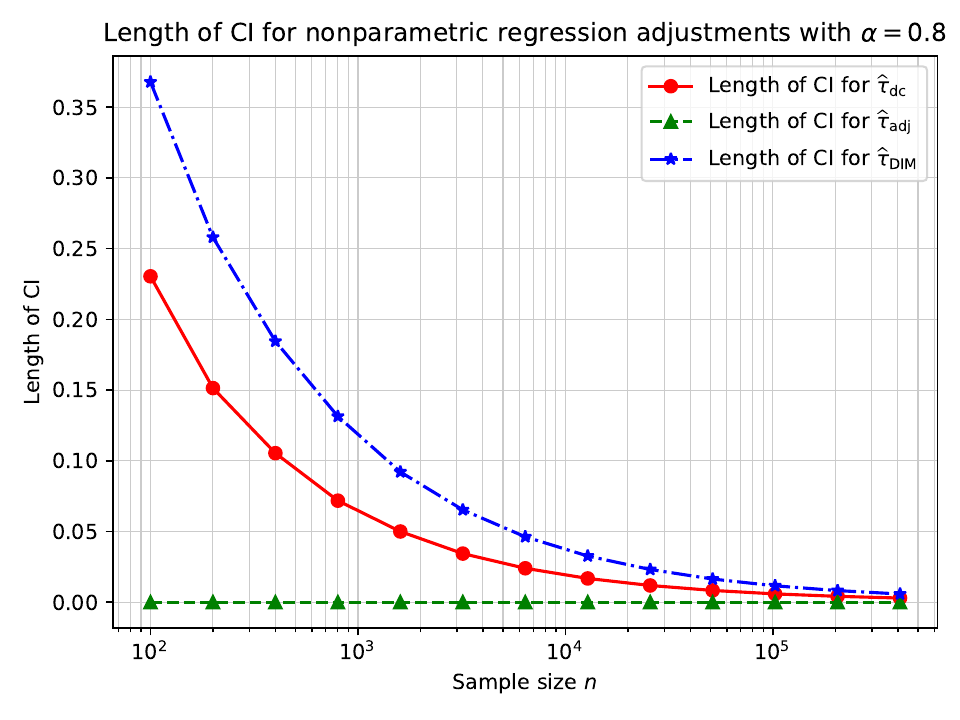}
  \\
(c) $\alpha = 0.6$ && (d) $\alpha = 0.8$ \\
\end{tabular}
\caption{Plots of the length of the confidence interval $2 z_\alpha \Exs \big[ \widehat{V}_{\diamond} / \sqrt{\numobs} \big]$ versus sample size
  $\numobs$.  Each curve corresponds to a different algorithm
  $\diamond \in \big\{ \mathrm{dc}, \mathrm{adj},
  \mathrm{DIM}\big\}$. Each marker corresponds to a Monte Carlo
  estimate based on the empirical average of $100$ independent
  runs. As indicated by the sub-figure titles, each panel corresponds
  to a different H\"{o}lder exponent $\alpha$. The sample size is given by logarithmic scale and the length is given by linear scale. Some of the curves may overlap with
  each other.}  
\label{fig:simulation-holder-length}
\end{figure}

\paragraph{Simulation results:}
In Figures~\ref{fig:simulation-holder},~\ref{fig:simulation-holder-coverage} and~\ref{fig:simulation-holder-length}, we present the simulation results for
the three estimators on the synthetic problem instances described
above. We consider four choices of H\"{o}lder exponents, with $\alpha
\in \{0.2, 0.4, 0.6, 0.8\}$. We estimate the mean-squared errors by
averaging $100$ independent Monte Carlo trials.

From Figure~\ref{fig:simulation-holder}, we can see that the
decorrelated method $\tauhatdc$ consistently outperforms other two
candidates. By estimating the slopes of the curves, we note that both
$\tauhatdc$ and $\tauhat_{\DIM}$ converges at an $\numobs^{-1/2}$
rate, while the decorrelated regression adjustment method achieves an
improved efficiency. The classical regression adjustment
$\tauhat_\adj$ exhibits a slower rate of convergence due to its large
bias. Its convergence rate becomes slower with smaller value of
$\alpha$. Indeed, since the classical regression adjustment method
becomes outcome regression in such a case, any biases in the estimated
functions become biases in the scalar estimation. Under our
construction, the bias is of order $\numobs^{- \frac{\alpha}{2 \alpha
    + 1}}$, which is consistent with the slopes in our plots. Finally,
we observe that the efficiency improvement of $\tauhatdc$ over
$\tauhat_{\DIM}$ kicks in faster with larger value of $\alpha$. This
is because the faster convergence rate of the estimator $\fhat$ leads
to a faster-converging high-order term in~\Cref{ThmMain}.

Turning to the coverage guarantees, in
Figure~\ref{fig:simulation-holder-coverage}, we observe that
$\tauhatdc$ and $\tauhat_{\DIM}$ both lead to valid confidence
intervals for all regimes and all sample sizes. Due to the
interpolation property of the non-parametric estimator, however, the
confidence intervals constructed from $\tauhat_\adj$ dramatically
fail. This is because we always have $\fhat_\treatsub (x_i) =
\outcome_i$ for the treated group and $\fhat_\controlsub (x_i) =
\outcome_i$ for the control group, so that the estimated variance is
always $0$. In contrast, by applying the idea of decorrelation to
variance estimation, as we have shown in Theorem~\ref{ThmInference},
$\Vhat$ is a reliable proxy for the uncertainty, and the finite-sample
coverage for $\tauhatdc$ is as good as $\tauhat_{\DIM}$. Finally, we
observe in Figure~\ref{fig:simulation-holder-length} that the lengths
of CIs for $\tauhatdc$ and $\tauhat_{\DIM}$ exhibit the same trends as
the MSE plot in Figure~\ref{fig:simulation-holder}, which is expected,
as both estimators are unbiased. As we have discussed, due to the
interpolation properties, CIs constructed from $\tauhat_\adj$ are
always singletons at the point estimators.


\section{Proofs}
\label{sec:proofs}

This section is devoted to the proofs of our main results, including
our non-asymptotic bound (\Cref{ThmMain}) in~\Cref{SecProofThmMain};
our asymptotic guarantee (\Cref{PropCLT}) in~\Cref{SecProofPropCLT}
and our confidence intervals (\Cref{ThmInference})
in~\Cref{SecProofThmInference}.  Note that the proofs of all the
corollaries stated in~\Cref{SecExamples} are deferred
to~\Cref{SecProofExamples}.

\subsection{Proof of~\Cref{ThmMain}}
\label{SecProofThmMain}

By definition of the DC estimator and its oracle version, we have
the decomposition
\mbox{$\tauhatdc - \tauoradc = \ErrTerm_1 - \ErrTerm_2$,} where
\begin{align*}
\ErrTerm_1 \defn \frac{1}{\numobs} \sumn \big(
\fstar_{\treatsub}(\covariate_i) - \fhat_{\fitsub}(\covariate_i) \big)
\cdot \big( \frac{ \smean_i}{\propensity_{\smeansub}} - 1 \big),
\mbox{and} \quad \ErrTerm_2 \defn \frac{1}{\numobs} \sumn
\big(\fstar_\controlsub(\covariate_i) -
\fhat_{\barfitsub}(\covariate_i) \big) \cdot \big(
\frac{\barsmean_i}{\propensity_{\barsmeansub}} - 1 \big).
\end{align*}
It suffices to establish high probability bounds on these two error
terms.  In particular, define the error probabilities
\begin{align}
\label{EqnDefnQ}  
q_1(\delta) \defn \Prob \left\{ \abss{\ErrTerm_1} \geq
\frac{\myerr_\treatsub (\propensity_{\fitsub}, \tfrac{\delta}{4})}{
  \propensity_{\smeansub}} \; \sqrt{\frac{\log (4 / \delta)}{\numobs}}
\right\}, \quad \mbox{and} \quad q_2(\delta) \defn \Prob \left\{
\abss{\ErrTerm_2} \geq
\frac{\myerr_\controlsub(\propensity_{\barfitsub},
  \tfrac{\delta}{4})}{\propensity_{\barsmeansub}} \; \sqrt{\frac{\log
    (4 / \delta)}{\numobs}} \right\}.
\end{align}
We will show that $\max \{q_1(\delta), q_2(\delta) \}
\leq \delta/2$, from which the claim follows by union bound.

Essential in our analysis are the independence properties of our
construction (cf.~\Cref{LemConstruction}): in particular, with
reference to $\ErrTerm_2$, the estimate $\fhat_\fitsub$ is independent
of the variables $\{\smean_i\}_{i=1}^\numobs$.  Note that an analogous
decoupling holds for $\ErrTerm_2$.

Let us show how this decoupling allows for easy control of
$\ErrTerm_1$. \Cref{LemConstruction} ensures that, conditioned on the
binary sequence $(\fit_i)_{i = 1}^\numobs$, the variables
$(\smean_i)_{i = 1}^\numobs$ are $\mathrm{i.i.d.}$ Bernoulli variables
with parameter $\propensity_{\smeansub}$. Consequently, we can apply
Hoeffding's inequality to obtain that
\begin{align*}
\Prob \Big\{ \abss{\ErrTerm_1} \geq t \mid (\fit_i)_{i = 1}^\numobs
\Big\} \leq 2 \exp \left( - \frac{2 t^2 \propensity_{\smeansub}^2
  \numobs}{\vecnorm{\fhat_{\fitsub} - \fstar_{\treatsub}}{\numobs}^2}
\right), \quad \mbox{for any $t > 0$.}
\end{align*}
By the assumed function estimation bound, we have
\begin{align*}
\Prob \left(\vecnorm{\fhat_{\fitsub} - \fstar_{\treatsub}}{\numobs}
\geq \myerr_\treatsub (\propensity_{\fitsub}, \delta / 4) \right) \leq
\delta / 4.
\end{align*}
We now put these two results together to bound the error probability
$q_1(\delta)$ from equation~\eqref{EqnDefnQ}. We can write
\begin{align*}
q_1(\delta) & \leq \Prob \left(\vecnorm{\fhat_{\fitsub} -
  \fstar_{\treatsub}}{\numobs} \geq \myerr_\treatsub
(\propensity_{\fitsub}, \tfrac{\delta}{4}) \right) + \Prob
\left(\vecnorm{\fhat_{\fitsub} - \fstar_{\treatsub}}{\numobs} <
\myerr_\treatsub (\propensity_{\fitsub}, \tfrac{\delta}{4}), ~
\mbox{and}~
\abss{\ErrTerm_1} > \tfrac{\vecnorm{\fhat_{\fitsub} -
    \fstar_{\treatsub}}{\numobs}}{\propensity_{\smeansub}}
\sqrt{\tfrac{\log (4 / \delta)}{\numobs}}\right) \\
& \leq (\delta/4) \; + \; \Exs \Big[ \Prob \Big( \abss{\ErrTerm_1} >
  \tfrac{\vecnorm{\fhat_{\fitsub} -
      \fstar_{\treatsub}}{\numobs}}{\propensity_{\smeansub}}
  \sqrt{\tfrac{\log (4 / \delta)}{\numobs}}\mid (\fit_i)_{i =
    1}^\numobs \Big) \Big] \\
& \leq \delta / 2.
\end{align*}
A similar argument can be used to prove the bound $q_2(\delta)$, from
which the overall claim follows.


\subsection{Proof of~\Cref{PropCLT}}
\label{SecProofPropCLT}
By definition of the oracle DC estimator, we can write
$\sqrt{\numobs}(\tauoradc-\taustar) = \frac{1}{\sqrt{\numobs}} \sumn
W_i$, where
\begin{align*}
W_i & \defn \residual_i(1) \frac{\smean_i}{\propensity_{\smeansub}} -
\residual_i(0)
\frac{\bar{\smean}_i}{\propensity_{\barsmeansub}}+\fstar_{\treatsub}(\covariate_i)-
\fstar_{\controlsub}(\covariate_i) -\taustar
\end{align*}
are independent random variables.  Consequently, we have
\begin{align*}
  \numobs \cdot \Exs \big[ (\tauoradc - \taustar)^2 \big] =
  \frac{1}{\numobs} \sumn \Exs[W_i^2] = \frac{1}{\numobs} \sum_{i =
    1}^\numobs \Big\{\frac{1 -
    \propensity_{\smeansub}}{\propensity_\smeansub} \residual^2_i(1) +
  \frac{1 - \propensity_{\barsmeansub}}{\propensity_{\barsmeansub}}
  \residual^2_i(0) + 2 \residual_i(0) \residual_i(1) \Big\}.
\end{align*}

The sum of third moments of $\noise_i$ can be bounded as
\begin{align*}
\sumn \Exs|\noise_i|^3 \leq \sumn \left \{
\frac{\residual^3_i(1)}{\propensity_{\smeansub}^3} +
\frac{\residual_i^3(0)}{\propensity_{\barsmeansub}^3} \right\} \leq 2
\numobs\frac{C^3}{\alpha^3},
\end{align*}
where in the last step, we used H\"{o}lder's inequality to convert the
fourth moment bound to a third moment bound.  Consequently, we have
\begin{align*}
\big( \numobs \Exs(\tauoradc-\taustar)^2 \big)^{-3/2} \sumn \Exs
\abss{\frac{\noise_i}{\sqrt{\numobs}}}^3 \le \frac{2L^3}{\alpha^3
  \AsympStd^3 \sqrt{\numobs}} = \order{\frac{1}{\sqrt{\numobs}}}.
\end{align*}
Applying Lyapunov's CLT guarantees that $\sqrt{\numobs} (\tauoradc -
\taustar) / \AsympStd \convdist \mathcal{N}\left(0, 1\right)
$. Further, combining with the consistency
condition~\eqref{EqnConsistent}, we can apply Slutsky's theorem to
conclude that $\sqrt{\numobs} (\tauhatdc - \taustar) / \AsympStd
\convdist \mathcal{N}\left(0, 1\right)$.


\subsection{Proof of~\Cref{ThmInference}}
\label{SecProofThmInference}

Recalling the definition~\eqref{EqnDefnAsympVar} of $\AsympVar$, we
have
\begin{align*}
\AsympVar & \defn \frac{1 -
  \propensity_{\smeansub}}{\propensity_{\smeansub}}
\vecnorm{\residual(1)}{\numobs}^2 + \frac{1 -
  \propensity_{\barsmeansub}}{\propensity_{\barsmeansub}}
\vecnorm{\residual(0)}{\numobs}^2 + 2
\inprod{\residual(1)}{\residual(0)}_\numobs \stackrel{(\star)}{\leq}
\frac{1}{\propensity_{\smeansub}} \vecnorm{\residual(1)}{\numobs}^2 +
\frac{1}{\propensity_{\barsmeansub}}\vecnorm{\residual(0)}{\numobs}^2,
\end{align*}
where the inequality $(\star)$ follows from Young's inequality
\mbox{$2 \inprod{\residual(1)}{\residual(0)}_\numobs \leq
  \vecnorm{\residual(1)}{\numobs}^2 +
  \vecnorm{\residual(0)}{\numobs}^2$.}  Consequently, it suffices to
prove that
\begin{align*}
\Vhat^2 = \frac{1}{\numobs} \sumn
\left\{\frac{\residual^2_i(1)}{\propensity_{\smeansub}} +
\frac{\residual^2_i(0)}{\propensity_{\barsmeansub}} \right \} +
o_{\mathbb{P}}(1).
\end{align*}
In particular, we prove the two convergence statements:
\begin{align*}
\mbox{(a)} \quad \underbrace{\frac{1}{\numobs \propensity_{\smeansub}
  }\sumn (\outcome_i - \fhat_{\fitsub} (\covariate_i) )^2 \smean_i -
  \vecnorm{\residual(1)}{\numobs}^2}_{E_1} \xrightarrow{\Prob} 0, \quad
\mbox{and, (b)} \quad \frac{1}{\numobs \propensity_{\barsmeansub}}
\sumn (\outcome_i - \fhat_{\barfitsub} (\covariate_i) )^2
\bar{\smean}_i - \vecnorm{\residual(0)}{\numobs}^2 \xrightarrow{\Prob}
0.
\end{align*}
By symmetry, it suffices to prove statement (a): namely, that $E_1
\xrightarrow{\Prob} 0$.

In order to do so, we begin with the decomposition $E_1 = \sum_{j=1}^3
\Term_j$, where
\begin{align*}
\Term_1 \defn \frac{1}{\numobs \propensity_{\smeansub} } \sumn
\residual^2_i(1) \smean_i - \vecnorm{\residual(1)}{\numobs}^2, \qquad
\Term_2 \defn \frac{1}{\numobs \propensity_{\smeansub}} \sumn
(\fhat_{\fitsub} (\covariate_i) - \fstar_\treatsub (\covariate_i) )^2
\smean_i, \quad \mbox{and} \\
\Term_3 \defn \frac{2}{\numobs \propensity_{\smeansub} } \sumn (
\fstar_\treatsub (\covariate_i) - \fhat_{\fitsub} (\covariate_i) )
\residual_i(1) \smean_i. \qquad \qquad \qquad \qquad
\end{align*}
It suffices that each of these three terms converges to zero in probability.

\paragraph{Analysis of $\Term_1$:}
From the fourth moment bound in equation~\eqref{EqnAsympVar}
and the lower bound $\propensity_\smeansub \geq \alpha$ implied
by equation~\eqref{EqnOverlap} we have
\begin{align*}
\Exs[\Term_1^2] \leq \frac{1}{\numobs^2 \propensity_\smeansub^2} \sumn
\residual_i^4(1) \leq \frac{C}{\numobs \alpha^2} \rightarrow 0,
\end{align*}
which implies that $\Term_1 \xrightarrow{\Prob} 0$.

\paragraph{Analysis of $\Term_2$:}
We have $|\Term_2| \leq \frac{1}{\propensity_{\smeansub}}
\vecnorm{\fhat_{\fitsub} - \fstar_{\treatsub}}{\numobs}^2
\xrightarrow{\Prob} 0$, as required.

\paragraph{Analysis of $\Term_3$:}
For the term $\Term_3$, applying the Cauchy--Schwarz inequality yields
\begin{align*}
\abss{\Term_3} \leq \frac{2}{\propensity_{\smeansub}}
\vecnorm{\fhat_{\fitsub} - \fstar_{\treatsub}}{\numobs} \cdot
\vecnorm{\residual(1)}{\numobs} \xrightarrow{\Prob} 0.
\end{align*}


\section{Discussion}
\label{sec:discussion}

In this paper, we proposed and analyzed a new decorrelation
procedure for regression adjustment in the design-based framework.  At
the core is a randomized procedure for constructing overlapping
subsets of data for the regression steps and averaging steps that
induces some desirable independence properties. We show that regression adjustment procedures based on this
decorrelation step are able to match the attractive properties of
oracle estimators.  We illustrated some applications of this
methodology to both high-dimensional linear regression (ordinary and
sparse), along with various non-parametric regression methods, thereby
obtaining improvements on the sample complexity.

The method of this paper is a simple ``plug-and-play'' approach, which
has the potential to improve the finite-sample performance of any
regression adjustment procedure. It also opens up some interesting
directions for future research.  Let us discuss a few to conclude.

First, while we established some desirable guarantees for
decorrelation-based estimators, further improvements are possible. As
noted previously, the oracle itself need not always have smaller
variance than the difference-in-mean estimator.  Thus, one interesting
open question is whether it is possible to achieve the ``no-harm''
properties (e.g.,~\cite{cohen2023no}) along the sharp guarantees given
in this paper.  our paper at the same time.

Second, note that asymptotic optimality in the current framework means
that only a vanishing fraction of data is used to fit the outcome
functions.  While the method is asymptotically optimal, this fact
means that higher-order terms could be slowly decaying.  Thus, it
would interesting to explore analogues of ``cross-fitting'' in the
finite population framework, without destroying the independence
structure.

Last, while this paper focuses on $\mathrm{i.i.d.}$ Bernoulli trials,
more complicated probabilistic structures arise in controlled
experiments, including Markov chains~\cite{farias2022markovian},
cluster randomization~\cite{Su2020cluster}, adaptively collected
data~\cite{hadad2021confidence}, and optimized treatment/control group
assignments~\cite{bertsimas2015power}.  It would be interesting to
extend our decorrelation-based framework to these more general
settings.

\subsection*{Acknowledgements}
This work was partially supported by NSF grant CCF-1955450, ONR grant
N00014-21-1-2842 and NSF grant DMS-2311072 to MJW; and NSF DMS-1945136 to PD.


\bibliographystyle{alpha}

\bibliography{causal, reference}


\appendix

\section{Construction of decorrelating sequences:
  Proof of~\Cref{LemConstruction}}
\label{AppConstruction}

Recall the procedure described following the statement
of~\Cref{LemConstruction}.  In this appendix, we prove that it
generates random variables that satisfy the conditions
of~\Cref{LemConstruction}.
  
\paragraph{Property (a):} We first show that $(\smean, \fit)$ are independent $\Ber(\propensity_\smeansub)$
and $\Ber(\propensity_\fitsub)$. We have
\begin{align*}
\Prob(\smean = 1) &= \Prob(\treat = 1) \cdot \Prob(Z \in \{1, 2\} \mid
\treat = 1) = \plainprop \cdot \frac{ \propensity_{\smeansub} }{
  \propensity_{\fitsub} + \propensity_{\smeansub} -
  \propensity_{\fitsub} \propensity_{\smeansub}} =
\propensity_{\smeansub},\\ \Prob(\fit = 1) &= \Prob(\treat = 1) \cdot
\Prob(Z \in \{1, 3\} \mid \treat = 1) = \plainprop \cdot \frac{
  \propensity_{\fitsub} }{ \propensity_{\fitsub} +
  \propensity_{\smeansub} - \propensity_{\fitsub}
  \propensity_{\smeansub}} = \propensity_{\fitsub},\\ \Prob(\fit =
\smean = 1) &= \Prob (\treat = 1) \cdot \Prob(Z = 1 \mid \treat = 1) =
\plainprop \cdot \frac{ \propensity_{\fitsub} \propensity_{\smeansub}
}{ \propensity_{\fitsub} + \propensity_{\smeansub} -
  \propensity_{\fitsub} \propensity_{\smeansub}} =
\propensity_{\fitsub} \propensity_{\smeansub}.
\end{align*}

\paragraph{Property (b):}  The same claim for $(\smeanbar, \fitbar)$
follows by symmetry of the construction.

\paragraph{Property (c):} The construction also ensures that $\max \{ \smean,
\fit \} \leq \treat$, and $\max \{ \smeanbar, \fitbar \} \leq 1-
\treat$.

\paragraph{Property (d):}
We have
\begin{align*}
\cov \{\smean, \smeanbar \} & = \Exs[\smean \smeanbar] - \Exs[\smean]
\Exs[\smeanbar] \; = \; - \propensity_\smeansub
\propensity_\barsmeansub,
\end{align*}
since $\smean \smeanbar = 0$, and $\Exs[\smean] =
\propensity_\smeansub$ and $\Exs[\smeanbar] =
\propensity_\barsmeansub$ by construction.


\section{H\'{a}jek estimators}
\label{app:hajek}

In this appendix, we describe the H\`{a}jek version of our estimators.
Define the integers $\numtilde_\treatsub \mydefn \sum_{i = 1}^\numobs
\treat_i$, $\numtilde_{\controlsub} \mydefn \numobs -
\numtilde_{\treatsub}$, $\numtilde_{\smeansub} \mydefn \sum_{i =
  1}^\numobs \smean_i$, and $\numtilde_{\barsmeansub} \mydefn \sum_{i
  = 1}^\numobs \barsmean_i$. We define the quantities
\begin{align*}
\tauhat_{\DIM}^\hajek,\quad
\tauhat_\adj^\hajek,\quad\tauoraadj^\hajek, \quad\tauhatdc^\hajek,
\quad \mbox{and} \quad \tauoradc^\hajek
\end{align*}
by simply replacing $\numobs_{\star}$ with $\numtilde_{\star}$ for
$\star \in \{\treatsub, \controlsub, \smeansub, \barsmeansub\}$ in the
definitions of $\tauhat_{\DIM}$, $\tauhat_\adj$, $\tauoraadj$,
$\tauhatdc$, and $\tauoradc$, respectively. In this appendix, we
discuss asymptotic properties of the H\'{a}jek estimators, and compare
them with the standard ones studied in the main text.

Note that the difference-in-means H\'{a}jek estimator
$\tauhat_{\DIM}^\hajek$ and the oracles $\tauoraadj^{\hajek},
\tauoradc^{\hajek}$ are no longer unbiased. Nevertheless, under
appropriate assumptions, asymptotic normality can still be derived. In
particular, defining the re-centered outcomes and residuals
\begin{align*}
  \widetilde{\outcome} (z) \mydefn \outcome (z) - \frac{1}{\numobs}
  \sum_{i = 1}^\numobs \outcome_i(z), \quad \widetilde{\residual} (z)
  \mydefn \residual(z) - \frac{1}{\numobs} \sum_{i = 1}^\numobs
  \residual_i(z), \quad \mbox{for $z \in \{0, 1\}$},
\end{align*}
we have
\begin{align*}
\frac{\sqrt{\numobs} }{v_\DIM^\hajek} (\tauhat_{\DIM}^\hajek -
\taustar ),~ \frac{ \sqrt{\numobs} }{v_{\adj}^\hajek}
(\tauoraadj^\hajek - \taustar),~ \frac{\sqrt{\numobs}
}{v_{\mathrm{dc}}^\hajek} ( \tauoradc^\hajek - \taustar ) \convdist
\mathcal{N} (0, 1),
\end{align*}
where the variances are given by
\begin{align*}
  \big( v_\DIM^\hajek \big)^2 &= \frac{1 - \plainprop}{\plainprop}
  \vecnorm{\widetilde{\outcome} (1)}{\numobs}^2 + \frac{\plainprop}{1
    - \plainprop} \vecnorm{\widetilde{\outcome} (0)}{\numobs}^2 + 2
  \inprod{\widetilde{\outcome} (1)}{\widetilde{\outcome}
    (0)}_\numobs,\\ \big( v_\adj^\hajek \big)^2 &= \frac{1 -
    \plainprop}{\plainprop} \vecnorm{\widetilde{\residual}
    (1)}{\numobs}^2 + \frac{\plainprop}{1 - \plainprop}
  \vecnorm{\widetilde{\residual} (0)}{\numobs}^2 + 2
  \inprod{\widetilde{\residual} (1)}{\widetilde{\residual}
    (0)}_\numobs,\\ \big( v_\dc^\hajek \big)^2 &= \frac{1 -
    \propensity_\smeansub}{\propensity_\smeansub}
  \vecnorm{\widetilde{\residual} (1)}{\numobs}^2 + \frac{1 -
    \propensity_{\barsmeansub}}{\propensity_{\barsmeansub}}
  \vecnorm{\widetilde{\residual} (0)}{\numobs}^2 + 2
  \inprod{\widetilde{\residual} (1)}{\widetilde{\residual}
    (0)}_\numobs.
\end{align*}
For the decorrelated H\'{a}jek estimator $\tauhatdc^\hajek$, we could
also establish non-asymptotic approximation results and asymptotic
normality similar to~\Cref{ThmMain} and~\Cref{PropCLT}, as long as the
functions $\fstar_\treatsub, \fstar_\controlsub$ can be consistently
estimated.  Compared to the variances of the standard versions
$\tauhat_{\DIM}$ and the oracles $\tauoraadj, \tauoradc$ discussed in
the main text, the variances of H\'{a}jek estimators re-center the
outcomes and residuals in the expression. This could lead to
potentially smaller asymptotic variances. However, such a variance
reduction effect can also be achieved by adding a constant shift in
regression adjustment. In particular, given a function class
$\funcClass$, define the class $\widetilde{\funcClass} \mydefn \big\{
f + c: f \in \funcClass, c \in \real \big\}$, for any vector $\outcome
\in \real^\numobs$, by defining
\begin{align*}
\fstar \mydefn \arg\min_{f \in \widetilde{\funcClass}} \vecnorm{f -
  y}{\numobs}, \quad \mbox{and} \quad \residual = y - \fstar,
\end{align*}
we have that ${\numobs}^{-1} \sum_{i = 1}^\numobs \residual_i =
0$. Consequently, when we use least-square function estimators and an
intercept term in included, our estimators are asymptotically
equivalent to the H\'{a}jek estimator.


\section{Proofs of corollaries}
\label{SecProofExamples}

\noindent In this section, we prove the three corollaries stated
in~\Cref{SecExamples}.


\subsection{Proof of~\Cref{cor:ols}}
\label{SecProofCorOLS}

Given the assumed covariate structure, we have
\begin{align*}
\vecnorm{\Xmat (\betahat_\fitsub - \betastar_\treatsub)}{\numobs}^2
=(\widehat{\beta}_{\fitsub}-\beta_{\treatsub}^*)^\top \tfrac{\Xmat^\top
  \Xmat}{\numobs} (\widehat{\beta}_{\fitsub}-\beta_{\treatsub}^*) =
\vecnorm{\widehat{\beta}_{\fitsub}-\beta_{\treatsub}^*}{2}^2.
\end{align*}
Consequently, in order to apply~\Cref{ThmMain}, it suffices to bound
the estimation error $\vecnorm{\betahat_\fitsub -
  \betastar_\treatsub}{2}$, as well as its counterpart
$\vecnorm{\betahat_{\barfitsub} - \betastar_{\controlsub}}{2}$.

Here is the key auxiliary result that allows us to
apply~\Cref{ThmMain}:
\begin{lemma}
\label{LemOLSError}  
Under the conditions of~\Cref{cor:ols}, we have
\begin{align}
\label{eq:ols-error}
\vecnorm{\widehat\beta_{\fitsub} - \beta_{\treatsub}^* }{2} \leq c
\sqrt{\frac{\LevScore}{\propensity_\fitsub} \myboundmax{2}
  \log(\usedim) \log(1/\delta)}
\end{align}
with probability at least $1 - \delta$.
\end{lemma}
\noindent Observe that~\Cref{cor:ols} follows by combining this lemma
with~\Cref{ThmMain}.  It remains to prove the lemma.


\subsubsection{Proof of~\Cref{LemOLSError}}

We adopt the shorthand $\kappa_\numobs = \max \limits_{i = 1, \ldots,
  \numobs} \frac{\|x_i\|_2}{\sqrt{\numobs}}$ for the remainder of this
proof.  Our proof makes use of two auxiliary results, which we begin
by stating.
\begin{lemma}
\label{lem:denominator}
With probability at least $1-\delta$, we have
\begin{align*}
\opnorm{\frac{1}{\numobs} \sumn \covariate_i \covariate_i^\top
  (\fit_i-\propensity_{\fitsub}) }\le c \sqrt{\LevScore
  \propensity_{\fitsub} \log(\usedim) \log(1/\delta)} + c \LevScore
\log(\usedim) \log(1/\delta).
\end{align*}
\end{lemma}
\begin{lemma}
\label{lem:numerator}
With probability at least $1-\delta$, we have
\begin{align*}
\vecnorm{\frac{1}{\numobs}\sumn \covariate_i \residual_i(1) (\fit_i -
  \propensity_{\fitsub})}{2}\le c\sqrt{ \LevScore \myboundmax{2}^2
  \propensity_{\fitsub}\log(1/\delta)} + \frac{c\sqrt{\LevScore}
  \mybound{\infty} \log(1/\delta)}{\sqrt{\numobs}}.
\end{align*}
\end{lemma}
See~\Cref{sec:proof-denominator} and~\Cref{sec:proof-numerator} for
the proof of these lemmas.

Taking our two auxiliary lemmas as given, we now proceed with the
proof of~\Cref{LemOLSError}.  Define the random matrix
\begin{align*}
 \widehat{\Sigma}_{\fitsub} \mydefn \frac{1}{\numobs
   \propensity_{\fitsub}}\sumn \covariate_i \covariate_i^\top \fit_i.
\end{align*}
When $\propensity_{\fitsub}\ge 4c\LevScore
\log(\usedim)\log(1/\delta)$, ~\Cref{lem:denominator} guarantees that
\begin{align*}
\opnorm{\widehat{\Sigma}_{\fitsub} - I } \le c\sqrt{\LevScore
  \log(\usedim)\log(1/\delta)/\propensity_{\fitsub}} < 1/2,
\end{align*}
which implies the bound
\begin{align}
\label{eqn:inverse-concentration}
\opnorm{\widehat{\Sigma}_{\fitsub}^{-1} - I}\le 2c\sqrt{\LevScore
  \log(\usedim)\log(1/\delta)/\propensity_{\fitsub}}.
\end{align}
When $\propensity_{\fitsub}\ge c\mu_{\numobs} \log(1/\delta)$,
by~\Cref{lem:numerator}, we have
\begin{align*}
\vecnorm{\frac{1}{\numobs\propensity_{\fitsub}}\sumn \covariate_i
  \residual_i(1) \fit_i - \frac{1}{\numobs}\sumn \covariate_i
  \residual_i(1) }{2} \le c\sqrt{ \LevScore \myboundmax{2}^2
  \log(1/\delta)/\propensity_{\fitsub}}.
\end{align*}
If we further assume that $\propensity_{\fitsub}\ge c\LevScore
\log(1/\delta)$, we have
\begin{align}
  \label{eqn:numerator-concentration}
\vecnorm{\frac{1}{\numobs\propensity_{\fitsub}} \sumn \covariate_i
  \residual_i(1) \fit_i }{2} \le 2 c \sqrt{\myboundmax{2}^2}.
\end{align}
Combining equations~\eqref{eqn:inverse-concentration}
and~\eqref{eqn:numerator-concentration}, we see that with probability
at least $1 - \delta$
\begin{align*}
\vecnorm{\widehat \beta_{\fitsub} - \betastar_{\treatsub}}{2} & =
\vecnorm{\widehat{\Sigma}_{\fitsub}^{-1} \left(\frac{1}{n
    \propensity_{\fitsub}}\sumn \covariate_i \residual_i(1) \fit_i
  \right)- \frac{1}{\numobs}\sumn \covariate_i \residual_i(1)
}{2} \\
& \le \opnorm{\widehat{\Sigma}_{\fitsub}^{-1} - I} \cdot
\vecnorm{\frac{1}{n \propensity_{\fitsub}}\sumn \covariate_i
  \residual_i(1) \fit_i}{2} + \vecnorm{\frac{1}{n
    \propensity_{\fitsub}}\sumn \covariate_i \residual_i(1) \fit_i -
  \frac{1}{\numobs}\sumn \covariate_i \residual_i(1) }{2} \\
& \leq c \sqrt{\LevScore \myboundmax{2}^2 \log(\usedim)
  \log(1/\delta)/\propensity_{\fitsub}},
\end{align*}
completing the proof of the claimed bound~\eqref{eq:ols-error}.


\subsubsection{Proof of Lemma~\ref{lem:denominator}}
\label{sec:proof-denominator}

We first state two standard results on the suprema
of empirical processes:

\begin{lemma}\cite[Theorem 3.27]{wainwright2019high}
\label{lemma:functional_matrix}
For an independent sequence $\{X_i\}_{i=1}^\numobs$, with probability
at least $1-\delta$, we have
\begin{multline*}
\sup_{\Fun \in \mathcal{F}} \abss{\frac{1}{\numobs}\sumn \Fun(X_i)}
\le \Exs \left[\sup_{\Fun \in \mathcal{F}}
  \abss{\frac{1}{\numobs}\sumn \Fun(X_i)}\right] \\ + c
\sqrt{\frac{\Exs[\sup_{\Fun \in \mathcal{F}} \numobs^{-1} \sumn
      f^2(X_i)] \log(1/\delta)}{\numobs}} + c \frac{\sup_{\Fun \in
    \mathcal{F}} \vecnorm{f}{\infty}\log(1/\delta)}{\numobs},
\end{multline*}
where $c$ is a universal constant.
\end{lemma}
\begin{lemma}\cite[Proposition C.3]{lei2018regression}
\label{lemma:exp_matrix}
Consider a sequence $\{V_i \}_{i=1}^\numobs$ of independent
$d$-dimensional random matrices, with $\Exs [V_i] = 0$ for all
$i$. Then we have
\begin{align*}
\left( \Exs \opnorm{\sumn V_i}^2 \right)^{1/2} \le
C(\usedim)^{1/2}\opnorm{\sumn \Exs [V_i^2]}^{1/2} + C(\usedim) \left(
\Exs \max_{1 \le i \le \numobs} \opnorm{V_i}^2 \right)^{1/2},
\end{align*}
where $C(\usedim) = 4 (2 + \log \usedim)$. Then
\end{lemma}

\vspace*{0.1in}

Using these two lemmas, we proceed with the proof
of~\Cref{lem:denominator}. Define the random matrices $V_i =
\covariate_i \covariate_i^\top (\fit_i - \propensity_{\fitsub})$, and
consider the variational representation
\begin{align*}
\opnorm{\frac{1}{\numobs}\sumn V_i } = \sup_{w \in \ball_2(1)}
\frac{1}{\numobs}\sumn w^\top V_i w.
\end{align*}
Defining the function $f_w(V) = w^\top V w$,
\Cref{lemma:functional_matrix} guarantees that, with probability
$1-\delta$, we have
\begin{multline}
\label{eq:functiona-bernstein-in-ols-cov-mat}  
\opnorm{\frac{1}{\numobs}\sumn V_i } \le \Exs
\left[\opnorm{\frac{1}{\numobs}\sumn V_i }\right] + c \Bigg(
\Exs\left[ \sup_{w \in \ball_2 (1)} \frac{1}{\numobs}\sumn
  (\inprod{\covariate_i}{w})^4 (\fit_i - \propensity_{\fitsub})^2
  \right] \frac{\log(1/\delta)}{\numobs} \Bigg)^{1/2} \\
+ c \sup_{w \in \ball_2 (1)} \vecnorm{f_w}{\infty}
\frac{\log(1/\delta)}{\numobs}.
\end{multline}
Now we bound the terms on the right-hand side of
equation~\eqref{eq:functiona-bernstein-in-ols-cov-mat}. First,
by~\Cref{lemma:exp_matrix}, we have
\begin{align*}
&\Exs\left[\opnorm{ \frac{1}{\numobs}\sumn V_i}\right] \le
  \Exs\left[\opnorm{ \frac{1}{\numobs}\sumn V_i}^2\right]^{1/2} \\
& \le \frac{C(\usedim)^{1/2}}{\numobs} \opnorm{\sumn \Exs \covariate_i
    \covariate_i^\top \covariate_i \covariate_i^\top
    (\fit_i-\propensity_{\fitsub})^2 }^{1/2}
  +\frac{C(\usedim)}{\numobs}\left(\Exs \max_{i = 1, \ldots, \numobs }
  \opnorm{\covariate_i\covariate_i^\top
    (\fit_i-\propensity_{\fitsub})}^2 \right)^{1/2} \\
& \le \frac{C(\usedim)^{1/2}}{\numobs} \opnorm{ \sumn
    \vecnorm{\covariate_i}{2}^2 \covariate_i \covariate_i^\top }^{1/2}
  \sqrt{\propensity_{\fitsub}(1-\propensity_{\fitsub})} +
  \frac{C(\usedim)}{\numobs}\left( \max_{1\le i \le
    \numobs}\vecnorm{\covariate_i}{2}^2 \right)^{1/2} \\
& \le C(\usedim)^{1/2} \sqrt{\LevScore
    \propensity_{\fitsub}(1-\propensity_{\fitsub})}+C(\usedim)
  \LevScore.
\end{align*}
As for the second term on the right-hand side of
equation~\eqref{eq:functiona-bernstein-in-ols-cov-mat}, we have
\begin{align*}
\Exs \left[\sup_{w \in \ball_2(1)} \frac{1}{\numobs} \sumn
  (\inprod{\covariate_i}{w})^4 (\fit_i -
  \propensity_{\fitsub})^2\right] \le \Exs\left[\sup_{w \in
    \ball_2(1)} \frac{1}{\numobs} \sumn \big( \inprod{\covariate_i}{w}
  \big)^2 \vecnorm{\covariate_i}{2}^2 \right] \leq \numobs \LevScore,
\end{align*}
and for the last term,
\begin{align*}
\vecnorm{f_w}{\infty} = \max_i \max_{w \in \ball_2(1)}
(\inprod{\covariate_i}{w})^2 = \max_i \vecnorm{\covariate_i}{2}^2 =
\numobs \LevScore.
\end{align*}
Putting together the pieces completes the proof.


\subsubsection{Proof of Lemma \ref{lem:numerator}}
\label{sec:proof-numerator}

Let $Z_i = \covariate_i \residual_i(1) (\fit_i-\propensity_{\fitsub})$ and
$f_w(z) = w^\top z$. By~\Cref{lemma:functional_matrix}, we have with
probability $1-\delta$,
\begin{multline}
\label{eq:functional-bernstein-in-ols-l2-norm-bound}  
\vecnorm{\frac{1}{\numobs}\sumn Z_i }{2} \le \Exs\left[ \sup_{w \in
    \ball_2 (1)} \abss{\frac{1}{\numobs}\sumn w^\top Z_i } \right]
\\
+ \sqrt{\Exs \sup_{w \in \ball_2 (1)} \numobs^{-1}\sumn (w^\top Z_i)^2
  \frac{\log(1/\delta)}{\numobs}} + \max_{i} \sup_{w \in \ball_2
  (1)}\abss{(w^\top \covariate_i) \residual_i(1)}
\frac{\log(1/\delta)}{\numobs}.
\end{multline}

Now we bound the terms on the right hand side of
equation~\eqref{eq:functional-bernstein-in-ols-l2-norm-bound}. First,
by Jensen's inequality, we have
\begin{multline*}
\Exs\Big[ \vecnorm{\frac{1}{\numobs}\sumn Z_i}{2} \Big] \leq
\sqrt{\Exs\Big[ \vecnorm{\frac{1}{\numobs}\sumn Z_i }{2}^2 \Big]} =
\sqrt{\numobs^{-2} \sumn \Exs [\vecnorm{Z_i}{2}^2]} \\ = \numobs^{-1}
\Big(\sum_{i = 1}^\numobs \vecnorm{\covariate_i}{2}^2 \residual_i(1)^2
\propensity_{\fitsub}(1-\propensity_{\fitsub}) \Big)^{1/2} \leq \sqrt{
  \LevScore \myboundmax{2}^2
  \propensity_{\fitsub}(1-\propensity_{\fitsub})}.
\end{multline*}
For the second term on the right hand side of
equation~\eqref{eq:functional-bernstein-in-ols-l2-norm-bound}, we have
\begin{align*}
\Exs \Big[ \sup_{w \in \ball_2 (1)} \numobs^{-1}\sumn (w^\top Z_i)^2
  \Big] \leq \Exs \Big[ \numobs^{-1}\sumn \vecnorm{Z_i}{2}^2 \Big]
\leq n\LevScore \myboundmax{2}^2
\propensity_{\fitsub}(1-\propensity_{\fitsub}).
\end{align*}
For the last term, we have
\begin{align*}
\max_{i} \abss{(w^\top \covariate_i) \residual_i(1)} \le \max_i
\vecnorm{\covariate_i}{2} \mybound{\infty} \le \sqrt{\numobs
  \LevScore} \mybound{\infty}.
\end{align*}
Putting together the pieces completes the proof
of~\Cref{lem:numerator}.


\subsection{Proof of~\Cref{CorLasso}}
\label{SecProofCorLasso}

We now turn the proof of our result on regression adjustment based on
sparse linear regression.  We begin by defining the restricted
eigenvalue condition.  For any index set $\suppset \subseteq \{1,2,
\cdots, \usedim\}$, define the cone
\begin{align*}
\cone_\alpha (\suppset) \mydefn \Big\{ \theta:
\vecnorm{\theta_{\suppset^C}}{1} \leq 3 \vecnorm{\theta_\suppset}{1}
\Big\}.
\end{align*}
We say that $\Xmat \in \real^{\numobs \times \usedim}$ satisfies the
$\resEigen$-RE condition if
\begin{align}
\label{EqnRE}
\vecnorm{\Xmat \theta}{\numobs}^2 \geq \resEigen
\vecnorm{\theta}{2}^2,& \quad \mbox{for all $\theta \in
  \cone_3(\suppset)$.}
\end{align}

\begin{lemma}
  \label{LemLassoErr}
Under the conditions of~\Cref{CorLasso}, we have
\begin{align}
\label{EqnLassoErr}
  \vecnorm{\Xmat (\betahat_{\fitsub} - \betastar)}{\numobs} \leq
  \frac{9 \covariate_\infty \outcome_\infty}{\propensity_{\fitsub}}
  \sqrt{\frac{k \log (\usedim / \delta)}{\resEigen \numobs}} + 12
  \regu_\numobs \sqrt{\frac{\sparse}{\resEigen}}
\end{align}
with probability at least $1 - \delta$.
\end{lemma}

\subsubsection{Proof of~\Cref{LemLassoErr}}

By the optimality condition of the convex
program~\eqref{eq:lasso-program-modified}, we have
\begin{align}
\label{eq:optimality-condition-lasso}  
\frac{1}{2 \numobs_{\fitsub}} \sum_{i = 1}^\numobs \fit_i \big(
\inprod{\covariate_i}{\betahat_{\fitsub}} - \outcome_i \big)^2 +
\regu_\numobs \vecnorm{\betahat_{\fitsub}}{1} \leq \frac{1}{2
  \numobs_{\fitsub}} \sum_{i = 1}^\numobs \fit_i \big(
\inprod{\covariate_i}{\betastar} - \outcome_i \big)^2 + \regu_\numobs
\vecnorm{\betastar}{1}.
\end{align}
Introducing the shorthand notation \mbox{$\SigHat_{\fitsub} \mydefn
  \frac{1}{\numobs_{\fitsub}} \sum_{i = 1}^\numobs \covariate_i
  \covariate_i^\top \fit_i$} and \mbox{$\Delhat_{\fitsub} \mydefn
  \betahat_{\fitsub} - \betastar$,} the optimality
condition~\eqref{eq:optimality-condition-lasso} can be equivalently
re-written as
\begin{align}
\label{eq:basic-ineq-lasso}  
  0 \leq \frac{1}{2} \Delhat_{\fitsub}^\top \SigHat_\fitsub
  \Delhat_{\fitsub} \leq \frac{1}{\numobs_{\fitsub}} \sum_{i =
    1}^\numobs \fit_i(\outcome_i - \inprod{\covariate_i}{\betastar})
  \inprod{\covariate_i}{\Delhat_{\fitsub}} + \regu_\numobs \big(
  \vecnorm{\betastar}{1} - \vecnorm{\betastar + \Delhat_{\fitsub}}{1}
  \big).
\end{align}
Since the vector $\betastar$ is supported on the set $\suppset$, we
have
\begin{align*}
\vecnorm{\betastar}{1} - \vecnorm{\betastar + \Delhat_{\fitsub}}{1} =
\vecnorm{\betastar_{\suppset}}{1} - \vecnorm{\betastar_\suppset +
  \Delhat_{\fitsub, \suppset}}{1} - \vecnorm{\Delhat_{\fitsub,
    \suppset^C}}{1} \leq \vecnorm{\Delhat_{\fitsub, \suppset}}{1} -
\vecnorm{\Delhat_{\fitsub, \suppset^C}}{1}.
\end{align*}
We therefore have the $\ell_1$-norm bound
\begin{align}
\label{eq:l1-basic-ineq-lasso}  
\vecnorm{\Delhat_{\fitsub}}{1} \leq 2 \vecnorm{\Delhat_{\fitsub,
    \suppset}}{1} + \frac{1}{\regu_\numobs \numobs
  \propensity_{\fitsub}}\vecnorm{\Delhat_{\fitsub}}{1} \cdot \vecnorm{
  \sum_{i = 1}^\numobs \fit_i(\outcome_i -
  \inprod{\covariate_i}{\betastar}) \covariate_i}{\infty}.
\end{align}
In order to control the $\vecnorm{\cdot}{\infty}$ norm term, we use
the following lemma.
\begin{lemma}
\label{lemma:linfty-conc-in-lasso-basic-ineq}
Under the conditions of~\Cref{CorLasso}, for any $\delta \in (0, 1)$,
and $\propensity_{\fitsub} \geq \log (\usedim /\delta)/{\numobs}$, we
have
\begin{align}
\label{eq:lemma-linfty-conc-in-lasso-basic-ineq-bound}  
\vecnorm{ \frac{1}{\numobs \propensity_{\fitsub}} \sum_{i = 1}^\numobs
  \fit_i(\outcome_i - \inprod{\covariate_i}{\betastar})
  \covariate_i}{\infty} \leq \frac{1}{\numobs} \vecnorm{\Xmat^T (y -
  \Xmat \betastar)}{\infty} + 8 \covariate_\infty \outcome_\infty
\sqrt{\frac{\log (\usedim /\delta)}{\propensity_\fitsub\numobs}}
\end{align}
with probability $1 - \delta$.
\end{lemma}
\noindent
See~\Cref{subsubsec:proof-lemma-linfty-conc-in-lasso-basic-ineq} for
the proof of this lemma.

For a regularization parameter satisfying the lower
bound~\eqref{eq:regularization-req-lasso}, with probability $1 -
\delta$, we have
\begin{align*}
\frac{1}{\numobs \propensity_{\fitsub}}\vecnorm{\Delhat_{\fitsub}}{1}
\cdot \vecnorm{ \sum_{i = 1}^\numobs \fit_i(\outcome_i -
  \inprod{\covariate_i}{\betastar}) \covariate_i}{\infty} \leq
\frac{\regu_\numobs}{2}\vecnorm{\Delhat_{\fitsub}}{1},
\end{align*}
and consequently, equation~\eqref{eq:l1-basic-ineq-lasso} leads to the
bound
\begin{align}
\label{eq:lasso-cone-condition}  
\vecnorm{\Delhat_{\fitsub, \suppset^C}}{1} =
\vecnorm{\Delhat_{\fitsub}}{1} - \vecnorm{\Delhat_{\fitsub,
    \suppset}}{1} \leq 3 \vecnorm{\Delhat_{\fitsub, \suppset}}{1},
\end{align}
showing that $\Delhat_{\fitsub} \in \cone_3(\suppset)$.

Now we apply the basic inequality~\eqref{eq:basic-ineq-lasso} again
using the facts derived above. For notational convenience, we define
the following set for $r > 0$.
\begin{align*}
\funcClass(r) \mydefn \cone_3(\suppset) \cap \Big\{ \theta \in
\real^\usedim ~: ~\vecnorm{\Xmat \theta}{\numobs} \leq r, \quad
\mbox{and} \quad \max_{i \in [\numobs]}
|\inprod{\covariate_i}{\theta}| \leq 2 \outcome_\infty \Big\}.
\end{align*}
Since $|\inprod{\covariate_i}{\betahat_{\fitsub}}| \leq
\outcome_\infty$ and $|\inprod{\covariate_i}{\betastar}| \leq
\outcome_\infty$ for any $i \in [\numobs]$, we have $\Delhat_{\fitsub}
\in \funcClass (\vecnorm{\Xmat \Delhat_{\fitsub}}{\numobs})$.
Consequently, with probability $1 - \delta$, the basic inequality
implies that
\begin{multline}
\vecnorm{\Xmat \Delhat_{\fitsub}}{\numobs}^2 \leq \sup_{\Delta \in
  \funcClass (\vecnorm{X^\top \Delhat_{\fitsub}}{\numobs})} \Big\{
\vecnorm{\Xmat \Delta}{\numobs}^2 - \Delta^\top
\widehat{\Sigma}_{\fitsub} \Delta \Big\} +
2\vecnorm{\Delhat_{\fitsub}}{1} \cdot \big( \frac{1}{\numobs}
\vecnorm{\Xmat^T (y - \Xmat \betastar)}{\infty} + \regu_\numobs \big)
\\
\label{eq:basic-ineq-lasso-in-emp-proc-form}    
+ 2 \sup_{\Delta \in \funcClass (\vecnorm{X^\top
    \Delhat_{\fitsub}}{\numobs})} \Big\{ \frac{1}{\numobs} \inprod{(y
  - \Xmat \betastar)}{\Xmat \Delta} - \frac{1}{\numobs_{\fitsub}}
\sum_{i = 1}^\numobs \fit_i(\outcome_i -
\inprod{\covariate_i}{\betastar}) \inprod{\covariate_i}{\Delta}
\Big\}.
\end{multline}
For any $r > 0$, define the stochastic process suprema:
\begin{align*}
Z_1(r) & \mydefn \sup_{\Delta \in \funcClass(r)} \Big\{ \vecnorm{\Xmat
  \Delta}{\numobs}^2 - \Delta^\top \SigHat_{\fitsub} \Delta \Big\},
\quad \mbox{and} \\
Z_2(r) & \mydefn \sup_{\Delta \in \funcClass(r)} \Big\{
\frac{1}{\numobs} \inprod{(y - \Xmat\betastar)}{\Xmat \Delta} -
\frac{1}{\numobs_{\fitsub}} \sum_{i = 1}^\numobs \fit_i(\outcome_i -
\inprod{\covariate_i}{\betastar}) \inprod{\covariate_i}{\Delta}
\Big\}.
\end{align*}
Equation~\eqref{eq:basic-ineq-lasso-in-emp-proc-form} can be
re-written as
\begin{align*}
\vecnorm{\Xmat \Delhat_{\fitsub}}{\numobs}^2 \leq Z_1
\big(\vecnorm{\Xmat \Delhat_{\fitsub}}{\numobs} \big) + 2 Z_2
\big(\vecnorm{\Xmat \Delhat_{\fitsub}}{\numobs} \big) +
2\vecnorm{\Delhat_{\fitsub}}{1} \cdot \big( \frac{1}{\numobs}
\vecnorm{ \Xmat^T(y - \Xmat \betastar)}{\infty} + \regu_\numobs \big).
\end{align*}

The rest of this section is devoted to bounding various terms on the
right hand side of~\Cref{eq:basic-ineq-lasso-in-emp-proc-form}. For
the $\ell_1$-norm term, we note that
\begin{align*}
\vecnorm{\Delhat_{\fitsub}}{1} \leq 4 \vecnorm{\Delhat_{\fitsub,
    \suppset}}{1} \leq 4 \sqrt{\sparse} \vecnorm{\Delhat_{\fitsub,
    \suppset}}{2} \leq 4 \sqrt{\sparse} \vecnorm{\Delhat_{\fitsub}}{2}
& \leq 4 \sqrt{\frac{\sparse }{\resEigen}} \vecnorm{\Xmat
  \Delhat_{\fitsub}}{\numobs}, \quad \mbox{and} \\
\frac{1}{\numobs}\vecnorm{\Xmat^T (y - \Xmat \betastar)}{\infty} +
\regu_\numobs &\leq 3 \regu_\numobs.
\end{align*}
The following two lemmas provide upper bounds on the empirical process
suprema involved in
equation~\eqref{eq:basic-ineq-lasso-in-emp-proc-form}.
\begin{lemma}
\label{lemma:re-lasso}
Under the conditions of~\Cref{CorLasso}, for any $r > 0$,
we have
\begin{align}
\label{eq:lemma-re-lasso-bound}     
Z_1(r) \leq \frac{4 r \outcome_\infty}{\propensity_{\fitsub}} \Big\{2
\covariate_\infty \sqrt{\frac{k \log \usedim}{\resEigen \numobs}} +
\sqrt{\frac{\log (1 / \delta)}{\numobs}} \Big\}
\end{align}
with probability $1 - \delta$.
\end{lemma}
\noindent See~\Cref{subsubsec:proof-lemma-re-lasso} for the proof.\\

Note that for any $\theta \in \cone_3(\suppset)$, we have
\begin{align*}
\resEigen \vecnorm{\theta}{2}^2 \leq \frac{1}{\numobs} \sum_{i =
  1}^\numobs (\inprod{\covariate_i}{\theta})^2 \leq
\covariate_\infty^2 \vecnorm{\theta}{1}^2 \leq 16 \covariate_\infty^2
\vecnorm{\theta_{\suppset}}{1}^2 \leq 16 \sparse \covariate_\infty^2
\vecnorm{\theta_{\suppset}}{2}^2 \leq 16 \sparse \covariate_\infty^2
\vecnorm{\theta}{2}^2,
\end{align*}
which implies that $\resEigen \leq 16 \sparse \covariate_\infty^2$. So
the bound~\eqref{eq:lemma-re-lasso-bound} can be simplified as
\begin{align*}
Z_1(r) \leq \frac{24 r \covariate_\infty
  \outcome_\infty}{\propensity_{\fitsub}} \sqrt{\frac{k \log (\usedim
    / \delta)}{\resEigen \numobs}}.
\end{align*}

\begin{lemma} \label{lemma:noise-emp-proc-lasso}
Under the conditions of~\Cref{CorLasso}, for any $r > 0$, we have
\begin{align}
\label{eq:lemma-noise-emp-proc-lasso-bound}  
Z_2(r) \leq \frac{\covariate_\infty \outcome_\infty
  r}{\propensity_{\fitsub}} \sqrt{\frac{\sparse \log (\usedim
    /\delta)}{\resEigen \numobs}}
\end{align}
probability $1 - \delta$.
\end{lemma}
\noindent See~\Cref{subsubsec:proof-lemma-noise-emp-proc-lasso} for
the proof of this lemma.

Taking these lemmas as given, let us now finish the proof
of~\Cref{LemLassoErr}.  Note that the basic
inequality~\eqref{eq:basic-ineq-lasso-in-emp-proc-form} implies that
\begin{align}
\label{eq:rescaled-basic-ineq-lasso}  
1 \leq \frac{Z_1 \big( \vecnorm{\Xmat \Delhat_{\fitsub}}{\numobs}
  \big)}{ \vecnorm{\Xmat \Delhat_{\fitsub}}{\numobs}^2} + 2\frac{Z_2
  \big( \vecnorm{\Xmat \Delhat_{\fitsub}}{\numobs} \big)}{
  \vecnorm{\Xmat \Delhat_{\fitsub}}{\numobs}^2} + 24
\frac{\regu_\numobs}{ \vecnorm{\Xmat \Delhat_{\fitsub}}{\numobs}}
\sqrt{\frac{\sparse}{\resEigen}}
\end{align}
By definition, the functions $r \mapsto Z_1(r)/r^2$ and $r \mapsto
Z_2(r)/r^2$ are both non-increasing functions of $r$. So the right
hand side of equation~\eqref{eq:rescaled-basic-ineq-lasso} is a
strictly decreasing function of $\vecnorm{\Xmat
  \Delhat_{\fitsub}}{\numobs}$. Define the scalar
\begin{align*}
r_\numobs \mydefn \frac{9 \covariate_\infty
  \outcome_\infty}{\propensity_{\fitsub}} \sqrt{\frac{k \log (\usedim
    / \delta)}{\resEigen \numobs}} + 24 \regu_\numobs
\sqrt{\frac{\sparse}{\resEigen}}.
\end{align*}
Define the events
\begin{align*}
\Event & \mydefn \Big\{
\mbox{equation~\eqref{eq:lemma-linfty-conc-in-lasso-basic-ineq-bound}
  holds, equations~\eqref{eq:lemma-re-lasso-bound},
  and~\eqref{eq:lemma-noise-emp-proc-lasso-bound} hold with $r =
  r_\numobs$} \Big\}, \quad \mbox{and} \\
\Event' & \mydefn \big\{ \vecnorm{\Delhat_{\fitsub}}{\numobs} >
r_\numobs \big\}.
\end{align*}
On the event $\Event \cap \Event'$, we have
\begin{multline*}
1 \leq \frac{Z_1 \big( \vecnorm{\Xmat \Delhat_{\fitsub}}{\numobs}
  \big)}{ \vecnorm{\Xmat \Delhat_{\fitsub}}{\numobs}^2} + 2\frac{Z_2
  \big( \vecnorm{\Xmat \Delhat_{\fitsub}}{\numobs} \big)}{
  \vecnorm{\Xmat \Delhat_{\fitsub}}{\numobs}^2} + 24
\frac{\regu_\numobs}{ \vecnorm{\Xmat \Delhat_{\fitsub}}{\numobs}}
\sqrt{\frac{\sparse}{\resEigen}} \\
< \frac{Z_1 (r_\numobs)}{r_\numobs^2} +2 \frac{Z_2(r_\numobs)}{
  r_\numobs^2} + 24 \frac{\regu_\numobs}{r_\numobs}
\sqrt{\frac{\sparse}{\resEigen}} \leq \frac{1}{r_\numobs}
\left(\frac{26 \covariate_\infty
  \outcome_\infty}{\propensity_{\fitsub}} \sqrt{\frac{k \log (\usedim
    / \delta)}{\resEigen \numobs}} + 24 \regu_\numobs
\sqrt{\frac{\sparse}{\resEigen}} \right) < 1,
\end{multline*}
which leads to
contradiction. Since~\Cref{lemma:linfty-conc-in-lasso-basic-ineq,lemma:re-lasso,lemma:noise-emp-proc-lasso}
imply that $\Prob(\Event) \geq 1 - \delta$, we have
\begin{align*}
    \vecnorm{\Xmat (\betahat_{\fitsub} - \betastar)}{\numobs} \leq r_\numobs,
\end{align*}
with probability $1 - \delta$, which completes the proof
of~\Cref{LemLassoErr}.


\subsubsection{Proof of Lemma~\ref{lemma:linfty-conc-in-lasso-basic-ineq}}
\label{subsubsec:proof-lemma-linfty-conc-in-lasso-basic-ineq}

Starting with the variational representation
\begin{align*}
\vecnorm{ \frac{1}{\numobs \propensity_{\fitsub}} \sum_{i = 1}^\numobs
  \fit_i(\outcome_i - \inprod{\covariate_i}{\betastar})
  \covariate_i}{\infty} = \max_{z \in \{\pm \coordinate_j: j \in
  [\usedim]\}} \frac{1}{\numobs \propensity_{\fitsub}} \sum_{i =
  1}^\numobs \fit_i(\outcome_i - \inprod{\covariate_i}{\betastar})
\inprod{\covariate_i}{z},
\end{align*}
we note that for each $z$, the terms in the summation are independent,
satisfying $\abss{\fit_i(\outcome_i - \inprod{\covariate_i}{
    \betastar}) \inprod{\covariate_i}{z}} \leq \covariate_\infty
\outcome_\infty$ almost surely. Applying Bernstein's inequality yields
\begin{align*}
\abss{\frac{1}{\numobs \propensity_{\fitsub}} \sum_{i = 1}^\numobs
  \fit_i(\outcome_i - \inprod{\covariate_i}{\betastar})
  \inprod{\covariate_i}{z} - \frac{1}{\numobs} (y - \Xmat
  \betastar)\Xmat z} \leq 2 \covariate_\infty \outcome_\infty \left\{
\sqrt{\frac{\log (1 / \delta)}{\propensity_{\fitsub} \numobs} } +
\frac{\log (1 / \delta)}{\propensity_{\fitsub} \numobs} \right\},
\end{align*}
with probability $1 - \delta$, where $c > 0$ is a universal constant.

Taking the union bound over $2 \usedim$ possible choices of the vector
$z$, we arrive at the bound
\begin{multline*}
\max_{z \in \{\pm \coordinate_j: j \in [\usedim]\}} \frac{1}{\numobs
  \propensity_{\fitsub}} \sum_{i = 1}^\numobs \fit_i(\outcome_i -
\covariate_i^\top \betastar) \inprod{\covariate_i}{z} \\
\leq \max_{z \in \{\pm \coordinate_j: j \in [\usedim]\}}
\frac{1}{\numobs} (y - \Xmat \betastar)\Xmat z + 2 \covariate_\infty
\outcome_\infty \left \{ \sqrt{\frac{\log (\usedim /
    \delta)}{\propensity_{\fitsub} \numobs} } + \frac{\log (\usedim /
  \delta)}{\propensity_{\fitsub} \numobs} \right \},
\end{multline*}
with probability $1 - \delta$. Taking $\propensity_{\fitsub} \geq
{\log (\usedim / \delta)}/{\numobs}$, we complete the proof of this
lemma.

\subsubsection{Proof of Lemma~\ref{lemma:re-lasso}}
\label{subsubsec:proof-lemma-re-lasso}

For any $h \in \funcClass(r)$, the terms $(\fit_i
\inprod{\covariate_i}{h})^2$ are independent random variables,
satisfying the almost-sure bound
\begin{align*}
  (\fit_i \inprod{\covariate_i}{h})^2 \leq 2 \outcome_\infty \cdot
  |\inprod{\covariate_i}{h}|,
\end{align*}
so that the summation satisfies
\begin{align*}
\sum_{i = 1}^\numobs (\fit_i \inprod{\covariate_i}{h})^4 \leq \sum_{i
  = 1}^\numobs \big( 2 \outcome_\infty \cdot |
\inprod{h}{\covariate_i}| \big)^2 = 4 \numobs \outcome_\infty^2
\vecnorm{\Xmat h}{\numobs}^2 \leq 4 \numobs \outcome_\infty^2 r^2,
\quad \mbox{for any $h \in \funcClass(r)$.}
\end{align*}
Invoking the functional Hoeffding inequality
(c.f.~\cite{wainwright2019high}, Theorem 3.7), we note that
\begin{align}
\label{eq:emp-proc-sup-conc-in-lasso}  
\Prob \Big[ Z_1(r) - \Exs [Z_1(r)] \geq t \Big] \leq \exp \left( -
\frac{\numobs \propensity_{\fitsub}^2 t^2}{16 r^2 \outcome_\infty^2}
\right), \quad \mbox{for any $t > 0$}.
\end{align}
It suffices to bound the expectation $\Exs [Z_1(r)]$. In doing so, we
define the symmetrized empirical process supremum
\begin{align*}
V_1(r) \mydefn \sup_{h \in \funcClass(r)} \frac{1}{\numobs
  \propensity_{\fitsub}} \sum_{i = 1}^\numobs \varepsilon_i \cdot
(\fit_i \inprod{\covariate_i}{h})^2,
\end{align*}
where $(\varepsilon_i)_{i = 1}^\numobs$ are $\mathrm{i.i.d.}$
Rademacher random variables independent of $(\fit_i)_{i =
  1}^\numobs$. Standard symmetrization arguments yield \mbox{$\Exs
  [Z_1(r)] \leq 2 \Exs [V_1(r)]$.}  Applying the contraction principle
(\cite{van1996weak}, Proposition A.1.10) to $V_1(r)$, we find that
\begin{align*}
\Exs[V_1(r)] \leq 2 \outcome_\infty \Exs \Big[ \sup_{h \in
    \funcClass(r)} \frac{1}{\numobs \propensity_{\fitsub}} \sum_{i =
    1}^\numobs \varepsilon_i \fit_i \inprod{\covariate_i}{h} \big].
\end{align*}
For any $h \in \funcClass(r)$, we note that
\begin{align*}
  \vecnorm{h}{1} \leq 4 \vecnorm{h_\suppset}{1} \leq 4 \sqrt{\sparse}
  \vecnorm{h_\support}{2} \leq 4 \sqrt{\sparse} \vecnorm{h}{2} \leq 4
  r \sqrt{\frac{\sparse}{\resEigen}},
\end{align*}
which leads to the inclusion
\begin{align}
\label{eq:lasso-funcclass-inclusion}
\funcClass(r) \subseteq 4 r \sqrt{\frac{\sparse}{\resEigen}} \cdot
\big\{ h: \vecnorm{h}{1} \leq 1 \big\} = 4 r
\sqrt{\frac{\sparse}{\resEigen}} \cdot \mathrm{conv} \big(\{ \pm
\coordinate_j: j \in [\usedim]\} \big).
\end{align}
Consequently, we can bound the Rademacher complexity as
\begin{align*}
  \Exs \Big[ \sup_{h \in \funcClass(r)} \frac{1}{\numobs
      \propensity_{\fitsub}} \sum_{i = 1}^\numobs \varepsilon_i \fit_i
    \inprod{\covariate_i}{h} \big] \leq 4 r
  \sqrt{\frac{\sparse}{\resEigen}} \Exs \Big[ \sup_{h \in \{\pm
      \coordinate_j: j \in [\usedim]\}} \frac{1}{\numobs
      \propensity_{\fitsub}} \sum_{i = 1}^\numobs \varepsilon_i \fit_i
    \inprod{\covariate_i}{h} \big].
\end{align*}
For each vector $h \in \{\pm \coordinate_j: j \in [\usedim]\}$,
applying Hoeffding's inequality yields
\begin{align*}
  \Prob \left( \abss{\frac{1}{\numobs} \sum_{i = 1}^\numobs
    \varepsilon_i \fit_i \inprod{\covariate_i}{h} } \geq t \right)
  \leq 2 \exp \left( - \frac{2 t^2 \numobs}{\covariate_\infty^2}
  \right), \quad \mbox{for any $t > 0$}.
\end{align*}
Invoking union bound over $2 \usedim$ possible choices of the vector
$h$, and substituting into the Rademacher complexity bound above, we
have
\begin{align*}
  \Exs \Big[ \sup_{h \in \funcClass(r)} \frac{1}{\numobs
      \propensity_{\fitsub}} \sum_{i = 1}^\numobs \varepsilon_i \fit_i
    \inprod{\covariate_i}{h} \big] \leq \frac{4 r
    \covariate_\infty}{\propensity_{\fitsub}} \sqrt{\frac{\sparse \log
      \usedim}{\resEigen \numobs}}.
\end{align*}
Combining with equation~\eqref{eq:emp-proc-sup-conc-in-lasso}
completes the proof of~\Cref{lemma:re-lasso}.


\subsubsection{Proof of Lemma~\ref{lemma:noise-emp-proc-lasso}}
\label{subsubsec:proof-lemma-noise-emp-proc-lasso}

By the inclusion relation~\eqref{eq:lasso-funcclass-inclusion}, we
have the upper bound
\begin{align*}
Z_2(r) \leq 4 r \sqrt{\frac{\sparse}{\resEigen}} \sup_{h \in \{\pm
  \coordinate_j: j \in [\usedim]\}} \frac{1}{\numobs} \sum_{i =
  1}^\numobs (\outcome_i - \inprod{\covariate_i}{\betastar})
\inprod{\covariate_i}{h} \cdot \Big(
\frac{\fit_i}{\propensity_{\fitsub}} - 1 \Big).
\end{align*}
For each $h \in \{\pm \coordinate_j: j \in [\usedim]\}$, applying the
Hoeffding bound yields
\begin{align*}
\Prob \left[ \frac{1}{\numobs} \sum_{i = 1}^\numobs (\outcome_i -
  \inprod{\covariate_i}{\betastar}) \inprod{\covariate_i}{h} \cdot
  \Big( \frac{\fit_i}{\propensity_{\fitsub}} - 1 \Big) > t \right]
\leq \exp \left( - \frac{2 t^2 \propensity_{\fitsub}^2
  \numobs}{\covariate_\infty^2 \outcome_\infty^2} \right), \quad
\mbox{for any $t > 0$}.
\end{align*}
Taking the union bound over $2 \usedim$ possible choices, with
probability $1 - \delta$, we have
\begin{align*}
Z_2(r) \leq 4 r \sqrt{\frac{\sparse}{\resEigen}} \sup_{h \in \{\pm
  \coordinate_j: j \in [\usedim]\}} \frac{1}{\numobs} \sum_{i =
  1}^\numobs (\outcome_i - \inprod{\covariate_i}{\betastar})
\inprod{\covariate_i}{h} \cdot \Big(
\frac{\fit_i}{\propensity_{\fitsub}} - 1 \Big) \leq
\frac{\covariate_\infty \outcome_\infty r}{\propensity_{\fitsub}}
\sqrt{\frac{\sparse \log (\usedim /\delta)}{\resEigen \numobs}},
\end{align*}
which completes the proof.


\subsection{Proof of~\Cref{CorNonPara}}
\label{SecProofCorNonPara}

We prove this corollary by applying~\Cref{ThmMain} with
an appropriate bound on the function estimation error.

\begin{lemma}
\label{LemNonParaEstimation}  
Under the conditions of~\Cref{CorNonPara}, for any $\delta \in (0,
1)$, we have
\begin{align}
\label{eq:nonpar-estimation-err-bound}  
\vecnorm{\fhat_{\fitsub} - \fstar}{\numobs} \leq \radius_{\numobs,
  \delta}
\end{align}
with probability at least $1 - \delta$.
\end{lemma}
The statement of the corollary follows by combining this bound
with~\Cref{ThmMain}.

\subsubsection{Proof of~\Cref{LemNonParaEstimation}}

Note that we have $\fstar \in \funcClass$ and that the set $\funcClass$ is convex.
Consequently, by the first-order conditions for optimality in the
convex program~\eqref{eq:nonpar-regression}, we have
\begin{align*}
\frac{1}{\numobs_{\fitsub}} \sum_{i = 1}^\numobs \fit_i(\outcome_i -
\fhat_{\fitsub}(\covariate_i) )(\fhat_{\fitsub }(\covariate_i) -
\fstar(\covariate_i) ) \geq 0.
\end{align*}
On the other hand, since $\fhat_{\fitsub} \in \funcClass$, the
first-order optimality condition for the Euclidean
projection~\eqref{EqnOracleChoice} defining $\fstar$ implies that
\begin{align*}
 \frac{1}{\numobs} \sum_{i = 1}^\numobs (\outcome_i -
 \fstar(\covariate_i)) (\fstar(\covariate_i) -
 \fhat_{\fitsub}(\covariate_i)) \geq 0.
\end{align*}
Summing up these two inequalities yields the basic inequality
\begin{align}
\label{eq:basic-ineq-isotonic}  
\frac{1}{\numobs_{\fitsub}} \sum_{i = 1}^\numobs
\fit_i(\fhat_{\fitsub}(\covariate_i) - \fstar(\covariate_i))^2 \leq
\frac{1}{\numobs} \sum_{i = 1}^\numobs (\outcome_i -
\fstar(\covariate_i)) (\fstar(\covariate_i) -
\fhat_{\fitsub}(\covariate_i)) \cdot \Big(1 -
\tfrac{\fit_i}{\propensity_{\fitsub}}\Big).
\end{align}
For any $r > 0$, define the set $\funcClass(r) \mydefn \Big\{ f -
\fstar \mid f \in \funcClass, \vecnorm{f - \fstar}{\numobs} \leq r
\Big\} \subset \real^\numobs$.  Playing in a key role in our analysis
are the following empirical process suprema
\begin{align*}
  Z_1(r)  \mydefn \sup_{h \in \funcClass(r)} \frac{1}{\numobs}
  \sum_{i = 1}^\numobs (\outcome_i - \fstar(\covariate_i) )h_i \cdot
  \Big(1 - \tfrac{\fit_i}{\propensity_{\fitsub}}\Big) \quad
  \mbox{and} \quad
Z_2(r) \mydefn \sup_{h \in \funcClass(r)} \frac{1}{\numobs} \sum_{i =
  1}^\numobs h_i^2 \cdot \Big(1 -
\tfrac{\fit_i}{\propensity_{\fitsub}}\Big) ,
\end{align*}
where $h_i = h(\covariate_i)$. Introducing the shorthand
$\Delhat_{\fitsub} \mydefn \fhat_{\fitsub} - \fstar$, the basic
inequality~\eqref{eq:basic-ineq-isotonic} implies that
\begin{align}
\label{eq:basic-ineq-isotonic-rescaled}  
1 \leq \frac{Z_1(
  \vecnorm{\Delhat_{\fitsub}}{\numobs})}{\vecnorm{\Delhat_{\fitsub}}{\numobs}^2}
+
\frac{Z_2(\vecnorm{\Delhat_{\fitsub}}{\numobs})}{\vecnorm{\Delhat_{\fitsub}}{\numobs}^2}.
\end{align}
Since each term in the empirical process is uniformly bounded, we can
apply a functional Hoeffding inequality
(e.g.,~\cite{wainwright2019high}, Theorem 3.7) to assert that
for $j \in \{0,1 \}$, we have
\begin{align}
\label{eq:functional-hoeffding-in-isotonic}    
\Prob \big( |Z_j(r) - \Exs [Z_j(r)]| \geq t \big) & \leq \exp \Big(
\frac{- \numobs \propensity_{\fitsub}^2 t^2 }{ 16 r^2} \Big), \quad
\mbox{for any $t > 0$.}
\end{align}

The following result provides control on the expectations:
\begin{lemma}
\label{lemma:emp-proc-isotonic}
Under the conditions of~\Cref{CorNonPara}, for any $r > 0$ and $\gamma
\in [0, r]$, we have
\begin{align*}
\max \Big\{ \Exs [Z_1(r)], \Exs [Z_2(r)] \Big\} \leq \frac{4
  \gamma}{\propensity_\fitsub} + \frac{32}{\propensity_\fitsub
  \sqrt{\numobs}}\int_{\gamma / 4}^{2r} \sqrt{\log N (t, \funcClass
  (r))} dt.
\end{align*}
\end{lemma}
\noindent See~\Cref{subsubsec:proof-lemma-emp-proc-isotonic} for the
proof. \\

Note that the deterministic inequality in Lemma~\ref{lemma:emp-proc-isotonic} holds for any $\gamma \in
[0, r]$. So we can always take the infimum over $\gamma$ on the right
hand side.

Taking this lemma as given, we now proceed to complete the proof of
equation~\eqref{eq:nonpar-estimation-err-bound}. Using the quantity
$\radius_{\numobs, \delta}$ defined in
equation~\eqref{eq:critical-radius}, we consider the following events:
\begin{align*}
\Event & \mydefn \Big\{ Z_1 (\radius_{\numobs, \delta}) + Z_2
(\radius_{\numobs, \delta}) \leq \frac{8 \gamma}{\propensity_\fitsub}
+ \frac{64}{\propensity_\fitsub \sqrt{\numobs}} \int_{\gamma /
  4}^{2\radius_{\numobs, \delta}} \sqrt{\log N(t, \funcClass(r))} dt +
\frac{8 \radius_{\numobs,\delta}}{\propensity_{\fitsub}}
\sqrt{\frac{\log (1 / \delta)}{\numobs}} \Big \}, \\
\Event' & \mydefn \big\{ \vecnorm{\Delhat_{\fitsub}}{\numobs} >
\radius_{\numobs, \delta} \big\}.
\end{align*}
By~~\Cref{lemma:emp-proc-isotonic} and
equation~\eqref{eq:functional-hoeffding-in-isotonic}, we have
$\Prob(\Event) \geq 1 - \delta$. On the other hand, note that the
function $r \mapsto Z_1(r) / r^2, ~r \mapsto Z_2(r) / r^2$ are
non-increasing in $r > 0$, on the event $\Event \cap \Event'$, we have
\begin{multline*}
1 \leq
\frac{Z_1(\vecnorm{\Delhat_{\fitsub}}{\numobs})}{\vecnorm{\Delhat_{\fitsub}}{\numobs}^2}
+ \frac{Z_2
  (\vecnorm{\Delhat_{\fitsub}}{\numobs})}{\vecnorm{\Delhat_{\fitsub}}{\numobs}^2}
\leq \frac{Z_1 (\radius_{\numobs, \delta})}{\radius_{\numobs,
    \delta}^2} + \frac{Z_2 (\radius_{\numobs,
    \delta})}{\radius_{\numobs, \delta}^2} \\
\leq \frac{1}{r_{\numobs, \delta}^2} \Big\{ \frac{8
  \gamma}{\propensity_\fitsub} + \frac{64}{\propensity_\fitsub
  \sqrt{\numobs}}\int_{\gamma / 4}^{2\radius_{\numobs, \delta}}
\sqrt{\log N (t, \funcClass(r))} dt \Big \} +
\frac{8}{\propensity_{\fitsub} \radius_{\numobs,\delta}}
\sqrt{\frac{\log (1 / \delta)}{\numobs}} < 1,
\end{multline*}
leading to a contradiction. Consequently, we must have $\Prob(\Event')
\leq \delta$, which completes the proof
of~\Cref{LemNonParaEstimation}.

\subsubsection{Proof of Lemma~\ref{lemma:emp-proc-isotonic}}
\label{subsubsec:proof-lemma-emp-proc-isotonic}

Let $(\varepsilon_i)_{i = 1}^\numobs$ be $\mathrm{i.i.d.}$ Rademacher
random variables independent of $(\fit_i)_{i = 1}^\numobs$. Define the
empirical process supremum \mbox{$V(r) \mydefn \sup_{h \in
    \funcClass(r)} \frac{1}{\numobs} \sum_{i = 1}^\numobs h_i
  \varepsilon_i$.}  We note that
\begin{align*}
\Exs[ Z_1(r) ] &\overset{(i)}{\leq} \Exs \Big[ \sup_{h \in
    \funcClass(r)} \frac{1}{\numobs} \sum_{i = 1}^\numobs
  \varepsilon_i(\outcome_i - \fstar_i) h_i \cdot \Big(1 -
  \frac{\fit_i}{\propensity_{\fitsub}}\Big) \Big] \overset{(ii)}{\leq}
\frac{2}{\propensity_{\fitsub}} \Exs [V(r)], \quad \mbox{and} \\
\Exs [ Z_2(r) ] &\overset{(i')}{\leq} \Exs \Big[ \sup_{h \in
    \funcClass(r)} \frac{1}{\numobs} \sum_{i = 1}^\numobs
  \varepsilon_i h_i^2 \cdot \Big(1 -
  \frac{\fit_i}{\propensity_{\fitsub}}\Big) \Big]
\overset{(ii')}{\leq} \frac{1}{\propensity_{\fitsub}} \Exs [V(r)],
\end{align*}
where in steps (i) and (i'), we use the symmetrization lemma
(\cite{wainwright2019high}, Theorem 4.2), and in steps (ii) and (ii'),
we use the contraction principle (\cite{van1996weak}, Proposition
A.1.10).

It suffices to bound the quantity $\Exs [V(r)]$. In order to do so, we
invoke the Dudley entropy bound (e.g., Theorem 5.1 of
\cite{wainwright2019high}), thereby finding that
\begin{align*}
\Exs[V(r)] \leq 2 ~\Exs \big[ \sup_{f,g: \vecnorm{f - g}{\numobs} \leq
    \delta} \inprod{f - g}{\varepsilon}_\numobs \big] +
\frac{16}{\sqrt{\numobs}} \int_{\delta / 4}^{2r} \sqrt{\log N \big(t,
  \funcClass(r) \big)} dt.
\end{align*}
For the first term, invoking Cauchy--Schwarz inequality yields
\begin{align*}
\Exs \big[ \sup_{f,g: \vecnorm{f - g}{\numobs} \leq \delta} \inprod{f
    - g}{\varepsilon}_\numobs \big] \leq \Exs \big[ \sup_{f,g:
    \vecnorm{f - g}{\numobs} \leq \delta} \vecnorm{f - g}{\numobs}
  \cdot \vecnorm{\varepsilon}{\numobs} \big] \leq \delta \cdot
\sqrt{\Exs [\vecnorm{\varepsilon}{\numobs}^2}] = \delta.
\end{align*}
Substituting back completes the proof of this lemma.

\section{Additional simulation details}
\label{app:additional-simulation}

In this section, we describe the construction of the vector
$\varepsilon \in \real^\numobs$ in Lei and
Ding~\cite{lei2018regression}, which is used in our simulation studies
in~\Cref{subsec:simu-ols}.

Given a problem dimension satisfying $\usedim < \numobs$, consider the data matrix $\Xmat \in \real^{\numobs \times \usedim}$ and the intercept version $\Xmat^+ \in \real^{\numobs \times (\usedim + 1)}$ (the first column of $\Xmat^+$ is an $\numobs$-dimensional all-one vector). Assume that
the matrices $\Xmat^\top \Xmat / \numobs \in \real^{\usedim \times
  \usedim}$ and $(\Xmat^+)^\top \Xmat^+ / \numobs \in \real^{\usedim \times
  \usedim}$ are non-singular, we define $H \mydefn \Xmat (\Xmat^\top
\Xmat)^{-1} \Xmat \in \real^{\numobs \times \numobs}$ and $H^+ \mydefn \Xmat^+ ((\Xmat^+)^\top
\Xmat^+)^{-1} \Xmat^+ \in \real^{\numobs \times \numobs}$. Let $h \in
\real^\numobs$ be the diagonal elements of $H$. We define
\begin{align*}
\widetilde{\varepsilon} \mydefn (I_\numobs - H^+) h \quad \mbox{and}
\quad \varepsilon \mydefn \sqrt{\numobs} \widetilde{\varepsilon} /
\vecnorm{\widetilde\varepsilon}{2}.
\end{align*}
According to the paper~\cite{lei2018regression}, Appendix J.1, the
resulting vector $\varepsilon$ maximizes the leading-order bias in the
Taylor expansion, subject to norm and orthogonality constraints.


\end{document}

%% file: final_macros.tex
\def\ball{\mathbb{B}}

\newcommand{\real}{\ensuremath{\mathbb{R}}}

\newcommand{\order}[1]{\ensuremath{\mathcal{O}\parenth{#1}}}

\newcommand{\Xmat}{\ensuremath{X}}

\newcommand{\parenth}[1]{\left( #1 \right)}

\newcommand{\abss}[1]{\left| #1 \right |}

\newcommand{\tauoradc}{\widehat{\tau}_{\dc.\ora}}

\newcommand{\tauoradctextup}{\widehat{\tau}_{\mathrm{dc}.\ora}}

\newcommand{\mydefn}{\ensuremath{:=}}

\newcommand{\defn}{:=}

\newcommand{\matsnorm}[2]{|\!|\!| #1 | \! | \!|_{{#2}}}
\newcommand{\vecnorm}[2]{\| #1\|_{#2}}

\newcommand{\opnorm}[1]{\ensuremath{\matsnorm{#1}{\tiny{\mbox{\textup{op}}}}}}

\newcommand{\inprod}[2]{\ensuremath{\langle #1 , \, #2 \rangle}}

\newcommand{\Exs}{\ensuremath{{\mathbb{E}}}}
\newcommand{\Prob}{\ensuremath{{\mathbb{P}}}}

\DeclareMathOperator{\cov}{cov}

\newtheoremstyle{named}{}{}{\itshape}{}{\bfseries}{.}{.5em}{\thmnote{#3's }#1}
\theoremstyle{named}

\theoremstyle{plain}

\newlength{\widebarargwidth}
\newlength{\widebarargheight}
\newlength{\widebarargdepth}

\makeatletter
\long\def\@makecaption#1#2{
        \vskip 0.8ex
        \setbox\@tempboxa\hbox{\small {\bf #1:} #2}
        \parindent 1.5em  
        \dimen0=\hsize
        \advance\dimen0 by -3em
        \ifdim \wd\@tempboxa >\dimen0
                \hbox to \hsize{
                        \parindent 0em
                        \hfil
                        \parbox{\dimen0}{\def\baselinestretch{0.96}\small
                                {\bf #1.} #2
                                }
                        \hfil}
        \else \hbox to \hsize{\hfil \box\@tempboxa \hfil}
        \fi
        }
\makeatother

\long\def\comment#1{}
\definecolor{battleshipgrey}{rgb}{0.52, 0.52, 0.51}
\definecolor{darkgray}{rgb}{0.66, 0.66, 0.66}
\definecolor{darkgreen}{rgb}{0.0, 0.2, 0.13}
\definecolor{darkspringgreen}{rgb}{0.09, 0.45, 0.27}
\definecolor{dukeblue}{rgb}{0.0, 0.0, 0.61}
\definecolor{olivedrab7}{rgb}{0.24, 0.2, 0.12}
\definecolor{darkblue}{rgb}{0.0, 0.0, 0.55}
\definecolor{darkscarlet}{rgb}{0.34, 0.01, 0.1}
\definecolor{candyapplered}{rgb}{1.0, 0.03, 0.0}
\definecolor{ao(english)}{rgb}{0.0, 0.5, 0.0}
\definecolor{applegreen}{rgb}{0.55, 0.71, 0.0}

\newcommand{\widgraph}[2]{\includegraphics[keepaspectratio,width=#1]{#2}}

\newcommand{\DIM}{\mathrm{DIM}}

\newcommand{\adj}{\mathrm{adj}}

\newcommand{\ora}{\mathrm{oracle}}

\newcommand{\dc}{\mathrm{dc}}

\newcommand{\fitsub}{\textup{\fit}}

\newcommand{\barfitsub}{{\bar{\textup{\fit}}}}

\newcommand{\smeansub}{\mathrm{\smean}}

\newcommand{\barsmeansub}{{\bar{\textup{\smean}}}}

\newcommand{\treatsub}{1}
\newcommand{\controlsub}{0}

%% file: regression_formal.bbl
\newcommand{\etalchar}[1]{$^{#1}$}
\begin{thebibliography}{MDWB23}

\bibitem[BHMM19]{belkin2019reconciling}
M.~Belkin, D.~Hsu, S.~Ma, and S.~Mandal.
\newblock Reconciling modern machine-learning practice and the classical
  bias--variance trade-off.
\newblock {\em Proceedings of the National Academy of Sciences},
  116(32):15849--15854, 2019.

\bibitem[BJK15]{bertsimas2015power}
D.~Bertsimas, M.~Johnson, and N.~Kallus.
\newblock The power of optimization over randomization in designing experiments
  involving small samples.
\newblock {\em Operations Research}, 63(4):868--876, 2015.

\bibitem[BLLT20]{bartlett2020benign}
P.~L. Bartlett, P.~M. Long, G.~Lugosi, and A.~Tsigler.
\newblock Benign overfitting in linear regression.
\newblock {\em Proceedings of the National Academy of Sciences},
  117(48):30063--30070, 2020.

\bibitem[BLZ{\etalchar{+}}16]{bloniarz2015lasso}
A.~Bloniarz, H.~Liu, C.~H. Zhang, J.~Sekhon, and B.~Yu.
\newblock Lasso adjustments of treatment effect estimates in randomized
  experiments.
\newblock {\em Proceedings of the National Academy of Sciences of the United
  States of America}, 113:7383--7390, 2016.

\bibitem[BRT19]{belkin2019does}
M.~Belkin, A.~Rakhlin, and A.~B. Tsybakov.
\newblock Does data interpolation contradict statistical optimality?
\newblock In {\em The 22nd International Conference on Artificial Intelligence
  and Statistics}, pages 1611--1619. PMLR, 2019.

\bibitem[CCD{\etalchar{+}}18]{chernozhukov2018double}
V.~Chernozhukov, D.~Chetverikov, M.~Demirer, E.~Duflo, C.~Hansen, W.~Newey, and
  J.~Robins.
\newblock Double/debiased machine learning for treatment and structural
  parameters.
\newblock {\em Econometrics Journal}, 21:C1--C68, 2018.

\bibitem[CF23]{cohen2023no}
P.~L. Cohen and C.~B. Fogarty.
\newblock No-harm calibration for generalized oaxaca--blinder estimators.
\newblock {\em Biometrika}, page asad036, 2023.

\bibitem[CMA21]{chang2021exact}
H.~Chang, J.~Middleton, and P.~M. Aronow.
\newblock Exact bias correction for linear adjustment of randomized controlled
  trials.
\newblock {\em arXiv preprint arXiv:2110.08425}, 2021.

\bibitem[CSW76]{cassel1976some}
C.~M. Cassel, C.~E. S{\"a}rndal, and J.~H. Wretman.
\newblock Some results on generalized difference estimation and generalized
  regression estimation for finite populations.
\newblock {\em Biometrika}, 63(3):615--620, 1976.

\bibitem[Fis35]{Fisher:1935}
R.~A. Fisher.
\newblock {\em The {D}esign of {E}xperiments}.
\newblock Edinburgh, London: Oliver and Boyd, 1st edition, 1935.

\bibitem[FLPZ22]{farias2022markovian}
V.~Farias, A.~Li, T.~Peng, and A.~Zheng.
\newblock Markovian interference in experiments.
\newblock {\em Advances in Neural Information Processing Systems}, 35:535--549,
  2022.

\bibitem[Fre08]{freedman2008regression_b}
D.~A. Freedman.
\newblock On regression adjustments to experimental data.
\newblock {\em Advances in Applied Mathematics}, 40:180--193, 2008.

\bibitem[GB21]{guo2021generalized}
K.~Guo and G.~Basse.
\newblock The generalized {O}axaca-{B}linder estimator.
\newblock {\em Journal of the American Statistical Association}, pages 1--13,
  2021.

\bibitem[GS12]{guntuboyina2012covering}
A.~Guntuboyina and B.~Sen.
\newblock Covering numbers for convex functions.
\newblock {\em IEEE Transactions on Information Theory}, 59(4):1957--1965,
  2012.

\bibitem[HHZ{\etalchar{+}}21]{hadad2021confidence}
V.~Hadad, D.~A. Hirshberg, R.~Zhan, S.~Wager, and S.~Athey.
\newblock Confidence intervals for policy evaluation in adaptive experiments.
\newblock {\em Proceedings of the National Academy of Sciences},
  118(15):e2014602118, 2021.

\bibitem[IR15]{imbens2015causal}
G.~W. Imbens and D.~B. Rubin.
\newblock {\em Causal Inference for Statistics, Social, and Biomedical
  Sciences: An Introduction}.
\newblock Cambridge: Cambridge University Press, 2015.

\bibitem[LD17]{li2017general}
X.~Li and P.~Ding.
\newblock General forms of finite population central limit theorems with
  applications to causal inference.
\newblock {\em Journal of the American Statistical Association},
  112:1759--1769, 2017.

\bibitem[LD19]{li2019rerandomization}
X.~Li and P.~Ding.
\newblock Rerandomization and regression adjustment.
\newblock {\em arXiv preprint arXiv:1906.11291}, 2019.

\bibitem[LD21]{lei2018regression}
L.~Lei and P.~Ding.
\newblock Regression adjustment in completely randomized experiments with a
  diverging number of covariates.
\newblock {\em Biometrika}, 108(4):815--828, 2021.

\bibitem[Lin13]{lin2013}
W.~Lin.
\newblock {Agnostic notes on regression adjustments to experimental data:
  Reexamining Freedman's critique}.
\newblock {\em Annals of Applied Statistics}, 7:295--318, 2013.

\bibitem[LMS22]{list2022using}
J.~List, I.~Muir, and G.~Sun.
\newblock Using machine learning for efficient flexible regression adjustment
  in economic experiments.
\newblock Technical report, The Field Experiments Website, 2022.

\bibitem[LYW23]{lu2023debiased}
X.~Lu, F.~Yang, and Y.~Wang.
\newblock Debiased regression adjustment in completely randomized experiments
  with moderately high-dimensional covariates.
\newblock {\em arXiv preprint arXiv:2309.02073}, 2023.

\bibitem[MDWB23]{mou2023kernel}
W.~Mou, P.~Ding, M.~J. Wainwright, and P.~L. Bartlett.
\newblock Kernel-based off-policy estimation without overlap: Instance
  optimality beyond semiparametric efficiency.
\newblock {\em arXiv preprint arXiv:2301.06240}, 2023.

\bibitem[MW85]{mackinnon1985some}
J.~G. MacKinnon and H.~White.
\newblock Some heteroskedasticity-consistent covariance matrix estimators with
  improved finite sample properties.
\newblock {\em Journal of Econometrics}, 29:305--325, 1985.

\bibitem[Ney23]{Neyman:1923}
J.~Neyman.
\newblock On the application of probability theory to agricultural experiments.
  essay on principles (with discussion). section 9 (translated). reprinted ed.
\newblock {\em Statistical Science}, 5:465--472, 1923.

\bibitem[SD21]{Su2020cluster}
F.~Su and P.~Ding.
\newblock Model-assisted analyses of cluster-randomized experiments.
\newblock {\em Journal of the Royal Statistical Society Series B: Statistical
  Methodology}, 83(5):994--1015, 2021.

\bibitem[Tsy08]{tsybakov2008introduction}
A.~B. Tsybakov.
\newblock {\em Introduction to Nonparametric Estimation}.
\newblock Springer Science \& Business Media, 2008.

\bibitem[vdVW96]{van1996weak}
A.~van~der Vaart and J.~Wellner.
\newblock {\em Weak convergence and empirical processes: with applications to
  statistics}.
\newblock Springer Science \& Business Media, 1996.

\bibitem[Wai19]{wainwright2019high}
M.~J. Wainwright.
\newblock {\em High-dimensional Statistics: A Non-asymptotic Viewpoint},
  volume~48.
\newblock Cambridge University Press, 2019.

\bibitem[WDTT16]{wager2016high}
S.~Wager, W.~Du, J.~Taylor, and R.~J. Tibshirani.
\newblock High-dimensional regression adjustments in randomized experiments.
\newblock {\em Proceedings of the National Academy of Sciences of the United
  States of America}, 113:12673--12678, 2016.

\bibitem[WGB18]{wu2018loop}
E.~Wu and J.~A. Gagnon-Bartsch.
\newblock {The LOOP estimator: Adjusting for covariates in randomized
  experiments}.
\newblock {\em Evaluation Review}, 42:458--488, 2018.

\bibitem[WGB21]{wu2021design}
E.~Wu and J.~A. Gagnon-Bartsch.
\newblock Design-based covariate adjustments in paired experiments.
\newblock {\em Journal of Educational and Behavioral Statistics},
  46(1):109--132, 2021.

\bibitem[WS20]{wang2020debiased}
Y.~Wang and R.~D. Shah.
\newblock Debiased inverse propensity score weighting for estimation of average
  treatment effects with high-dimensional confounders.
\newblock {\em arXiv preprint arXiv:2011.08661}, 2020.

\bibitem[ZBH{\etalchar{+}}21]{zhang2021understanding}
C.~Zhang, S.~Bengio, M.~Hardt, B.~Recht, and O.~Vinyals.
\newblock Understanding deep learning (still) requires rethinking
  generalization.
\newblock {\em Communications of the ACM}, 64(3):107--115, 2021.

\end{thebibliography}
